\documentclass[12pt,a4paper,oneside]{article}
\usepackage[top=3cm, bottom=3cm, left=2cm, right=2cm]{geometry}
\usepackage{setspace} 
\usepackage{geometry, booktabs, subcaption}
\usepackage{titlesec}
\usepackage[T1]{fontenc}
\usepackage[utf8]{inputenc}
\usepackage{authblk}
\usepackage{amssymb,amsmath}
\usepackage{graphics}

\RequirePackage{amsthm,amsmath,amsfonts,mathrsfs,amssymb}
\RequirePackage[authoryear]{natbib}
\RequirePackage[colorlinks,citecolor=blue,urlcolor=blue,linkcolor=blue]{hyperref}
\usepackage{cleveref}
\RequirePackage{graphicx}

\usepackage{latexsym}
\usepackage{comment}
\usepackage[inline]{enumitem}
\usepackage{graphicx}
\usepackage{placeins}
\usepackage{bbm}
\usepackage{bm}
\usepackage{subcaption}
\usepackage{soul}

\usepackage{xcolor}
\hypersetup{
    colorlinks=true,
    linkcolor=blue,
    citecolor=blue,
    urlcolor=blue
}

\usepackage{tikz}

\theoremstyle{thmstyleone}%

\theoremstyle{thmstyletwo}%
\theoremstyle{thmstylethree}%

\renewcommand{\mid}{\ensuremath{\,|\,}}

\def\simind{\stackrel{\mbox{\scriptsize{\rm ind}}}{\sim}}
\def\simiid{\stackrel{\mbox{\scriptsize{\rm iid}}}{\sim}}

\usepackage[normalem]{ulem}

\usepackage{amsthm, dsfont, bm, bbm, placeins}
\usepackage{bibunits}
\usepackage{float}

\def\simind{\stackrel{\mbox{\scriptsize{ind}}}{\sim}}
\def\simiid{\stackrel{\mbox{\scriptsize{iid}}}{\sim}}
\newcommand{\ddr}{\mathrm{d}}
\newcommand{\prob}{\mathrm{Pr}}

\allowdisplaybreaks  

\newtheorem{theorem}{Theorem}

\newtheorem{definition}{Definition}
\newtheorem{remark}{Remark}
\newtheorem{property}{Property}

\newtheorem{lemma}{Lemma}

\newtheorem{myalgorithm}{Algorithm}

\usepackage[normalem]{ulem}

\title{Repulsive Mixture Model with Projection Determinantal Point Process}

\author[1]{Ziyi Song\thanks{Ziyi Song completed part of this work while visiting the Department of Economics, Management and Statistics at the University of Milano-Bicocca, Milano, Italy. He gratefully acknowledges the kind support received during his visit and ever since.}}
\author[2]{Federico Camerlenghi}
\author[1]{Weining Shen}
\author[3]{Michele Guindani}
\author[2]{Mario Beraha}
\affil[1]{\normalsize{Department of Statistics, University of California, Irvine, CA 92697, US}}
\affil[2]{\normalsize{Department of Economics, Management and Statistics, University of Milano--Bicocca, 20126 Milano, Italy}}
\affil[3]{\normalsize{Department of Biostatistics, UCLA Fielding School of Public Health, Los Angeles, CA 90095, US}}

\begin{document}

\maketitle

\begin{abstract}
In many scientific domains, clustering aims to reveal interpretable latent structure that reflects relevant subpopulations or processes.
Widely used Bayesian mixture models for model-based clustering often produce overlapping or redundant components because priors on cluster locations are specified independently, hindering interpretability.
 To mitigate this, repulsive priors have been proposed to encourage well-separated components, yet existing approaches face both computational and theoretical challenges. We introduce a fully tractable Bayesian repulsive mixture model by assigning a projection Determinantal Point Process (DPP) prior to the component locations. Projection DPPs induce strong repulsion and allow exact sampling, enabling parsimonious and interpretable posterior clustering. Leveraging their analytical tractability, we derive closed-form posterior and predictive distributions. These results, in turn, enable two efficient inference algorithms: a conditional Gibbs sampler and the first fully implementable marginal sampler for DPP-based mixtures. We also provide strong frequentist guarantees, including posterior consistency for density estimation, elimination of redundant components, and contraction of the mixing measure. Simulation studies confirm superior mixing and clustering performance compared to alternatives in misspecified settings. Finally, we demonstrate the utility of our method on event-related potential functional data, where it uncovers interpretable neuro-cognitive subgroups. Our results support the projection DPP mixtures as a theoretically sound and practically effective solution for Bayesian clustering.
\end{abstract}

\textbf{Keywords}: Bayesian nonparametrics; Clustering; Frequentist consistency; MCMC; Palm theory; Random measures 

\doublespacing     

\section{Introduction}\label{sec: introduction}

In many scientific domains, particularly in biomedical studies, clustering is used not merely to group similar observations but to uncover interpretable latent structure that corresponds to meaningful subpopulations or phenomena. For such goals, Bayesian mixture models have proven central to this task, but they often face a major shortcoming. When priors over component atoms are specified independently, models tend to yield redundant or overlapping components, resulting in overfitting and poor interpretability \citep{rousseau2011asymptotic,miller2014inconsistency,xu2016bayesian,quinlan2018density}.

To mitigate these issues, recent research has explored repulsive mixtures to enhance interpretability of the posterior clustering. Pioneered by \cite{petralia2012repulsive}, repulsive priors explicitly discourage mixture components from clustering too closely. 
Several recent contributions focused on modelling \citep{xu2016bayesian, xie2020bayesian, bianchini2020determinantal, cremaschi2025repulsion, ghilotti2025bayesian} and computational \citep{xu2016bayesian, beraha2022mcmc} aspects of repulsive mixtures, as well as on the associated distributional theory \citep{beraha2025bayesian}.
However, fitting a repulsive mixture model is known to pose substantial challenges: algorithms either require to design model-specific split-merge reversible jump moves \citep{xu2016bayesian,bianchini2020determinantal}, or rely on birth-death moves with potentially slow mixing \citep{beraha2022mcmc, cremaschi2025repulsion}, or require the numerical approximation of a large number of normalizing constants \citep{xie2020bayesian}.
Moreover, frequentist guarantees on the posterior of repulsive mixtures are lacking, with the only exception of \cite{petralia2012repulsive} and \cite{xie2020bayesian} that establish consistency results for the posterior density estimate, but not for the mixing measure, which is indeed a natural object of inference since repulsive mixtures are employed mainly for clustering.

In this paper, we contribute to the growing literature on repulsive mixtures by proposing a fully tractable model that assumes a projection Determinantal Point Process (DPP) prior on the component locations. DPPs \citep{macchi1975coincidence,hough2009zeros,lavancier2015determinantal} have been successfully utilized in repulsive mixtures, thanks to their inherent repulsion properties and well-developed probabilistic foundations 
\citep[see][]{xu2016bayesian, bianchini2020determinantal, ghilotti2025bayesian}.
Within the class of DPPs, projection DPPs are particularly tractable: they enjoy simple exact sampling algorithms \citep{lavancier2015determinantal,lavancier2023simulation}, and are the most repulsive DPPs for a fixed intensity \citep{moller2021couplings}, making them an excellent candidate as repulsive priors.

In doing so, we make several contributions. First, leveraging the analytical tractability of projection DPPs, we derive straightforward closed form expressions for the posterior and predictive distributions. 
Our theoretical framework builds upon the Palm-calculus approach developed in \cite{beraha2025bayesian}. Unlike their formulation, however, our framework yields fully tractable expressions. These translate into two practical inference algorithms: a conditional Gibbs sampler and the first fully implementable marginal sampler for DPP-based mixtures, eliminating the need for split-merge reversible jump or birth-death moves, and substantially improving sampling efficiency. 
The basic version of the marginal algorithm does require the numerical approximation of a potentially complex integral, similarly to what is done in \cite{xie2020bayesian}; however, we show  that introducing an auxiliary variable method, akin to the one employed by Algorithm 8 in \cite{neal2000markov}, yields a practical and efficient marginal sampler. 
We illustrate the efficiency gains of our approach through simulations comparing our conditional Gibbs sampler with the split-merge reversible jump algorithm by \cite{xu2016bayesian}, who use an $L$-ensemble (a special class of DPPs) as a prior, and  the birth-death sampler of \cite{beraha2022mcmc} under a Gaussian DPP prior. Considering two posterior estimands of the clustering structure - namely, the number of clusters and the entropy of the partition - our sampler achieves several-fold increases in  effective sample size (ESS).
In addition, we show that the marginal algorithm yields superior mixing performance at the expense of increased computational costs, a trade-off commonly observed  in  nonparametric mixtures \citep[see, e.g.,][]{favaroteh2013}.

We further provide a detailed investigation of the frequentist properties of our repulsive mixture models. For multivariate Gaussian mixtures, with repulsive priors on locations, and Gaussian mixture data-generating processes, we establish posterior contraction rates for the density estimate in the $L_1$ metric; the weight associated to the ``extra components'' in the model; and the mixing measure in the $1$-Wasserstein distance. 
These rates coincide with the ones previously obtained for non-repulsive mixtures, indicating that there is no penalty in assuming a repulsive prior. Conversely, our simulations show that our model consistently recovers well-separated, interpretable clusters even under some model misspecification. 
Taken together, these results support the use of our repulsive mixture priors as a sensible default option for model-based clustering, when the true data generating process is unknown.

We showcase the practical value of our approach in a dataset of event-related potential (ERP) waveforms from the ERP CORE project \citep{kappenman2021erp}. ERPs represent time-locked voltage patterns obtained by averaging electroencephalographic (EEG) recordings, yielding temporally precise indices of neural processing. Adopting a two-step approach, we first apply functional principal component analysis to reduce the dimensionality of the functional data and then perform clustering on the scores. 
Our model groups the scores into four well separated, interpretable,  neuro-cognitive clusters. In contrast, a standard Dirichlet process mixture yields eight clusters, four of which are singletons, obscuring latent structure due to insufficient repulsion. In section \ref{appendix: one-step_method} of the Supplementary material, we also present  a joint model for dimensionality reduction and clustering, in the spirit of \cite{ghilotti2025bayesian}, showing that the repulsive prior integrates naturally in that setting as well.

The remainder of the paper is organized as follows. Section~\ref{sec: pDPP_mixture_models} introduces the projection DPPs and specifies our hierarchical repulsive mixture model. In Section~\ref{sec: theorems}, we develop a tractable Bayesian analysis by deriving closed-form posterior and predictive distributions using the properties of projection DPPs. Section~\ref{sec: algorithms} presents the conditional and marginal Markov Chain Monte Carlo (MCMC) algorithms for posterior inference. Section~\ref{sec: thms_asymptotic_consistency} investigates the asymptotic properties of the proposed model, establishing posterior consistency for density estimation, component selection, and mixing measure contraction. Section~\ref{sec: Simulations} contains simulation studies assessing both computational efficiency and clustering performance. Section~\ref{sec: ERPs_data_analysis} applies our method to the real ERP data, demonstrating the model’s ability to uncover interpretable neuro-cognitive clusters. Section~\ref{sec: conclusion_discussion} concludes with a discussion and future directions. Supplementary material provide additional theoretical backgrounds, results, proofs, discussions, and algorithmic details.


\section{Mixture models with projection DPPs}\label{sec: pDPP_mixture_models}

Before introducing the model, we recall below the definition of projection Determinantal point process (DPP) models.

\subsection{Projection DPPs}

Let $\Phi = \{\phi_1, \ldots, \phi_m\}$ be a finite point process on a compact set $\Theta \subset \mathds{R}^d$. Let $K: \Theta \times \Theta \to \mathds{C}$ be a continuous complex-valued covariance function such that $\int_{\Theta} K(x, x) \mathrm{d} x < +\infty$. Then, $\Phi$ is a DPP with kernel $K$ if, for all $k \in \mathds{N} = \{1,2,\cdots\}$, the $k$-th order joint intensity function exists and is given by
\begin{equation}\label{eq: kth_order_joint_intensity}
    \rho(\phi_1, \ldots, \phi_k) = \det \left\{ K(\phi_i, \phi_j) \right\}_{i, j=1}^k, \qquad \phi_1, \ldots, \phi_k \in \Theta ,
\end{equation}
for pairwise distinct points, where  $\left\{ K(\phi_i, \phi_j) \right\}_{i, j=1}^k$ denotes the $k \times k$ matrix with $(i,j)$-th entry $K(\phi_i, \phi_j)$, and $\det\{ \cdot\}$ denotes the determinant. 
Intuitively, $\rho(\phi_1, \ldots, \phi_k) \, \mathrm{d}\phi_1 \ldots \mathrm{d}\phi_k$ represents the probability that $\Phi$ has a point in an infinitesimally small region around $\phi_i$ of volume $\mathrm d \phi_i$ for each $i = 1, \ldots, k$. 
See \citet{hough2009zeros, lavancier2015determinantal, baccelli2024random} for further theoretical details. 
By Mercer's theorem,  $K$ admits a spectral representation, $K(x, y) = \sum_{j = 1}^{\infty} \lambda_j \, \varphi_j(x) \, \overline{\varphi_j(y)}$, $(x, y) \in \Theta \times \Theta$, where eigenfunctions $(\varphi_j)_{j \geq 1}$ form an orthonormal basis of the space of square integrable functions, and the eigenvalues $(\lambda_j)_{j \geq 1}$ satisfy $0 \leq \lambda_j \leq 1$ and $\sum_{j=1}^{\infty} \lambda_j < +\infty$.

In this work, we focus on DPPs whose kernel satisfies $\int_{\Theta} K(x, z) K(z, y) \ddr z = K(x, y)$, i.e., on \emph{projection DPPs}. Equivalently, the kernel $K$ has exactly $m$ nonzero eigenvalues, all equal to 1. Moreover, the number of points in the process is almost surely equal to $m$. We write $\Phi \sim \mathrm{pDPP}_{\Theta} (K)$.
Among the class of DPPs, projection DPPs are the most repulsive ones \citep{moller2021couplings}, thus providing an excellent candidate for Bayesian repulsive mixture models.  See also Section~\ref{appendix: describe_repulsiveness_DPPs} in the Supplementary material, which formalizes the trade-off between repulsiveness and intensity. 
Moreover, unlike general DPPs, projection DPPs admit exact sampling algorithms that do not require use of the spectral decomposition 
\citep{hough2006determinantal, lyons2014determinantal}, a feature we exploit for posterior simulation. Rigorous definitions and derivations of (projection) DPPs are provided in Section~\ref{section: preliminaries} of the Supplementary material.

\subsection{A Projection DPP mixture model}\label{sec:mix_model}

Let $\bm{Y} = (Y_1, \ldots, Y_n)$, with $Y_i \in \mathds{R}^{d}$, denote the observed data. We assume each observation is generated from an $m$-component mixture model, where $m$ is fixed.
Let $\{f(\cdot \mid  \theta): \theta \in \Theta \subset \mathds{R}^d\}$, where each $f(\cdot \mid \theta)$ denotes a kernel density parameterized by $\theta$.
For simplicity of presentation, we begin by assuming that $\theta$ is a location parameter. We aim to promote \emph{a priori} separation of cluster-specific $\theta$ values, to facilitate the interpretability of posterior estimates.
We will extend our treatment to the case of location-scale mixtures in Sections~\ref{sec: algorithms} and~\ref{sec: thms_asymptotic_consistency} below. The mixture model can be written as
\begin{equation}\label{eq:mixture_df}
    \begin{aligned}
        Y_i \mid P &\simiid \int_{\Theta} f(\cdot \mid \theta) \, P(\ddr \theta) \\
        P & \sim \mathscr{P}
    \end{aligned}
\end{equation}
where the prior process, $\mathscr{P}$, for the mixing measure, $P$,  is defined to encode repulsion as follows. First, we assume that $P(\cdot) = G(\cdot) / G(\Theta)$, i.e., $P$ arises as the normalization of a finite random measure, a standard construction in Bayesian nonparametrics, following \cite{regazzini2003distributional} and \cite{james2009posterior}. Then, we define
\begin{equation}\label{eq:g_def}
    G = \sum_{h=1}^{m} s_h \delta_{\phi_h}, \quad \{\phi_1, \ldots, \phi_m\} \sim \mathrm{pDPP}_{\Theta} (K), \quad s_h \simiid \pi
\end{equation}
where $K$ is a projection kernel and  $\pi$  is a density on the positive real line,  belonging to the class of infinitely divisible (ID) distributions, see \citet{favaro2011class} for further details. For computational simplicity, we will assume that  $\pi$ is the density of a gamma distribution $\mathrm{Ga}(a_s, 1)$, with $a_s > 0$, 

Any projection kernel $K$ in \eqref{eq:g_def} leads to a valid repulsive model; for simplicity and specificity, we consider here the kernel analyzed in  \cite{lavancier2023simulation}. To this end, we assume that $\Theta$ is a $d$-dimensional rectangle. For specific choices of $\Theta$, we refer to the discussion in the examples of Sections~\ref{sec: Simulations} and~\ref{sec: ERPs_data_analysis}. Define the affine transformation $T(x) = Ax + b$ mapping $[-1/2, 1/2]^d \rightarrow \Theta$. Letting $m = (1 + 2 \ell)^d$ and $L = \{-\ell, \ldots, 0, \ldots, \ell\}$, we assume
\begin{equation}\label{eq: transformed_projection_kernel}
    K(x,y) = \frac{1}{|\operatorname{det}(A)|} \sum_{j \in L^d} \cos \left(2 \pi j^T \left(A^{-1} \left(x-y\right)\right)\right), \qquad (x,y) \in \Theta \times \Theta.
\end{equation}
See \citet{lavancier2023simulation}, as well as to  Section~\ref{section: preliminaries} of the Supplementary material, for additional details. Section~\ref{section: preliminaries} of the Supplementary material also discusses key properties of the kernel, including its stationarity, homogeneous intensity, and repulsion as characterized by the pair correlation function.
Other examples of projection kernels can be found in \citet{BorodinSoshnikov2003JanossyI, Johansson2001DiscreteOPE, hough2009zeros}.

\section{Posterior and Predictive Characterization of projection DPPs}
\label{sec: theorems}

Model \eqref{eq:mixture_df} is equivalent to assuming the hierarchical formulation, $Y_i \mid \theta_i \simind f(\cdot \mid \theta_i)$, $\theta_i \mid P \simiid P$, and $P\sim \mathscr{P}$. Therefore, in this section, we focus specifically on the prior model; that is, we consider realizations $\{\theta_1, \ldots, \theta_n\}$ of size $n$ from $P$, $\theta_i \mid P \simiid P$, where $P = G/ G(\Theta)$ with $G$ distributed as in \eqref{eq:g_def}. 
The study of this \textit{latent} model allows us to characterize 
the prior assumed for the mixture model, as well as to derive several analytical results which will serve as the backbone of the novel algorithms for posterior inference detailed in Section \ref{sec: algorithms}.
In particular, the properties of the projection DPP prior allow us to obtain closed-form expressions for both conditional and predictive distributions, yielding an exact and efficient inference procedure. 
Our approach shares high-level motivation with the general framework of \citet{beraha2025bayesian}. 
However, their results are extremely general and do not provide computable forms for posterior or predictive quantities. 
In contrast, we develop a theoretically novel and tractable construction that enables full marginalization and closed-form inference, something not previously achieved in repulsive mixture models.

\begin{figure}[t]
\centering
\includegraphics[width=\textwidth]{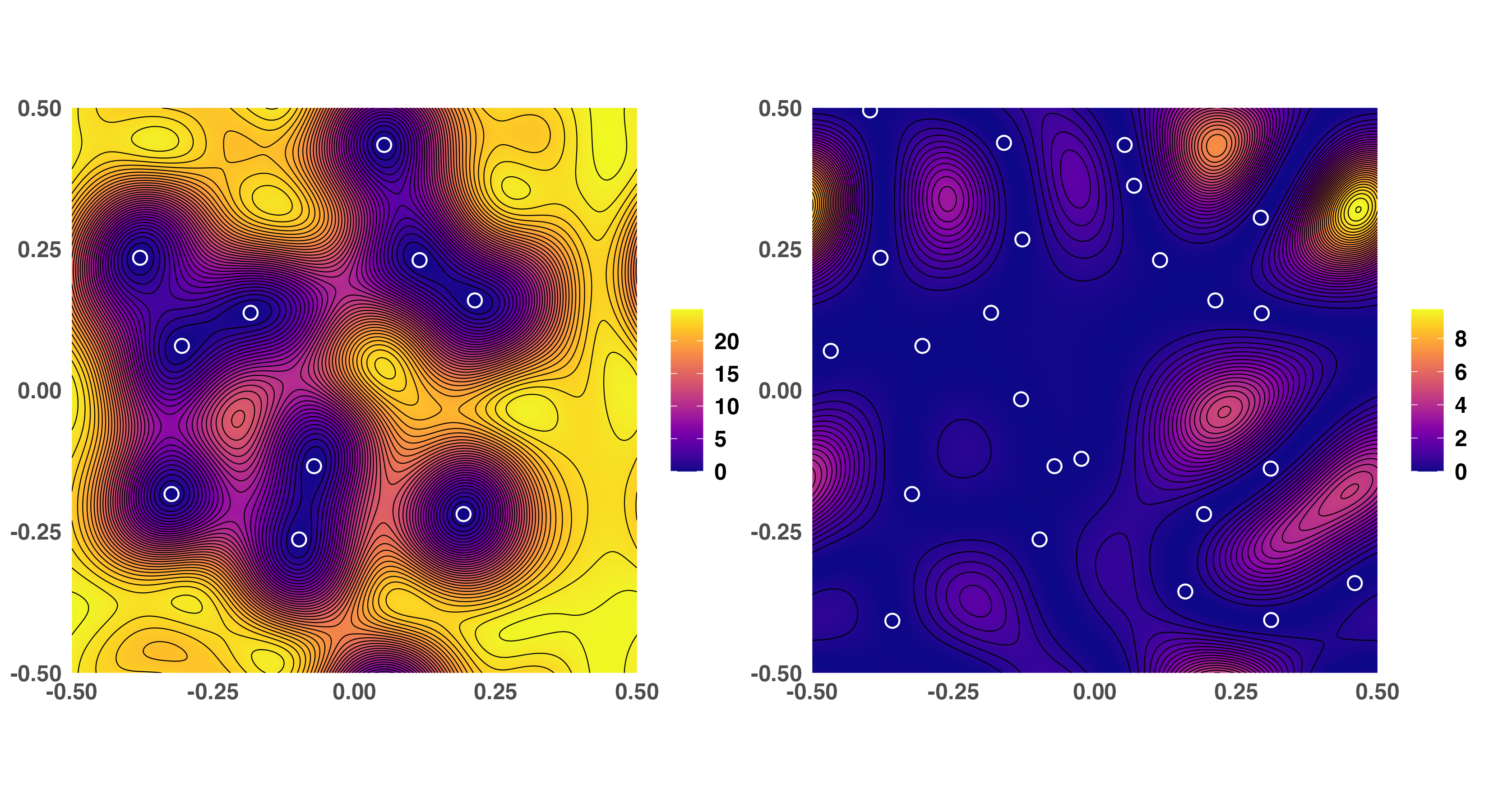}
\caption{Intensity $K^!_{\bm{\theta}^*}(x,x)$ on $[-1/2,1/2]^2$ using the projection DPP kernel $K$ defined in~\eqref{eq: transformed_projection_kernel} with $\ell=2$ and $m = 25$. Left: intensity of the 11th point conditional on a realization of the first 10 points (shown as white circles). Right: intensity of the 25th point conditional on a realization of the first 24 points. The reduced intensity near existing atoms illustrates the repulsive nature of $\Phi^!_{\bm{\theta}^*}$.}
\label{figure: conditional_intensity_contour_plot}
\end{figure}

Let us introduce some notation. Given a realization $\{\theta_1, \ldots, \theta_n\}$, let $\theta^*_1, \ldots, \theta^*_k$  denote the unique values in the vector, $k\leq n$. Each value occurs with frequency $n_h$, such that $\sum_{h=1}^{k}n_h = n$.  
Given the set of unique values, $\bm{\theta}^* = \{\theta^*_1, \ldots, \theta^*_k\}$, we can use  this realization to define a new kernel,
\begin{equation}\label{eq: conditional_DPP_kernel}
    K^!_{\bm{\theta}^*}(\phi_i, \phi_j) := K(\phi_i, \phi_j) - \bm{k}_{\bm{\theta}^*}^T (\phi_i) \bm{K}_{\bm{\theta}^*, \bm{\theta}^*}^{-1} \bm{k}_{\bm{\theta}^*}(\phi_j), \qquad \phi_i,\phi_j \in \Theta\backslash \bm{\theta}^*,
\end{equation}
where $\bm{k}_{\bm{\theta}^*}^T (\phi_i) = \left(K(\phi_i, \theta^*_1), \ldots, K(\phi_i, \theta^*_k)\right)$ and $\bm{K}_{\bm{\theta}^*, \bm{\theta}^*} = \left( K(\theta^*_t, \theta^*_r)\right)_{1 \leq t, r \leq k}$.
The kernel $K^!_{\bm{\theta}^*}$ also induces a DPP, denoted by $\Phi^!_{\bm{\theta}^*}$, following standard notation in the Palm theory of point processes,  since it corresponds to the \emph{reduced Palm version} of the original process $\Phi$ at $\bm{\theta}^*$ \citep{baccelli2024random}.
Informally, this corresponds to the conditional law of $\Phi$ given that $\theta^*_1, \ldots, \theta^*_k $ are points of $ \Phi$. See Section \ref{appendix: proofs} of the Supplementary material for further details.
In particular,  it is straightforward to check that  $\int_{\Theta\backslash \bm{\theta}^*} K^!_{\bm{\theta}^*}(x,z) K^!_{\bm{\theta}^*}(z,y) \mathrm d z = K^!_{\bm{\theta}^*}(x,y)$ and $\int_{\Theta\backslash \bm{\theta}^*} K^\prime(x,x) \mathrm dx = m - k$, implying that the process $\Phi_{\bm{\theta}^*}^!$, restricted to the set $\Theta \backslash \bm{\theta}^*$, has cardinality $m-k$, while the original projection DPP $\Phi$ has cardinality $m$. Moreover, the intensity function $K^!_{\bm{\theta}^*}(x,x)$ is no longer constant on $\Theta\backslash \bm{\theta}^*$, so that $\Phi_{\bm{\theta}^*}^{!}$ is an inhomogeneous projection DPP. In particular, the projection DPP $\Phi^!_{\bm{\theta}^*}$ tends to avoid placing additional points \emph{near} the locations of those already present in $\bm{\theta}^*$.  This behavior is illustrated in Figure~\ref{figure: conditional_intensity_contour_plot}, which shows contour plots of the intensity $K^!_{\bm{\theta}^*}$ for increasing sizes of $\bm{\theta}^*$. For $m=25$, The left panel displays the intensity of the 11th point conditional on a realization of the first 10 points, while the right panel shows the intensity of the 25th point conditional on a realization of the preceding 24.

Following \cite{james2009posterior}, we find it useful to introduce an auxiliary variable $u \mid \bm{s} \sim \operatorname{Ga}(n, \sum_{h=1}^{m} s_h)$, which facilitates the characterization of the posterior and predictive distributions.
The following theorem provides a description of the former. 

\begin{theorem}\label{thrm: theorem_1}
Conditional on $\bm{\theta} = \left\{\theta_1, \ldots, \theta_n\right\}$, and $u$, the distribution of the unnormalized mixing measure $G$ equals the distribution of $\sum_{h=1}^{k} s^{*}_h \delta_{\theta^*_h} + \sum_{h=1}^{m-k} s^\prime_h \delta_{\phi^\prime_h}$,
where:
\begin{itemize}
\item[(i)] The first term represents a discrete random measure with fixed atoms and independent random jumps, with density $f_{s^*_h} (s) \propto s^{n_h} \mathrm{e}^{-u s} \pi(s)$, for $h = 1, \ldots, k$. 
\item[(ii)] The second term is a discrete random measure supported on the atoms $\bm{\phi}^{\prime} = \left\{\phi_1^{\prime}, \ldots, \phi_{m-k}^{\prime}\right\}$, drawn from the reduced Palm DPP $\Phi_{\bm{\theta}^{*}}^{!}$ with kernel as in \eqref{eq: conditional_DPP_kernel}. The associated jumps $s_{h}^{\prime}$ are i.i.d. with exponentially tilted density $\pi^\prime(s) := \mathrm{e}^{-u s} \pi(s) / \psi(u)$, where $\psi(u) := \int_{\mathds{R}_+}  \mathrm{e}^{-u s} \pi(s) \mathrm{d} s$.
\end{itemize}
Moreover, the posterior of $u$ given $\theta_1, \ldots, \theta_n$ has density w.r.t to the Lebesgue measure
\begin{equation}\label{eq: condition_density_u}
f_{u \mid \bm{\theta}} (u) \propto u^{n-1} {\psi(u)}^{m-k} \prod_{h=1}^{k} \kappa(u, n_h), \qquad u > 0, 
\end{equation}
where $\kappa(u, n_h) := \int_{\mathds{R}_{+}} \mathrm{e}^{-us} s^{n_h} \pi(s) \mathrm{d} s$.
\end{theorem}

After observing the sample $\bm{\theta}^*$, the posterior distribution over $\Theta$ decomposes into two parts:
$(i)$ a discrete component supported on $\bm{\theta}^*$, with weights proportional to the number of allocations; $(ii)$ a random component comprising new, unobserved atoms $\bm{\phi}^{\prime}$, sampled from the reduced Palm version $\Phi_{\bm{\theta}^*}^!$ of the projection DPP $\Phi$. It is important to reiterate that the original projection DPP $\Phi$ is homogeneous with a constant intensity $K$ on $\Theta$. In contrast, the reduced Palm $\Phi_{\bm{\theta}^{*}}^{!}$ exhibits a distinctly inhomogeneous structure on $\Theta\backslash \bm{\theta}^{*}$. As shown in Figure~\ref{figure: conditional_intensity_contour_plot}, its intensity $K^!_{\bm{\theta}^*}(x,x)$ vanishes at the conditioning locations and notably increases  as $x$ moves away from the points in $\bm{\theta}^{*}$. This spatial modulation of the intensity explicitly reflects the local repulsion induced by the presence of the conditioning points  $\bm{\theta}^{*}$. 

In the case of gamma distributed weights, \eqref{eq: condition_density_u} simplifies to $f_{u \mid \bm \theta} (u) \propto u^{(n-1)} \, (1 + u)^{-(a_s m + n)}, u > 0$. Notably, this conditional density depends only on the fixed values of $a_s, m$, and $n$. Thus, its form does not vary with the specific configuration of $\bm\theta$. The resulting structural simplicity simplifies posterior computations and, as discussed later, facilitates an efficient MCMC implementation.

We now characterize the predictive distribution of a new realization $\theta_{n+1}$, conditional on the observed realization $\bm{\theta}$.  

\begin{theorem}\label{thrm: theorem_2}
Conditionally on $\bm{\theta}$ and the auxiliary variable $u$, the predictive distribution of a new latent variable $\theta_{n+1}$ is given, for $k\leq m-1$, by
\begin{equation}\label{eq: predictive_distribution_1}
\begin{split}
\operatorname{Pr}(\theta_{n+1} \in \mathrm{d}\theta \mid \bm{\theta}, u) & \propto \sum_{h=1}^{k} \frac{\kappa(u, n_h + 1)}{\kappa(u, n_h)} \delta_{\theta_{h}^{*}} (\mathrm{d}\theta)  +  \frac{\kappa(u,1)}{\psi(u)} K^!_{\bm \theta^*}(\theta, \theta) \ddr \theta,
\end{split}
\end{equation}
and, for $k=m$ (i.e., all $m$ components are already occupied), by
\begin{equation}\label{eq: eq: predictive_distribution_2}
\operatorname{Pr}(\theta_{n+1} \in \mathrm{d}\theta \mid \bm{\theta}, u) \propto \sum_{h=1}^{m} \frac{\kappa(u, n_h + 1)}{\kappa(u, n_h)} \delta_{\theta_{h}^{*}} (\mathrm{d}\theta) .
\end{equation}
\end{theorem}
By the properties of the reduced Palm measure $\Phi_{\bm{\theta}^*}^{!}$, under our projection DPP kernel $K$, in \eqref{eq: predictive_distribution_1} we have $K^{!}_{\bm{\theta}^*}(\theta,\theta)=0$ for $\theta\in \bm{\theta}^*$. In addition, the probability to generate a new $\theta$ is proportional to $\int_{\Theta} K^!_{\bm \theta^*}(\theta, \theta) \mathrm{d} \theta = m - k$. These features suggest an interpretation of the predictive distribution through a metaphor akin to the traditional Chinese restaurant process. In this analogy, customers seated at the same table eat the same dish, and each dish is served exclusively at one table. Initially, all $m$ tables of the restaurant are stored in a warehouse and the dining hall $\Theta$ is empty. The first customer necessarily brings in a new table from the warehouse and places it at a random position $\theta_1^*\in \Theta$. As customers arrive, they may either occupy an existing table in the hall or bring in a new table from the warehouse and assign it a location. For example, the second customer has two choices: either join the same table as the first customer with probability proportional to $\kappa(u_1, 2)/\kappa(u_1, 1)$, or bring a new table from the warehouse with probability proportional to $\left(m - 1\right)\kappa(u_1,1) / \psi(u_1)$. If a new location is chosen, this is selected to maintain a certain repulsive distance from $\theta_1^*$, owing to the repulsiveness of the projection DPP. Thus, the probability distribution for the second customer's table location becomes proportional to 
\[
\operatorname{Pr}(\theta_2 \in \mathrm{d} \theta \mid \theta_1^*) \propto K(\theta,\theta) - K^2(\theta,\theta_1^*) / K(\theta_1^*, \theta_1^*) \mathrm{d} \theta, \qquad \theta \in \Theta \backslash \{\theta_1^*\} ,
\]
that is, the intensity of the reduced Palm version of the original process, conditional on knowing the location $\theta_1^*$; see \eqref{eq: conditional_DPP_kernel}. The metaphor unfolds as each new customer arrives at time $n+1$.   They may either sit at one of the $k$ already existing tables in the dining hall, with probability proportional to $\kappa(u_n, n_h + 1) / \kappa(u_n, n_h),$ for $ h = 1, \ldots, k$, or they may bring in a new table, with probability proportional to $\left(m - k\right)\kappa(u_n,1) / \psi(u_n)$. In the latter case, to preserve repulsiveness relative to the existing $k$ tables, the location $ \theta = \theta_{k+1}^{*}$ is selected  with probability proportional to 
\[
\operatorname{Pr}(\theta_{n+1} \in \mathrm{d} \theta \mid \theta_1^*, \ldots, \theta_{k}^*) \propto K(\theta,\theta) -  \bm{k}_{\bm{\theta}^{*}}^T (\theta) \bm{K}_{\bm{\theta}^{*},\bm{\theta}^{*}}^{-1} \bm{k}_{\bm{\theta}^{*}}(\theta) \mathrm{d} \theta, \qquad \theta \in \Theta \backslash \{\theta_1^*, \ldots, \theta_{k}^*\} .
\]
As the restaurant fills and all $m$ tables from the warehouse have been brought into the hall, no further tables can be added. From this point on, each new customer must sit at one of the existing $m$ tables, with the probability of choosing table $h$ proportional to $\kappa(u, n_h + 1) / \kappa(u, n_h)$, for $h = 1, \ldots, m$.


\section{MCMC strategies for projection DPPs.}
\label{sec: algorithms}

In this section, we present two MCMC algorithms for posterior inference under the proposed method: a conditional and a marginal algorithm, following the terminology established by \citet{papaspiliopoulos2008retrospective}. 
For ease of presentation, we assume our observed sample is drawn from a multivariate Gaussian mixture model, i.e., $y_i \simind \mathcal{N}_d (\cdot \mid \theta_i, \Delta_i), (\theta_i, \Delta_i) \simiid P$, for $i = 1, \ldots, n$, where the mixing measure is given by
\[
P(\cdot) = \frac{G(\cdot)}{G(\Theta \times \mathds{V})}, \qquad G = \sum_{h=1}^{m} s_h \delta_{(\phi_h, \Sigma_h)},
\]
where we further assume that the repulsive projection DPP prior is only imposed on the location atoms, $\phi_h$'s, while the cluster-specific covariance matrices, $\Sigma_h$'s, are i.i.d. draws from an Inverse-Wishart prior, $\Sigma_h \simiid \operatorname{InvWi}(\tau, \Omega)$; the jumps are specified by a generic density on the positive real line, $s_h \simiid \pi(\cdot)$, for $h = 1, \ldots, m$.

Generating samples from a projection DPP plays an essential role in both the conditional and marginal algorithms. Our approach relies on a fundamental baseline algorithm, which is  justified in \citet{hough2006determinantal} and \citet{lyons2014determinantal}, and detailed in   Algorithm~\ref{algo: projection_DPP_sampling} in the Supplementary material. This algorithm employs a sequential rejection sampling method for simulating a set of points $\{\phi_1, \ldots, \phi_m\}$ from a projection DPP $\Phi \sim \mathrm{pDPP}_{\Theta}(K)$ using the kernel $K$ defined in~\eqref{eq: transformed_projection_kernel}, for a fixed cardinality $m$. Although Algorithm~\ref{algo: projection_DPP_sampling} is tailored specifically to this kernel, it can be readily adapted to sample from other projection DPPs, including the reduced Palm version $\Phi_{\bm{\theta}^{*}}^{!}$ with the kernel in~(\ref{eq: conditional_DPP_kernel}), as required by Theorem~\ref{thrm: theorem_1}. As such, it serves as a key building block in our posterior sampling algorithms. For a comprehensive discussion on advanced sampling techniques applicable to general (projection) DPPs, readers may refer to \citet{lavancier2023simulation}.

For the remainder of this section, we focus on the multivariate case ($d \geq 2$) for illustrative purposes; these algorithms can be easily simplified to univariate data ($d = 1$).

\subsection{Conditional algorithm for posterior inference}

Given $\theta_1, \ldots, \theta_n$, define the cluster allocation variables $\bm c = (c_1, \ldots, c_n)$ such that $\theta_i = \theta_j$ if and only if $c_i = c_j$. Then, it is possible to partition the set of atoms $\bm{\phi} = \left\{\phi_1, \ldots, \phi_m\right\}$ into two disjoint subsets: those associated with at least one observation, and those not. Let's indicate $\bm \phi^{(a)} = \{\phi_{c_1}, \ldots, \phi_{c_n}\} = \{\theta^*_1, \ldots, \theta^*_k\}$, and $\bm{\phi}^{(na)} = \bm{\phi} \setminus \bm{\phi}^{(a)}$. Following the terminology of \citet{griffin2011posterior} and \citet{argiento2022infinity}, $\bm{\phi}^{(a)}$ and $\bm{\phi}^{(na)}$ denote the set of active and non-active atoms, respectively. Correspondingly, the jumps $\bm{s} = \left\{s_1, \ldots, s_m\right\}$ and the covariance matrices $\bm \Sigma = \left\{\Sigma_1, \ldots, \Sigma_m\right\}$ are  also partitioned into $\bm{s}^{(a)}$, $\bm{s}^{(na)}$, $\bm \Sigma^{(a)}$, and $\bm \Sigma^{(na)}$. The number of unique values $k$ equals the cardinality of the active-atom set, $|\bm c|$, and the frequency of each unique value is $n_h = \sum_{i=1}^n \mathds{I}(c_i=h)$, $h=1, \ldots, k$, where $\mathds{I} (\cdot)$ denotes the indicator function.  

In light of the notation above, for describing the computational algorithms, it is convenient to identify the $s^*_h$'s in Theorem~\ref{thrm: theorem_1} with $s^{(a)}_h$, the $s^\prime_h$'s  with $s^{(na)}_h$, and the $\phi^\prime_h$'s with $\phi^{(na)}_h$'s. Then, we can describe the conditional MCMC algorithm as follows:

\begin{myalgorithm}[Conditional algorithm]\label{algo: algorithm_1} 
\leavevmode
\begin{enumerate}
\item Sample the auxiliary variable $u \mid \sum_{h=1}^{m} s_h \sim \operatorname{Ga}(n, \sum_{h=1}^{m} s_h)$ .
\item Sample the allocation indicators independently for each observation $y_i$, with probability $\operatorname{Pr}(c_i = h \mid \bm{y}, \bm{s}, \bm{\phi}, \bm{\Sigma}) \propto s_h \, \mathcal{N}_d (y_i \mid  \phi_h, \Sigma_h)$, for $h = 1, \ldots, m$. Given the sampled vector $\bm c$, identify the active and non-active components of $\bm \phi$.
\item For the non-active part, $\sum_{h=1}^{m-|\bm{c}|} s^{(na)}_h \delta_{(\phi^{(na)}_h, \Sigma_h^{(na)})}$:
    \begin{enumerate}
    \item Sample non-active jumps i.i.d. from the tilted density $\pi^\prime(s) \propto \mathrm{e}^{-su} \pi(s)$, that is, $s_h^{(na)} \simiid \operatorname{Ga}(a_s, 1 + u)$.
    \item Sample non-active location atoms $\phi_h^{(na)}$ from the inhomogeneous projection DPP $\Phi^!_{\bm{\phi}^{(a)}}$ with kernel $K^!_{\bm\phi^{(a)}}$ defined in~(\ref{eq: conditional_DPP_kernel}), using Algorithm~\ref{algo: conditional_nonactive_atoms_samplings}.
    \item Sample the non-active covariances independently as $\Sigma_h^{(na)} \simiid \operatorname{InvWi}( \tau, \Omega)$.
    \end{enumerate}
\item For the active part, $\sum_{h=1}^{|\bm{c}|} s^{(a)}_h \delta_{(\phi^{(a)}_h, \Sigma_h^{(a)})}$: 
    \begin{enumerate}
    \item Sample the active jumps independently with density proportional to $s^{n_h} \mathrm{e}^{-su} \pi(s)$, that is,  $s_h^{(a)} \simiid \operatorname{Ga}(n_h + a_s, 1 + u)$.
    \item Update the active location atoms $\bm{\phi}^{(a)} = \{\phi_1^{(a)}, \ldots, \phi_{|\bm{c}|}^{(a)}\}$ from the density given by
    \[
    \begin{split}
    f \left(\bm{\phi}^{(a)} \mid \bm{y}, \bm{c}, \bm{\Sigma}^{(a)}, \bm{\phi}^{(na)}\right) &\propto \prod_{h=1}^{|\bm{c}|} \prod_{i: c_i = h} \mathcal{N}_d \left(y_i \mid \phi_h^{(a)}, \Sigma_h^{(a)} \right) \\
        & \qquad \times \det\left\{K\left(\phi_t, \phi_r \right)_{\phi_t, \phi_r \in \bm{\phi}^{(a)} \cup \bm{\phi}^{(na)}}\right\}, \qquad \bm{\phi}^{(a)} \in \Theta^{|\bm{c}|} ,
    \end{split}
    \]
    using a Metropolis-Hastings (MH) step, detailed in Algorithm~\ref{algo: active_atoms_MH_updating}.
\item Sample the active covariances independently from
\[
\Sigma_h^{(a)} \simiid \operatorname{InvWi}\left(n_h + \tau, \Omega + \sum_{i: c_i = h}\left(y_i - \phi_h^{(a)}\right)\left(y_i - \phi_h^{(a)}\right)^T\right) .
\] 
    \end{enumerate}
\item Combine active and non-active sets, $\bm{\phi} = \bm{\phi}^{(a)} \cup \bm{\phi}^{(na)}, \bm{\Sigma} = \bm{\Sigma}^{(a)} \cup \bm{\Sigma}^{(na)}, \bm{s} = \bm{s}^{(a)} \cup \bm{s}^{(na)} ,$ and update the mixing measure, accordingly, $G(\cdot) = \sum_{h=1}^{m-|\bm{c}|} s^{(na)}_h \delta_{(\phi^{(na)}_h, \Sigma_h^{(na)})} (\cdot) \cup \sum_{h=1}^{|\bm{c}|} s^{(a)}_h \delta_{(\phi^{(a)}_h, \Sigma_h^{(a)})} (\cdot)$.
\end{enumerate}
\end{myalgorithm}
To update the set of active location atoms $\bm{\phi}^{(a)}$, we refer to Algorithm~\ref{algo: active_atoms_MH_updating} in the Supplementary material for details. Briefly, Algorithm~\ref{algo: active_atoms_MH_updating} employs a MH strategy to update each $\phi_h^{(a)}$, conditioning on the remaining active atoms $\bm{\phi}^{(a)} \backslash \{\phi_h^{(a)}\}$ and all the non-active atoms $\bm{\phi}^{(na)}$, for $h = 1, \ldots, |\bm{c}|$. The proposal distribution is a symmetric mixture of two components: with high probability, it proposes local moves via a Gaussian centered at the current value of $\phi_h^{(a)}$; with lower probability, it proposes distant moves by sampling from a reduced Palm projection DPP $\Phi^!_{\bm{\phi}\backslash \{\phi_h^{(a)}\}}$, which is conditioned on all remaining atoms and inherently avoids placing new atoms near existing ones.

\subsection{Marginal algorithm for posterior inference}


Alternatively, we can avoid explicitly updating the mixing measure $G$, as in the conditional algorithm, by integrating it out, thus yielding a marginal algorithm. Using the predictive distribution in Theorem~\ref{thrm: theorem_2}, akin to a collapsed Gibbs sampler \citep{muller2015bayesian}, we alternate between sampling the auxiliary variable $u$ given  $\bm{y}$ and $\{(\theta_i, \Delta_i)\}_{i=1}^{n} $, and updating the observation-specific parameters $\{(\theta_i, \Delta_i)\}_{i=1}^{n}$ given $u$ and  $\bm{y}$. However, under gamma-distributed weights, the distribution of $u \mid \bm{y}, \{(\theta_i, \Delta_i)\}_{i=1}^{n} $ in \eqref{eq: condition_density_u} does not change across MCMC iterations; hence, we omit this sampling step to lower computational effort and accelerate convergence. To update $\{(\theta_i, \Delta_i)\}_{i=1}^{n} \mid u, \bm{y}$, we perform a Gibbs scan, sampling $(\theta_i, \Delta_i)$ from its full conditional distribution,
\[
\operatorname{Pr}\left((\theta_i, \Delta_i) \in \left(\mathrm{d} \theta , \mathrm{d} \Delta\right) \mid (\bm{\theta}, \bm{\Delta})_{-i}, u, \bm{y} \right) \propto \operatorname{Pr}\left((\theta_i, \Delta_i) \in \left(\mathrm{d} \theta , \mathrm{d} \Delta\right) \mid (\bm{\theta}, \bm{\Delta})_{-i}, u \right) \times \mathcal{N}_d (y_i \mid \theta, \Delta) ,
\]
where $(\bm{\theta}, \bm{\Delta})_{-i} = \left\{(\theta_1, \Delta_1), \ldots, (\theta_{i-1}, \Delta_{i-1}), (\theta_{i+1}, \Delta_{i+1}), \ldots, (\theta_n, \Delta_n)\right\}$. Let $\bm{\theta}_{-i}^{*} = \{\theta_j^*: j = 1, \ldots, q\}$ denote the $q$ unique values in $\bm{\theta}_{-i} = \left\{\theta_1, \ldots, \theta_{i-1}, \theta_{i+1}, \ldots, \theta_n\right\}$; similarly, for $\bm{\Delta}_{-i}^{*}$. Let $n_j^{(-i)}$ denote the frequency of the $j$-th unique value in $\bm{\theta}_{-i}$, such that $\sum_{j=1}^{q} n_j^{(-i)} = n - 1$. By Theorem~\ref{thrm: theorem_2}, the full conditional distribution can be obtained as, 
\begin{align*}
\begin{split}
\operatorname{Pr}\left((\theta_i, \Delta_i) \in \left(\mathrm{d} \theta , \mathrm{d} \Delta\right) \mid (\bm{\theta}, \bm{\Delta})_{-i}, u, \bm{y} \right) & \propto \sum_{j = 1}^{q} \left(n_j^{(-i)} + a_s\right) 
\, \mathcal{N}_d (y_i \mid \theta_j^*, \Delta_j^*) 
\, \delta_{(\theta_j^*, \Delta_j^*)} \left(\mathrm{d} \theta , \mathrm{d} \Delta\right) \\
& + a_s \,  K^!_{\bm \theta^*_{-i}}(\theta, \theta)\,   \operatorname{InvWi}(\Delta \mid \tau, \Omega) \, \mathcal{N}_d (y_i \mid \theta, \Delta) \,\mathrm{d} \theta \mathrm{d} \Delta ,
\end{split}
\end{align*}
since $\kappa(u, n^{(-i)}_j + 1) / \kappa(u, n^{(-i)}_j) = (n_j^{(-i)}+ a_s) / (u + 1)$, and $\kappa(u,1) / \psi(u) = a_s / (u + 1)$. Hence, the sampling scheme follows the following marginal MCMC algorithm:

\begin{myalgorithm}[Marginal algorithm]\label{algo: marginal_algorithm_Neal2} 
\leavevmode
\begin{enumerate}
\item[(1)] For each observation $i = 1, \ldots, n$, sequentially update $(\theta_i, \Delta_i)$ conditionally on the remaining values $(\bm{\theta}, \bm{\Delta})_{-i}$.  Let $(\theta_i, \Delta_i)$ either:
\begin{itemize}
\item coincide with one of the $q$ unique existing pairs $(\theta_j^*, \Delta_j^*)$, with probability proportional to $\left(n_j^{(-i)} + a_s\right) \mathcal{N}_d (y_i \mid \theta_j^*, \Delta_j^*), j = 1, \ldots, q$, or
\item assume a new value, say $(\theta, \Delta)$, with probability proportional to 
\begin{align*}
\begin{split}
&a_s \int_\mathds{V} \int_\Theta K^!_{\bm{\theta}^*_{-i}}(\theta,\theta) \, \operatorname{InvWi}(\Delta \mid \tau, \Omega)\,  \mathcal{N}_d (y_i \mid \theta, \Delta) \mathrm{d} \theta \mathrm{d} \Delta \\
& = \frac{a_s |\Omega|^{\tau/2} \Gamma((1+\tau)/2)}{\pi^{d/2} \Gamma((1+\tau-d)/2)}\,  \int_{\Theta} \frac{K^!_{\bm{\theta}^*_{-i}}(\theta,\theta)}{|\Omega + (y_i - \theta)(y_i - \theta)^T|^{(1+\tau)/2}} \,\mathrm{d} \theta .
\end{split}
\end{align*}
\end{itemize}
In the latter case, sample a new pair $(\theta_i, \Delta_i)$, as follows:
\begin{enumerate}
    \item[(a)] Sample $\Delta_i \sim \operatorname{InvWi}(\Delta \mid \tau, \Omega)$;
    \item[(b)] Sample $\theta_i$ from $\operatorname{Pr}(\theta_i \in \mathrm{d}\theta \mid \bm{\theta}_{-i}^{*}, \Delta_i, y_i) \propto K^!_{\bm \theta^*_{-i}}(\theta, \theta) \, \mathcal{N}_d (y_i \mid \theta, \Delta_i) \, \mathrm{d} \theta$, using standard rejection sampling and a uniform proposal on $\Theta$.
\end{enumerate}
\end{enumerate}
\item[(2)] After sequentially updating $\bm{\theta} = \left\{\theta_1, \ldots, \theta_n\right\}$ and $\bm{\Delta} = \left\{\Delta_1, \ldots, \Delta_n\right\}$, we recover the allocation indicators $\bm{c}$, and then update the unique component values $\bm{\theta}^* = \{\theta_1^*, \ldots, \theta_{|\bm{c}|}^*\}$ and $\bm{\Delta}^* = \{\Delta_1^*, \ldots, \Delta_{|\bm{c}|}^*\}$ in a reshuffling step:
\begin{enumerate}
\item[(a)] Sample $\bm{\theta}^*$ from a joint distribution on $\Theta^{|\bm{c}|}$, with probability proportional to 
\[
\prob(\bm{\theta}^*) \propto \det\{K(\theta^*_t, \theta^*_r)\}_{t, r =1}^{|\bm{c}|} \times 
\prod_{h=1}^{|\bm{c}|} \prod_{i: c_i = h} \mathcal{N}_d (y_i \mid \theta^*_h, \Delta_h^*), \quad \bm{\theta}^* \in \Theta^{|\bm{c}|} , 
\]
using MH steps, following Algorithm~\ref{algo: active_atoms_MH_updating} from the conditional MCMC algorithm. To sequentially update each $\theta_h^*$, for $ h = 1, \ldots, |\bm{c}|$, employ the symmetric proposal distribution,  $0.9 \,  \mathcal{N}_d (\theta_h^*, 0.01\,I_d) + 0.1 \, \Phi_{\bm{\theta}^* \backslash \{\theta_h^*\}}^!$, where the reduced Palm version $\Phi_{\bm{\theta}^* \backslash \{\theta_h^*\}}^!$ is a projection DPP of cardinality $m - |\bm{c}| + 1$. 
\item[(b)] Independently sample each $\Delta_h^*$ from
\[
\Delta_h^* \sim \operatorname{InvWi}\left(\Delta \mid n_h + \tau, \Omega + \sum_{i: c_i = h} (y_i - \theta_h^*)(y_i - \theta_h^*)^T \right), \qquad h = 1, \ldots, |\bm{c}|.
\]
\end{enumerate}
\end{myalgorithm}

The reshuffling step (2) in Algorithm~\ref{algo: marginal_algorithm_Neal2} is a standard technique in MCMC algorithms for mixture models, which has been shown to accelerate convergence and enhance mixing efficiency \citep{neal2000markov}. However, the integral
\[
\int_{\Theta} \frac{K^!_{\bm{\theta}^*_{-i}}(\theta,\theta)}{|\Omega + (y_i - \theta)(y_i - \theta)^T|^{(1+\tau)/2}} \mathrm{d} \theta
\]
in Algorithm~\ref{algo: marginal_algorithm_Neal2} can become computationally expensive, particularly for multivariate data. To circumvent this challenge, we adopt the auxiliary variable strategy proposed in Algorithm 8 of \citet{neal2000markov} to avoid direct evaluation of the integral. The resulting modified algorithm is presented as Algorithm~\ref{algo: marginal_algorithm_Neal8} in the Supplementary Material.

\section{Asymptotic consistency properties}\label{sec: thms_asymptotic_consistency}

The finite-sample properties in Section~\ref{sec: theorems} demonstrate that our model admits tractable posterior and predictive characterizations. In this section, we turn to the large-sample behavior and show that the model also exhibits strong asymptotic guarantees. In particular, under the Gaussian repulsive mixture discussed in Section \ref{sec: algorithms}, we establish three key results: posterior consistency for the density, elimination of redundant mixture components, and posterior contraction of the mixing measure in the 1-Wasserstein distance. 

We assume the true data-generating density $f_0$ is a finite Gaussian mixture with $k_0$ components, $f_0 := \sum_{h =1}^{k_0} w_{0, h} \, \mathcal{N}_d (\phi_{0, h}, \Sigma_{0, h})$, where $k_0$ is fixed, and we assume that the component locations are adequately separated: there exists $r > 0$ such that $\|\phi_{0, h} - \phi_{0, j}\| \geq r$, where $\|\cdot\|$ is the standard Euclidean norm, and the covariance eigenvalues lie in a bounded interval, $[\lambda_\ast, \lambda^\ast]$,  where $0 < \lambda_\ast < \lambda^\ast < \infty$. Given data $y_{1:n} = (y_1, \ldots, y_n)$, denote by $\Pi(\cdot)$ and $\Pi(\cdot \mid y_{1:n})$ the prior and posterior distributions on the space of densities on $\Theta \times \mathds{V}$, and let $\mathds{P}_0$ be the probability measure with density $f_0$, and $\mathds{P}_0^{(n)}$ for its $n$-fold product measure.
\begin{theorem}\label{thm:l1_consistency}
    Under the assumptions above, there exists a sufficiently large $M > 0$ such that
    \[
        \Pi\left(\|f_{P}- f_0\|_1 \geq M \sqrt{\log n / n} \mid y_{1:n}\right) \rightarrow 0
    \]
    in $\mathds{P}_0^{(n)}$-probability, where $f_P := \sum_{h=1}^{m} w_h \, \mathcal{N}_d(\phi_h, \Sigma_h)$, with $\sum_{h=1}^{m}w_h=1$, is the density induced from the repulsive mixture model considered in Section~\ref{sec: algorithms}.
\end{theorem}
Theorem~\ref{thm:l1_consistency} shows that the posterior density converges to the true density. However, in mixture models, density consistency alone does not guarantee that the posterior will remove redundant components. The next result, which follows directly from Theorem 1 in \cite{rousseau2011asymptotic}, establishes that the ``extra components'' in the mixture are emptied out a posteriori.
\begin{theorem}\label{thm:extra_components}
    If $a_s < d/2$, then for every $\varepsilon > 0$,
    \[
    \Pi\left( \exists\, J \subset \{1,\ldots,m\}, |J| = m-k_0\colon \sum_{h \in J} w_h < n^{-1/2 + \varepsilon} \mid y_{1:n}\right) \rightarrow 1
    \]
    in $\mathds{P}_0^{(n)}$-probability.
\end{theorem}
Under the proposed Gaussian mixture specification with gamma-distributed unnormalized weights, assumptions A2-A5 in \cite{rousseau2011asymptotic} hold trivially, whereas assumption A1 follows from our Theorem \ref{thm:l1_consistency}. Intuitively, the posterior discards the superfluous $m - k_0$ components by shrinking their weights toward zero.

Finally, we study the asymptotic behavior of the mixing measure $P$ itself. Leveraging Theorems~\ref{thm:l1_consistency} and~\ref{thm:extra_components}, we obtain a posterior contraction rate for $P$ in the 1-Wasserstein distance. Let $\Xi:=\{(x,v): x \in \Theta, \lambda_{\ast} I_d \preceq v \preceq \lambda^{\ast}I_d\}$, where $\preceq$ represent the Loewner ordering for positive semi-definite matrices, implying that the all the eigenvalues of $\nu$ must lie in $[\lambda_*, \lambda^*]$. Then, we define the metric $\eta\left((x,v),(x',v')\right):=\|x-x'\|+\|v-v'\|_F$, where $\|\cdot\|_F$ indicates the Frobenius norm. Let $W_1$ be the 1-Wasserstein distance induced by $\eta$ on probability measures over $\Xi$. 
\begin{theorem}\label{thm:W1_contraction} Let $P_0 = \sum_{h=1}^{k_0} w_{0,h} \, \delta_{(\phi_{0,h}, \Sigma_{0, h})}$ be the true mixing measure and $P = \sum_{h=1}^m w_h\,  \delta_{(\phi_h, \Sigma_h)}$ be the mixing measure induced by the model; then, for any $\varepsilon>0$, there exists $C<\infty$ such that
\[
    \Pi\big(W_1(P, P_0)\le C \, n^{-1/2+\varepsilon} \mid y_{1:n}\big)\to 1 
\]
in $\mathds{P}_0^{(n)}$-probability.
\end{theorem}
Proofs of Theorems~\ref{thm:l1_consistency},~\ref{thm:extra_components}, and~\ref{thm:W1_contraction} are provided in Section~\ref{appendix: asymptotic_consistency_properties} of the Supplementary material.

\section{Simulations}\label{sec: Simulations}

\subsection{Simulation 1: MCMC mixing}\label{sec: Simulation_1}

In this section, we evaluate and compare the mixing efficiency and convergence of the three proposed MCMC Algorithms~\ref{algo: algorithm_1},~\ref{algo: marginal_algorithm_Neal2},~\ref{algo: marginal_algorithm_Neal8}, which we refer to as \emph{pDPP-Cond}, \emph{pDPP-Marg-A}, and \emph{pDPP-Marg-B}, respectively. All three algorithms explore the posterior space of the same model. We generate a sample of $n = 300$ observations from a mixture of three univariate Student's $t$-distributions,
\begin{equation}\label{eq: simulation1_dataset_1dim}
y_1, \ldots, y_n \simiid \frac{1}{3}\,  t_6 (-4, 1) + \frac{1}{3} \, t_6 (0, 1) + \frac{1}{3} \, t_6 (4, 1) ,  
\end{equation}
where each component has 6 degrees of freedom, a scale parameter of 1, and respective locations centered at -4, 0, and 4. We carry out posterior inference using the same hyperparameters for each algorithm, setting $\ell = 5, a_s = 0.1$, and $\Sigma_h \simiid \mathrm{InvGa}(1,3)$. We set $\Theta$ to be the interval $[R_L, R_U]$ to define the projection DPP prior, with bounds $R_L = \bar{y} - 3 \zeta$ and $R_U = \bar{y} + 3 \zeta$, where $\bar{y}$ is the empirical mean of the data, and $\zeta = \max |y_i - \bar{y}|$ is the radius of the empirical range. For all simulations in Section~\ref{sec: Simulations}, we run each MCMC chain for 10,000 iterations, discarding the first 5,000 as burn-in, without thinning.

The trace plots in Figure~\ref{figure: simulation_mcmc_mixing_efficiency} in the Supplementary Material illustrate the evolution of the estimated number of clusters and the entropy of the partitions across iterations. If a sample of $n$ observations are grouped into $K$ clusters as $C_1, \ldots, C_K$, the partition entropy is defined as $H(C_1, \ldots, C_K) := \sum_{j=1}^{K} \frac{|C_j|}{n} \log \left(\frac{|C_j|}{n}\right)$, where $|C_j|$ is the cardinality of the cluster $C_j$. All three algorithms provide good estimates of the number of clusters in the mixture, and explore a reasonable posterior range. Notably, \emph{pDPP-Marg-A} and \emph{pDPP-Marg-B} display faster mixing convergence, particularly evident in the partition entropy trace shown in Figure~\ref{figure: simulation_mcmc_mixing_efficiency} (d). In contrast, the conditional sampler \emph{pDPP-Cond} exhibits slower mixing. This result aligns with our expectations, as marginal samplers integrate out component-specific latent parameters and enable more global updates, whereas conditional samplers rely on localized updates and consequently tend to mix more slowly.

Table~\ref{table: Simulation_1_ESS} in the Supplementary Material reports the effective sample size (ESS) per iteration and per second for each algorithm or method, focusing on the estimated number of clusters and the partition entropy. As shown in the first three rows (above the horizontal line) in Table~\ref{table: Simulation_1_ESS}, in terms of ESS per iteration, the marginal algorithms \emph{pDPP-Marg-A} and \emph{pDPP-Marg-B} are more statistically efficient than the conditional algorithm. While all three algorithms are implemented in C++, considering computational speed, \emph{pDPP-Cond} achieves the highest ESS per second due to its low computational cost. Among the three, \emph{pDPP-Marg-B} offers the best trade-off between statistical and computational efficiency, consistent with discussions of Neal's Algorithm 8 in \citet{neal2000markov} and \citet{muller2015bayesian}. It is worth noting that the marginal samplers appear to reach convergence much earlier, within the first 150 iterations as shown in Figure~\ref{figure: simulation_mcmc_mixing_efficiency} (d), suggesting that, in practice, their efficiency is higher than what is reported in Table~\ref{table: Simulation_1_ESS}, when  the burn-in period is considered. Moreover, as discussed, \emph{pDPP-Marg-B} represents the first practically accessible marginal sampler for Bayesian mixture models with a DPP prior. We consequently adopt \emph{pDPP-Marg-B} for all subsequent analyses, particularly the high-dimensional real data analysis in Section~\ref{sec: ERPs_data_analysis}.

\subsection{ Simulation 2: clustering repulsiveness}\label{Sec: Simulation_2}

In this subsection, we assess and compare the clustering repulsiveness of the proposed method against three alternative approaches: (1) the Dirichlet Process Mixture model (DPM); (2) the mixture of L-ensembles DPP of \citet{xu2016bayesian}; and (3) the mixture of Gaussian DPP of \citet{beraha2025bayesian}. The latter two approaches also utilize repulsive mixture models, with an unknown and random number of atoms, but differ notably in their latent atom structures. 
Specifically, the mixture of L-ensembles DPP assumes a specific subclass of DPPs, known as L-ensembles, to model the latent atom locations defined within a finite state space \citep{kulesza2012determinantal}. The model employs a split-merge reversible jump MCMC scheme for inference on the unknown number of latent atoms. 
The mixture of Gaussian DPP introduced by \citet{beraha2025bayesian} assumes a Gaussian DPP prior on the latent atoms and leverages a conditional Metropolis-within-Gibbs sampler that avoids the challenging task of designing problem-specific reversible jump proposals \citep{geyer1994simulation,beraha2022mcmc}. 

We implement the Gaussian DPPs method using the conditional algorithm of \citet{beraha2025bayesian} and the authors' publicly available \href{https://github.com/mberaha/interacting_mixtures}{Github code}. For a fair comparison, we calibrate their hyperparameters using the global repulsiveness measurement for DPPs proposed by \citet{moller2021couplings}. We further demonstrate that this model inherently exhibits a lower degree of repulsiveness than ours; additional details are provided in Section~\ref{appendix: Gaussian_DPPs_prior} of the Supplementary Material. Simulations for the L-ensembles DPP model are conducted using the authors' online \href{https://onlinelibrary.wiley.com/doi/full/10.1111/biom.12482}{R code}. The DPM is implemented using the R package \emph{BNPmix} \citep{corradin2021bnpmix}. For our proposed method, we employ the \emph{pDPP-Marg-B} sampler under the same hyperparameter settings described in Section~\ref{sec: Simulation_1}. 

\begin{figure}[t]
\centering
\includegraphics[width=\textwidth]{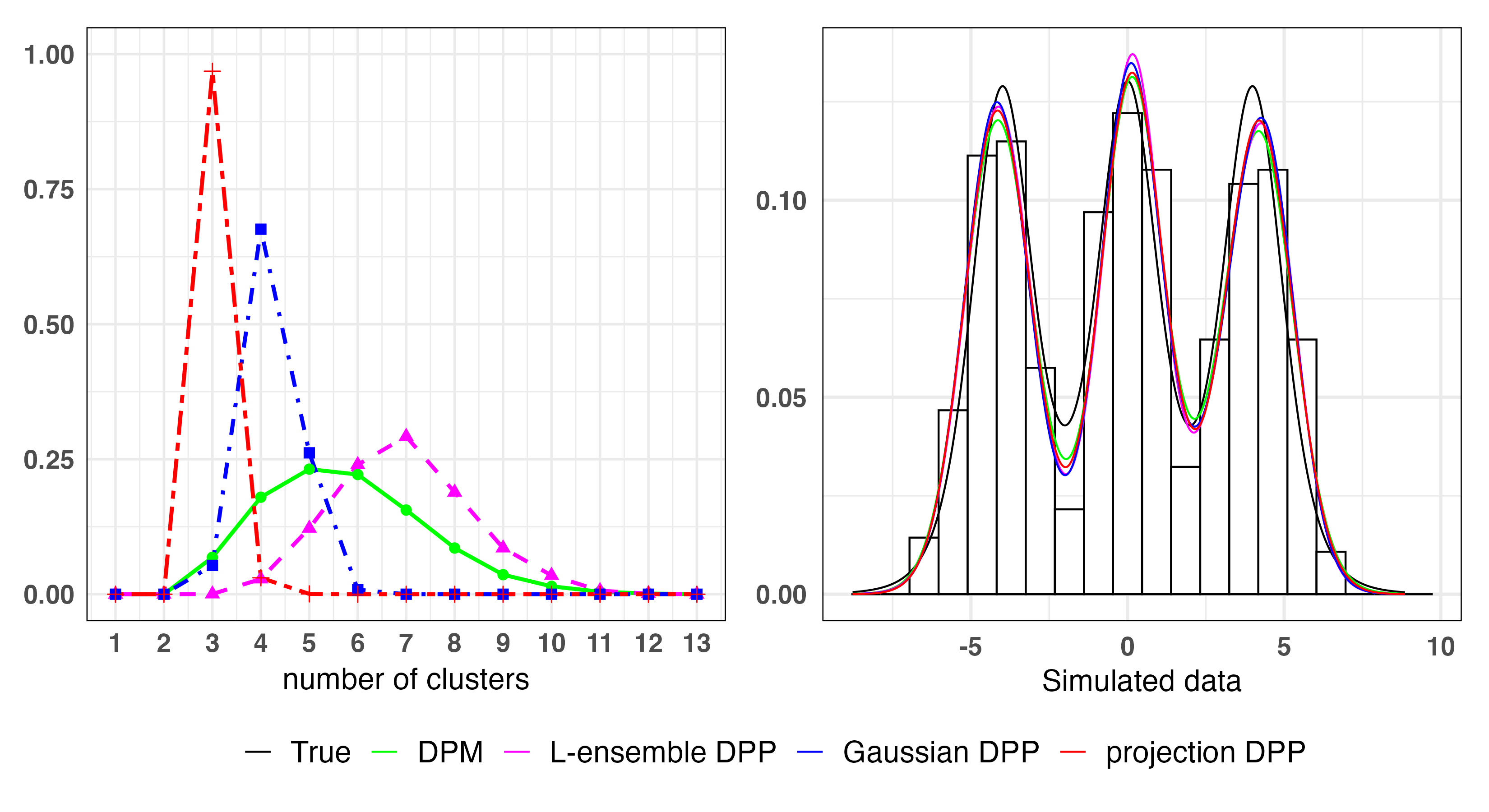}
\caption{Simulation study 2: posterior inference based on data simulated from~(\ref{eq: simulation1_dataset_1dim}). The left panel shows posterior distributions of the estimated number of clusters, and the right panel displays Bayesian mixture density estimates together with the true mixture density. See Section~\ref{Sec: Simulation_2} for details.}
\label{figure: simulation2_dim1}
\end{figure}

Figure~\ref{figure: simulation2_dim1} presents results based on data simulated from~(\ref{eq: simulation1_dataset_1dim}), showing the true data-generating density together with Bayesian mixture density estimates (right panel), as well as  posterior distributions of the number of clusters under the four alternative models (left panel). The Bayesian density estimate refers to the posterior expectation of the mixture density evaluated on a fixed grid of points. All four models yield similar and accurate density estimates, but differ substantially in the posterior distributions of the number of clusters. More specifically, the DPM, which is inherently non-repulsive, and the L-ensembles DPP allocate substantial posterior mass to much larger numbers of clusters, so to  better match the observed data density. In contrast, both the proposed projection DPP and the Gaussian DPP assign the majority of posterior probability to far fewer clusters due to their repulsive properties. In particular, our method correctly identifies three clusters. These results demonstrate that our proposed method achieves a sparser representation with interpretable clusters while maintaining a good fit to the density estimate, making it a preferable prior model for applications where such parsimony is desired.
   
\begin{figure}[t]
\centering
    \begin{subfigure}{0.495\textwidth}
        \centering
        \includegraphics[width=\textwidth]{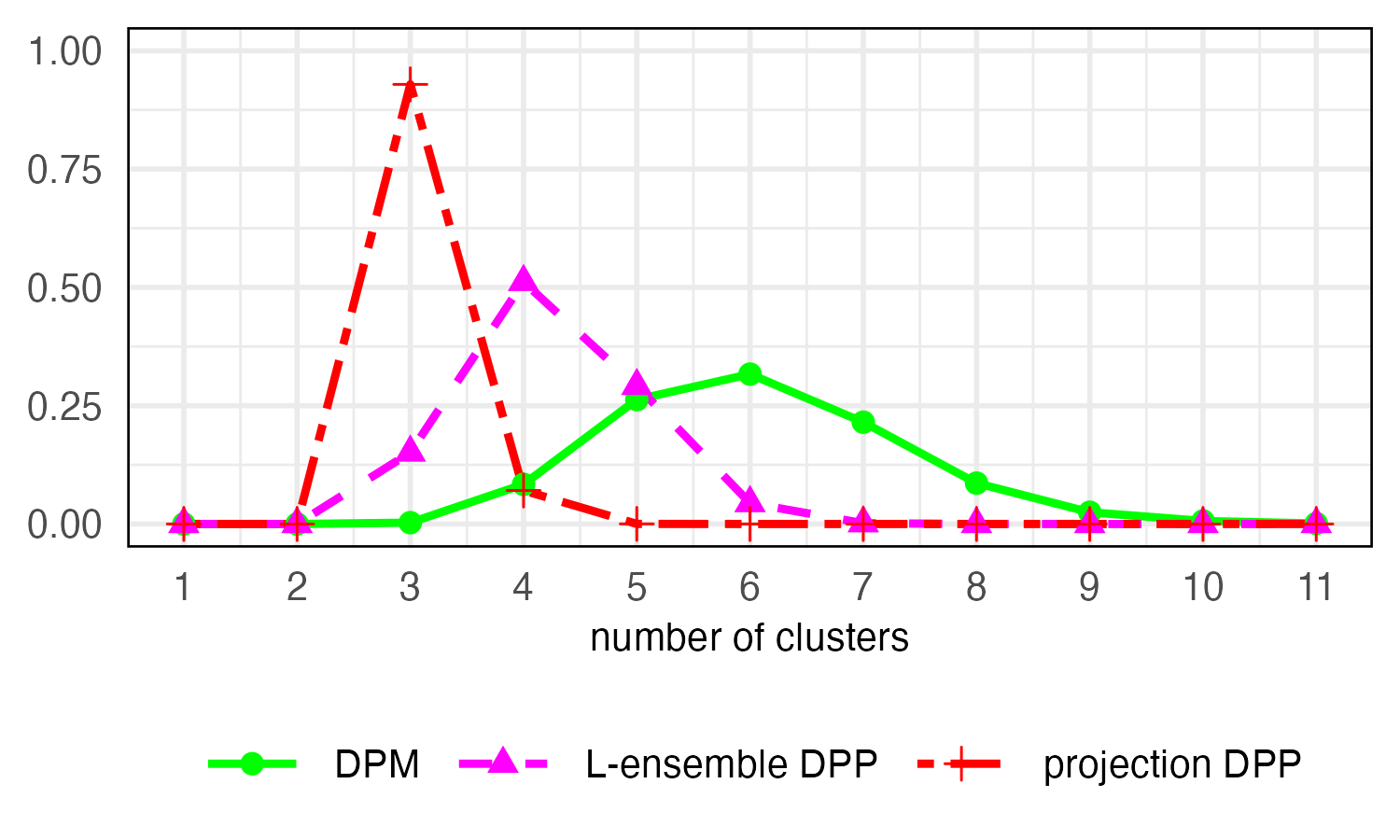}
        \caption{$d = 2$}
    \end{subfigure}
    \begin{subfigure}{0.495\textwidth}
        \centering
        \includegraphics[width=\textwidth]{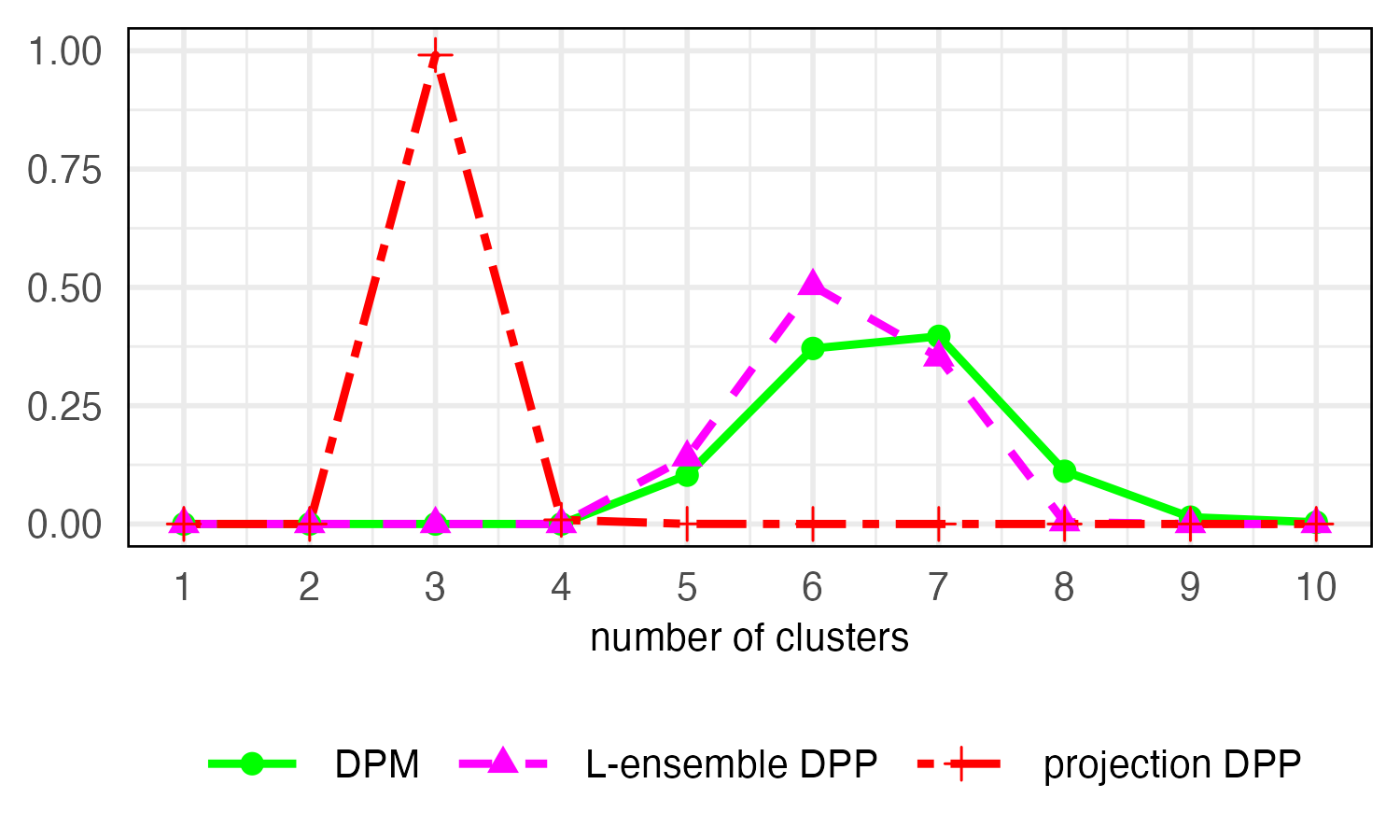}
        \caption{$d = 4$}
    \end{subfigure}
\caption{Simulation study 2: posterior distributions of the estimated number of clusters in the multivariate setting, based on data simulated from~(\ref{eq: simulation1_dataset}). Results are shown for $d=2$ (left) and $d=4$ (right). See Section~\ref{Sec: Simulation_2} for details.}
\label{figure: simulation2_mtv_clust}
\end{figure}

For the multivariate setting with dimension $d \ge 2$, we generate $n=300$ independent observations $y_i=(y_{i1},\ldots,y_{id})^\top$ from a mixture of three multivariate Student’s $t$ distributions,
\begin{equation}\label{eq: simulation1_dataset}
y_i \simiid \frac{1}{3} \,  t_6 (\bm{\mu}_1, I_d) + \frac{1}{3}\, t_6 (\bm{\mu}_2, I_d) + \frac{1}{3}\, t_6 (\bm{\mu}_3, I_d) , \qquad i = 1, \ldots, n,  
\end{equation}
where each component has 6 degrees of freedom and an identity scale matrix $I_d$. We set the component locations as follows: (i) for $d=2$, $\bm{\mu}_1 = (-4,4)^T, \bm{\mu}_2 = (0,0)^T$, and $\bm{\mu}_3 = (4,4)^T$; (ii) for $d = 4$, $\bm{\mu}_1 = (-4,-4,4,4)^T, \bm{\mu}_2 = (0,0,0,0)^T$, and $\bm{\mu}_3 = (4,4,4,4)^T$. For the proposed method, we set $a_s = 0.1$ and $\Sigma_h \simiid \operatorname{InvWi}(\tau = d+2, \Omega = I_d)$. We let $\ell = 1$, resulting in  $m = (2\ell+1)^d$ atoms in total. We define $\Theta$ as the space $\otimes_{e=1}^{d} [\bar{y}_e - 2.5 \zeta_e, \bar{y}_e + 2.5 \zeta_e]$,  covering  the simulated data, where $\bar{y}_e = \frac{1}{n} \sum_{i=1}^{n} y_{ie}$ and $\zeta_e = \max_{i=1,\ldots,n} |y_{ie} - \bar{y}_{e}|$ for $e = 1, \ldots, d$. Figure~\ref{figure: simulation2_mtv_clust} reports the posterior distribution of the number of clusters inferred from the three methods for dimensions $d = 2$ and $d = 4$. 
In both multivariate scenarios, our method assigns the highest posterior probability to the correct number of three clusters, highlighting its advantage over competing approaches in accurately capturing the true data structure.

\section{Application to ERP data.}\label{sec: ERPs_data_analysis}

To illustrate the method, we used data from the ERP CORE project \citep{kappenman2021erp}, comprising ERP recordings from 34 neurotypical young adults (ages 18–30) collected under several experimental paradigms. We focused on the Active Visual Oddball Task, which is commonly used to elicit the P3 component, a positive deflection in the EEG signal that appears around 300 ms after a rare or significant stimulus and is associated with attention and stimulus evaluation. In this task, the letters A, B, C, D, and E were presented in random order. At the start of each block, one letter was designated the target and the remaining four served as non-targets; on each trial, participants indicated whether the displayed letter was the target for that block. Following standard ERP methodology \citep{Luck2014}, subject-level ERPs were obtained by averaging EEG signals time-locked to target trials, yielding one representative waveform per participant.

\begin{figure}[t!]
    \centering
    \begin{subfigure}{0.5\textwidth}
        \centering
        \includegraphics[width=\textwidth]{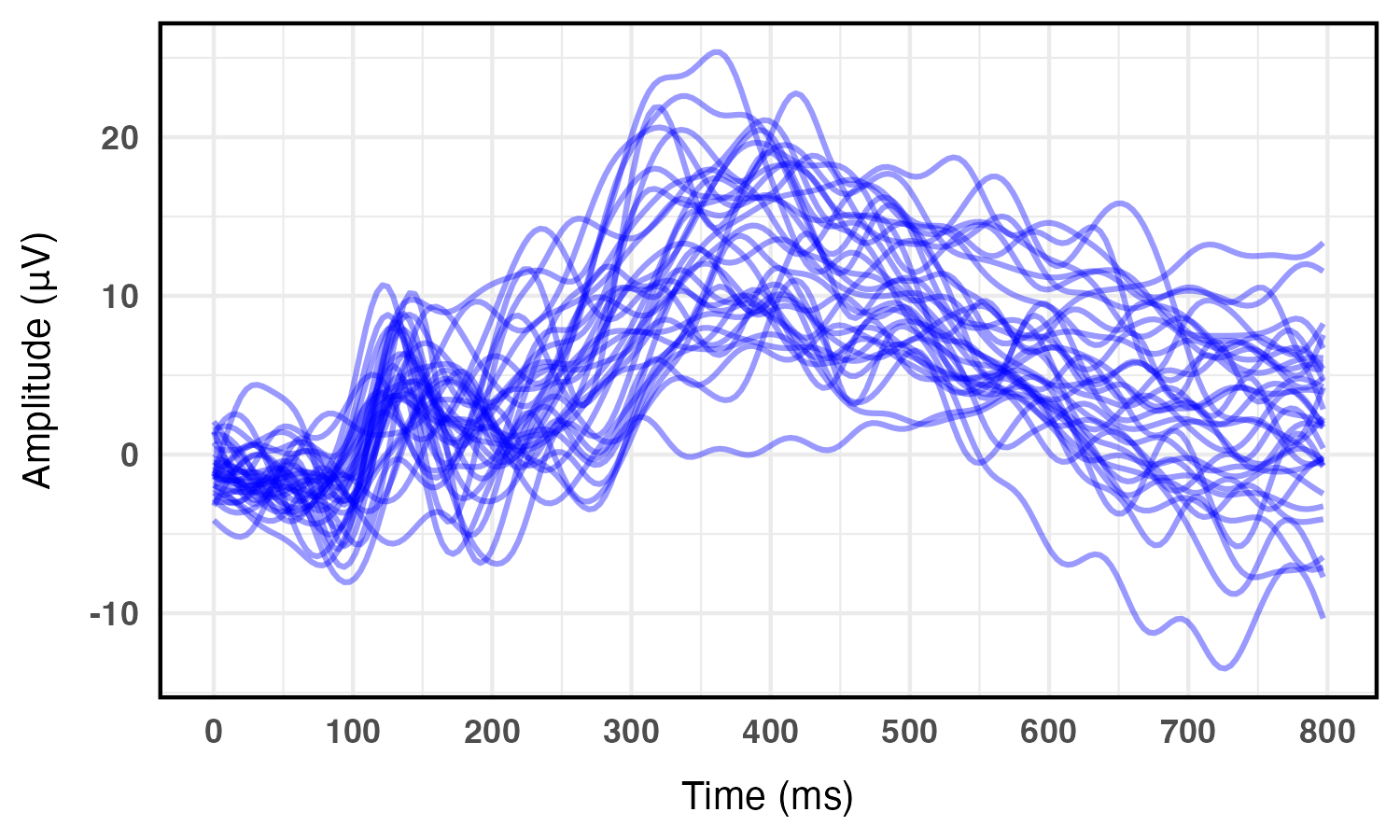}
        \caption{ERP waveforms}
    \end{subfigure}
    \begin{subfigure}{0.49\textwidth}
        \centering
        \includegraphics[width=\textwidth]{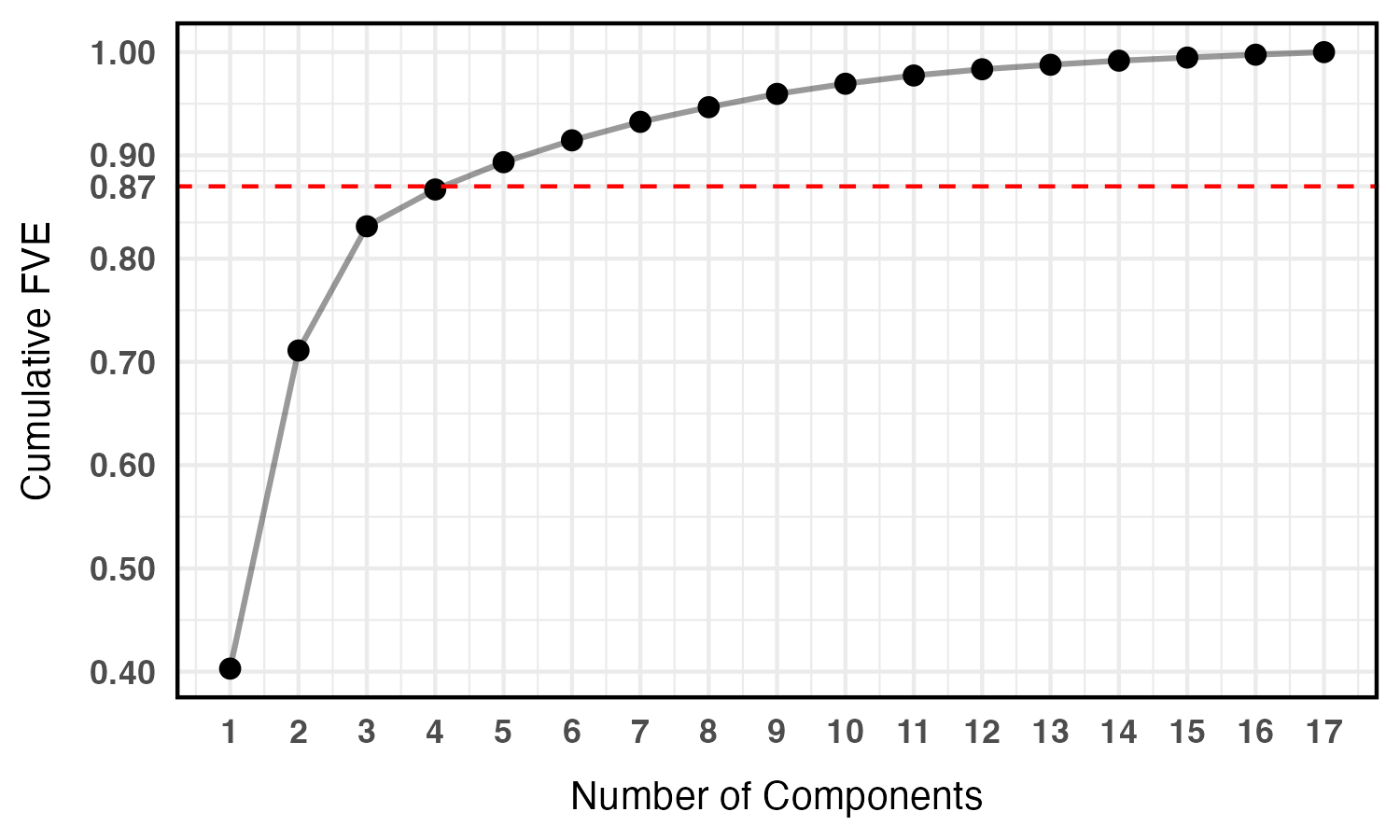}
        \caption{Fraction of Variance explained by fPCA}
    \end{subfigure}
    \caption{Real data analysis: (a) ERP waveforms from 34 subjects recorded at the Pz electrode between 0 and 800 ms in an Active Visual Oddball Task (b) Fraction of variance explained (FVE) by the functional PCA of the target ERPs. The first four principal components, accounting for 87\% of total variance, were retained for clustering. See Section~\ref{sec: ERPs_data_analysis}.}
    \label{fig: ERP_CORE_data}
\end{figure}

Recordings were acquired from 128 electrodes and epoched from $-200$ to $800$ ms relative to stimulus onset. We centered our analysis on Pz, a midline parietal site where the visual oddball P3 is typically maximal. To avoid confounding from pre-stimulus activity, analyses were restricted to the $0$–$800$ ms post-stimulus interval. The 34 waveforms used for clustering are shown in panel (a) of Figure~\ref{fig: ERP_CORE_data}. By convention, time $0$ ms marks stimulus onset, and the ERP waveform traces the subsequent neural processing over time.
Features such as P3 peak amplitude (typically observed between 300 and 600 ms in this task) are known to reflect cognitive  processes related to attention and stimulus evaluation, whereas peak latency reflects processing speed; variations in waveform shapes can indicate individual variability in cognitive functioning.
    
Building on this biological and cognitive significance, clustering the ERPs allows us to identify subgroups with shared neural response patterns that plausibly reflect distinct attentional strategies and stimulus-evaluation dynamics. Accordingly, our analysis aims to uncover interpretable structure in the ERP waveforms, offering a framework for understanding variability across subjects.


As a preprocessing step, we applied functional principal component analysis (fPCA) to the 34 ERP waveforms to obtain a low-dimensional representation that preserves the main sources of variability across subjects. The right panel (b) of Figure~\ref{fig: ERP_CORE_data} shows the proportion of variance explained by the leading fPCA components. The first four components, which together account for 87\% of the total variation, were retained as a low-dimensional embedding for clustering. Clustering was then performed on the resulting $d=4$ dimensional score vectors to study subject-level heterogeneity in ERP responses.

We applied the proposed projection DPP mixture model to the principal component vectors $y_i \in \mathds{R}^d$ for $i = 1, \ldots, n=34$. The hyperparameters were set as $a_s = 0.5$ and $\Sigma_h \simiid \mathrm{InvWi}(\tau, \Omega)$, with $\tau = d+2$ and $\Omega$ equal to the sample covariance of the fPCA scores. We fixed $\ell=1$, yielding $m=(2\ell+1)^d$ atoms, and defined $\Theta=\otimes_{e=1}^{d} [\bar{y}_e - 1.5 \zeta_e,; \bar{y}_e + 1.5 \zeta_e]$, covering all the score vectors, where $\bar{y}_e$ and $\zeta_e$ are as in Section~\ref{Sec: Simulation_2}. We run 10,000 MCMC iterations, discarding the first 5,000 as burn-in. Posterior clustering estimates were obtained from the MCMC samples via the SALSO greedy search algorithm \citep{dahl2022search}, which minimizes the posterior expectation of the Variation of Information loss.

\begin{figure}[t!]
\centering
\includegraphics[width=\textwidth]{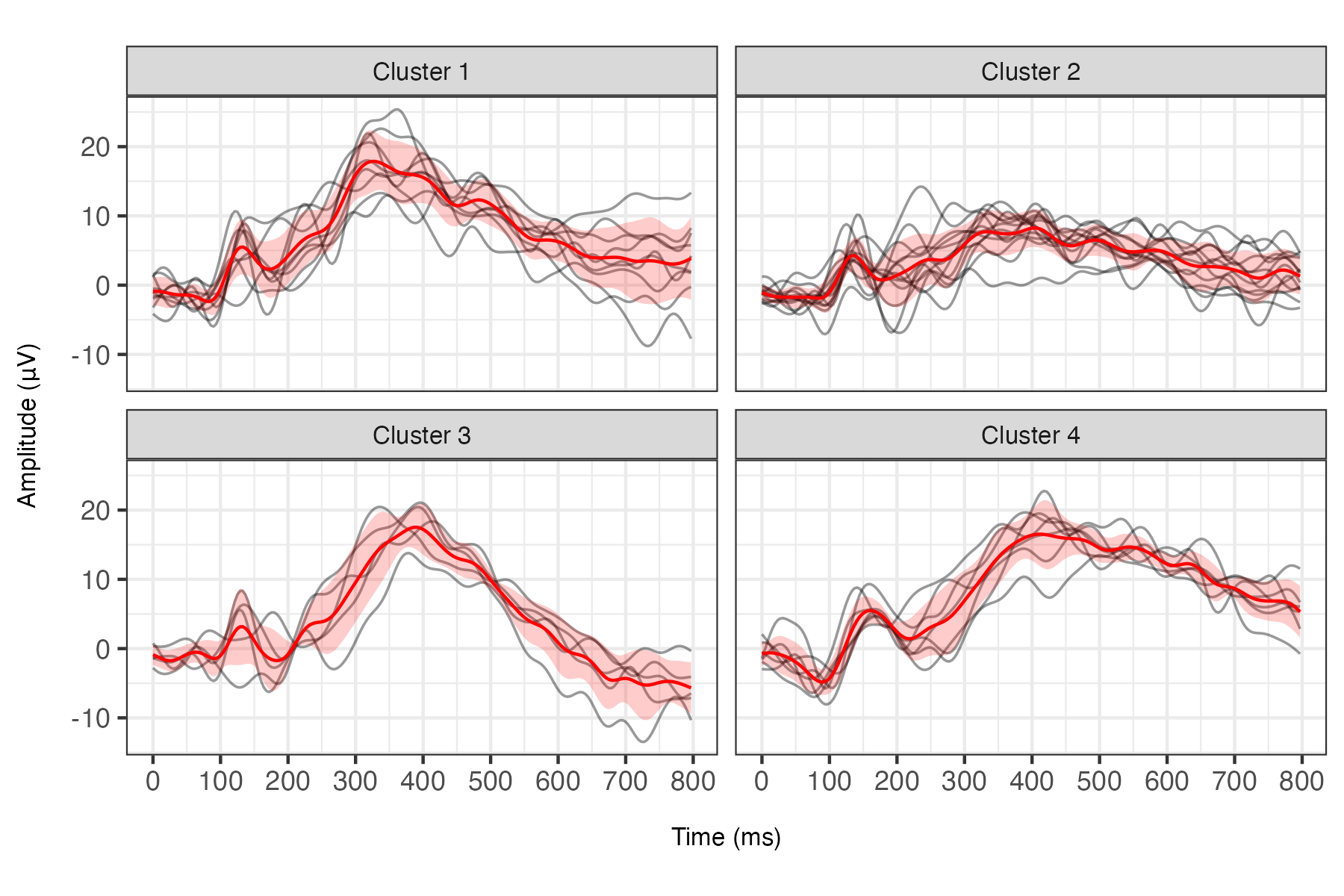}
\caption{Real data analysis (projection DPP): clustering of 34 ERPs using the mixture model of projection DPP applied to the functional PCA scores. The four cluster sizes are 9, 13, 5, and 7. The red curve represents the averaged ERP waveform within each cluster, and the red shaded area indicates the $\pm$ 1 standard deviation bandwidth. See Section~\ref{sec: ERPs_data_analysis}.}
\label{figure: fPCA_pDPP_34subjects}
\end{figure}

Figure~\ref{figure: fPCA_pDPP_34subjects} presents the clustering result. Our method identified four well-separated clusters of sizes 9, 13, 5, and 7, with no singleton or  overly small clusters. This illustrates the repulsive property of the projection DPP prior, which discourages redundant clusters. For each cluster, the figure also shows the mean ERP waveform (red curve) and its $\pm 1$ standard deviation envelope (red shading).
The resulting clusters reveal substantial neuro-cognitive heterogeneity. Cluster 1 is characterized by a single prominent positive-going wave. The ERP reaches a large amplitude 
quite early, with a broad peak roughly around $300-400$ ms. This peak corresponds to one of the most studied ERP components, the P300 component, and, more specifically, to the so-called P3b subcomponent, which is commonly linked to the evaluation of task-relevant stimuli. 
In this cluster, the pronounced P3b suggests that these individuals allocated more attention and cognitive resources to the target; the brain was actively ``updating context'' upon each target appearance \citep{Luck2014}. 
Cluster 2 shows a small positive deflection emerging roughly around 200 ms after the stimulus, which peaks by 250 ms and then tapers off. The amplitude of this positive peak is modest, especially with respect to other clusters. 
These reduced responses typically reflect less sustained attention, reduced motivation, or a preference for minimal-effort strategies. Cluster 3's average waveform is distinguished by a prominent P3b that peaks later in time and a prolonged positive reducing wave thereafter. Unlike Clusters 1 and 2, the major positive deflection in Cluster 3 occurs around 400-450 ms post-stimulus, i.e. with a noticeable delay. This pattern suggests a thorough but slower processing of the target, often coupled with prolonged evaluation or memory-related activity. These participants devote considerable attention to the oddball task and likely update their mental context robustly, but at a more relaxed pace, with their brains 
internally ``double-checking'' and possibly encoding the event into memory. However, the waveform then returns quickly toward baseline, suggesting that individuals disengage promptly once the stimulus has been classified.  The average ERP waveform for Cluster 4 shows a characteristic sequence of components: an early negative dip, a mid-latency positive surge around 150 ms, and a long-lasting late positive deflection extending roughly to 600-700 ms. The initial negative-going deflection corresponds to the so-called N1 component, which reflects the initial sensory processing of the stimulus.
The follow-up surge is associated with the P2 component, often associated with further evaluation of the stimulus and initial attention allocation beyond the very early sensory registration. 
The most prominent feature of Cluster 4 is the broad, sustained positive waveform that begins a few hundred milliseconds after stimulus onset and extends toward 600-700 ms. This suggests that this cluster represents individuals that combine early detection with sustained and prolonged attention. In summary, the clusters showcase distinct response profiles: some individuals showed rapid and decisive engagement (Cluster 1), others minimal and attenuated responses (Cluster 2), some engaged in early control followed by transient evaluation (Cluster 3), and others combined early perceptual selection with the most sustained evaluative processing (Cluster 4). These findings highlight the strength of the proposed method in uncovering interpretable clusters from complex brain data, offering insights for subsequent neuro-cognitive analysis, and, possibly, targeted interventions.

\begin{figure}[t!]
\centering
\includegraphics[width=\textwidth]{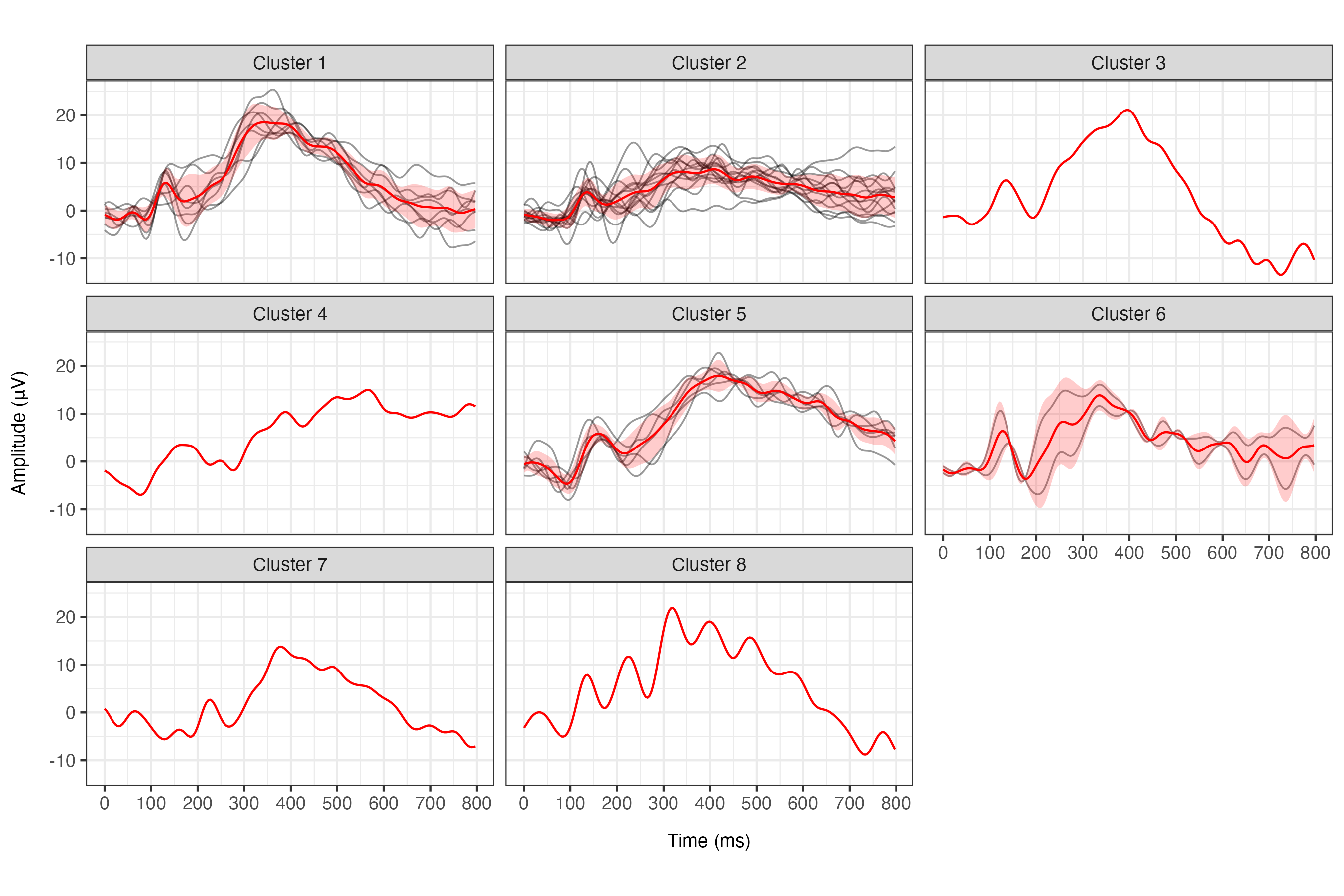}
\caption{Real data analysis (DPM model): clustering of 34 ERPs using a DPM model on the functional PCA scores. The model yields eight clusters of sizes 8, 14, 1, 1, 6, 2, 1, and 1, respectively. See details in Section~\ref{sec: ERPs_data_analysis}.}
\label{figure: fPCA_DPM_mass_0_1}
\end{figure}

For comparison, we also applied the DPM model to the same set of fPCA score vectors for clustering the 34 ERP waveforms, using the \texttt{PYdensity()} function from the R package \texttt{BNPmix}. The DPM consistently produced partitions that were difficult to interpret across a range of values for the total mass parameters. For example, with the total mass parameter set at 0.1, a value that should encourage a small number of clusters, the DPM model still yielded a clustering with many singleton groups and a group of only two waveforms (Figure~\ref{figure: fPCA_DPM_mass_0_1}). This clustering structure proves difficult to interpret and offers limited scientific insight. As detailed in Section~\ref{appendix: ERPs_clust_fPCA_DPM} of the Supplementary Material (Figures~\ref{figure: fPCA_DPM_mass0_0_1}--\ref{figure: fPCA_DPM_mass0_2}), increasing the total mass parameter merely increases the number of clusters without altering the underlying findings: a few large clusters accompanied by many small or singleton ones. This behavior is characteristic of the DPM, which tends to favor such partitions. Although the DPM may capture subtle waveform heterogeneity, the resulting partitions are challenging to interpret. These limitations underscore the need for an alternative clustering strategy that yields a smaller number of more balanced and interpretable clusters, motivating the development of our proposed repulsive clustering framework.

\section{Discussion}\label{sec: conclusion_discussion}

Bayesian repulsive mixture models have recently emerged as a powerful tool for model-based clustering. Compared to traditional mixture models, repulsive mixtures produce fewer and more interpretable clustering, thus enhancing the practical applicability of Bayesian nonparametric models in various domains.
However, the uptake of repulsive mixtures from practitioners remains limited, due to the complexity of the associated posterior inference, which results in slowly mixing algorithms that may produce unreliable summaries.
In this article, we have proposed the use of a projection determinantal point process (DPP) as a repulsive prior for the cluster-specific parameters.
The analytical tractability of projection DPPs allows us to derive
simple posterior and predictive characterizations, leading to exact and efficient algorithms for posterior inference that outperform recent competitors in terms of sampling efficiency. 
Moreover, the use of projection DPPs allowed us to provide strong frequentist guarantees for the posterior distribution.
Specifically, in the well-specified setting, we have shown that our model yields consistent density estimates, that extra mixture components (if any) {are ``emptied out'' a posteriori, and that the mixing measure contracts to the true one in the 1-Wasserstein distance. The contraction rates obtained for these behaviors match the ones previously obtained for standard (non-repulsive) mixtures, demonstrating that there is no downside in assuming our repulsive prior even if the model is well-specified. Together with the strong empirical performance on both simulated datasets in a misspecified setting and the real data analysis of ERP curves, our theoretical analysis entails that projection DPP mixtures can be safely used as a default option for model-based clustering.

A limitation of our approach is that, contrary to what happens when assuming more general DPP priors, the number of components $m$ needs to be fixed by the practitioner. This requirement is shared by other common approaches in the literature \citep{miller2018mixture,fruhwirth2019handbook}.  Theorem \ref{thm:extra_components} provides some reassurance: it guarantees that extra components (if any) will be emptied out a posteriori in the large sample regime; our experiments confirm this behavior in practice. Still, one needs to choose $m$ to be at least the number of clusters, which is typically unknown.
Despite this limitation, we view our theoretical analysis and algorithms as foundational steps toward more general DPP priors that do not require fixing $m$. Indeed, as shown in \cite{hough2006determinantal}, any DPP can be represented as a mixture of projection DPPs, where the mixing occurs over a countable number of independent Bernoulli random variables.
Then, conditional on the Bernoulli variables, our Theorems \ref{thrm: theorem_1} and \ref{thrm: theorem_2} yield the same conditional distribution of the mixing measure and the predictive distribution of a new observation, thus providing the main ingredients for new algorithms that do not require reversible jump moves.

A natural extension of our framework is to allow mixture atoms to depend on covariates, thereby enabling the response distribution to vary flexibly with observed features. This would connect our proposal with the literature on dependent Dirichlet processes \citep{quintana2022dependent}, where clustering structures also depend on covariates. In the context of functional data, our current analysis is based on multivariate representations from principal component scores. Future work can directly model the functional curves themselves. This could likely be achieved by developing functional point process models with repulsive structure \citep{ghorbani2021functional}, allowing inference on curve-level heterogeneity. We plan to address these extensions in future work.


\section{Disclosure statement}\label{disclosure-statement}

The authors have no conflicts of interest to declare.

\section{Acknowledgements}

M.B. and F.C. gratefully acknowledge support from the Italian Ministry of Education, University and Research (MUR), ``Dipartimenti di Eccellenza'' grant 2023-2027. F.C. is supported by the European Union-Next Generation EU funds, component M4C2, investment 1.1., PRIN-PNRR 2022 (P2022H5WZ9).

\bibliographystyle{chicago}
\bibliography{references}

\begin{thebibliography}{}

\bibitem[\protect\citeauthoryear{Argiento and De~Iorio}{Argiento and De~Iorio}{2022}]{argiento2022infinity}
Argiento, R. and M.~De~Iorio (2022).
\newblock {Is infinity that far? A Bayesian nonparametric perspective of finite mixture models}.
\newblock {\em The Annals of Statistics\/}~{\em 50\/}(5), 2641--2663.

\bibitem[\protect\citeauthoryear{Baccelli, B{\l}aszczyszyn, and Karray}{Baccelli et~al.}{2024}]{baccelli2024random}
Baccelli, F., B.~B{\l}aszczyszyn, and M.~Karray (2024).
\newblock Random measures, point processes, and stochastic geometry.
\newblock \url{https://hal.inria.fr/hal-02460214v2}.

\bibitem[\protect\citeauthoryear{Beraha, Argiento, Camerlenghi, and Guglielmi}{Beraha et~al.}{2025}]{beraha2025bayesian}
Beraha, M., R.~Argiento, F.~Camerlenghi, and A.~Guglielmi (2025).
\newblock Bayesian mixture models with repulsive and attractive atoms.
\newblock {\em J. R. Stat. Soc. Ser. B\/}.

\bibitem[\protect\citeauthoryear{Beraha, Argiento, M{\o}ller, and Guglielmi}{Beraha et~al.}{2022}]{beraha2022mcmc}
Beraha, M., R.~Argiento, J.~M{\o}ller, and A.~Guglielmi (2022).
\newblock {MCMC computations for Bayesian mixture models using repulsive point processes}.
\newblock {\em J. Comp. Graph. Statist.\/}~{\em 31\/}(2).

\bibitem[\protect\citeauthoryear{Bianchini, Guglielmi, and Quintana}{Bianchini et~al.}{2020}]{bianchini2020determinantal}
Bianchini, I., A.~Guglielmi, and F.~A. Quintana (2020).
\newblock Determinantal point process mixtures via spectral density approach.
\newblock {\em Bayesian Analysis\/}~{\em 15}, 187--214.

\bibitem[\protect\citeauthoryear{Borodin and Soshnikov}{Borodin and Soshnikov}{2003}]{BorodinSoshnikov2003JanossyI}
Borodin, A. and A.~Soshnikov (2003).
\newblock {Janossy Densities I. Determinantal Ensembles}.
\newblock {\em Journal of Statistical Physics\/}~{\em 113\/}(3-4), 595--610.

\bibitem[\protect\citeauthoryear{Coeurjolly, Mazoyer, and Amblard}{Coeurjolly et~al.}{2021}]{coeurjolly2021monte}
Coeurjolly, J.-F., A.~Mazoyer, and P.-O. Amblard (2021).
\newblock {Monte Carlo integration of non-differentiable functions on [0, 1] $\iota$, $\iota$= 1,…, d, using a single determinantal point pattern defined on [0, 1] d}.
\newblock {\em Electronic Journal of Statistics\/}~{\em 15\/}(2), 6228--6280.

\bibitem[\protect\citeauthoryear{Corradin, Canale, and Nipoti}{Corradin et~al.}{2021}]{corradin2021bnpmix}
Corradin, R., A.~Canale, and B.~Nipoti (2021).
\newblock {BNPmix: An R package for Bayesian nonparametric modeling via Pitman-Yor mixtures}.
\newblock {\em J. Stat. Soft.\/}~{\em 100}, 1--33.

\bibitem[\protect\citeauthoryear{Cremaschi, Wertz, and De~Iorio}{Cremaschi et~al.}{2025}]{cremaschi2025repulsion}
Cremaschi, A., T.~M. Wertz, and M.~De~Iorio (2025).
\newblock Repulsion, chaos, and equilibrium in mixture models.
\newblock {\em J. R. Stat. Soc. Ser. B\/}~{\em 87\/}(2), 389--432.

\bibitem[\protect\citeauthoryear{Dahl, Johnson, and M{\"u}ller}{Dahl et~al.}{2022}]{dahl2022search}
Dahl, D.~B., D.~J. Johnson, and P.~M{\"u}ller (2022).
\newblock Search algorithms and loss functions for bayesian clustering.
\newblock {\em Journal of Computational and Graphical Statistics\/}~{\em 31\/}(4), 1189--1201.

\bibitem[\protect\citeauthoryear{Favaro, Hadjicharalambous, and Pr{\"u}nster}{Favaro et~al.}{2011}]{favaro2011class}
Favaro, S., G.~Hadjicharalambous, and I.~Pr{\"u}nster (2011).
\newblock On a class of distributions on the simplex.
\newblock {\em Journal of Statistical Planning and Inference\/}~{\em 141\/}(9), 2987--3004.

\bibitem[\protect\citeauthoryear{Favaro and Teh}{Favaro and Teh}{2013}]{favaroteh2013}
Favaro, S. and Y.~W. Teh (2013).
\newblock {MCMC for Normalized Random Measure Mixture Models}.
\newblock {\em Statistical Science\/}~{\em 28\/}(3), 335 -- 359.

\bibitem[\protect\citeauthoryear{Fruhwirth-Schnatter, Celeux, and Robert}{Fruhwirth-Schnatter et~al.}{2019}]{fruhwirth2019handbook}
Fruhwirth-Schnatter, S., G.~Celeux, and C.~P. Robert (2019).
\newblock {\em Handbook of mixture analysis}.
\newblock CRC press.

\bibitem[\protect\citeauthoryear{Geyer and M{\o}ller}{Geyer and M{\o}ller}{1994}]{geyer1994simulation}
Geyer, C.~J. and J.~M{\o}ller (1994).
\newblock Simulation procedures and likelihood inference for spatial point processes.
\newblock {\em Scandinavian journal of statistics\/}, 359--373.

\bibitem[\protect\citeauthoryear{Ghilotti, Beraha, and Guglielmi}{Ghilotti et~al.}{2025}]{ghilotti2025bayesian}
Ghilotti, L., M.~Beraha, and A.~Guglielmi (2025).
\newblock Bayesian clustering of high-dimensional data via latent repulsive mixtures.
\newblock {\em Biometrika\/}~{\em 112\/}(2), asae059.

\bibitem[\protect\citeauthoryear{Ghorbani, Cronie, Mateu, and Yu}{Ghorbani et~al.}{2021}]{ghorbani2021functional}
Ghorbani, M., O.~Cronie, J.~Mateu, and J.~Yu (2021).
\newblock Functional marked point processes: a natural structure to unify spatio-temporal frameworks and to analyse dependent functional data.
\newblock {\em Test\/}~{\em 30\/}(3), 529--568.

\bibitem[\protect\citeauthoryear{Ghosal and Van Der~Vaart}{Ghosal and Van Der~Vaart}{2001}]{ghosal2001entropies}
Ghosal, S. and A.~W. Van Der~Vaart (2001).
\newblock Entropies and rates of convergence for maximum likelihood and bayes estimation for mixtures of normal densities.
\newblock {\em Annals of Statistics\/}, 1233--1263.

\bibitem[\protect\citeauthoryear{Griffin and Walker}{Griffin and Walker}{2011}]{griffin2011posterior}
Griffin, J.~E. and S.~G. Walker (2011).
\newblock Posterior simulation of normalized random measure mixtures.
\newblock {\em Journal of Computational and Graphical Statistics\/}~{\em 20\/}(1), 241--259.

\bibitem[\protect\citeauthoryear{Ho and Nguyen}{Ho and Nguyen}{2016}]{ho2016strong}
Ho, N. and X.~Nguyen (2016).
\newblock On strong identifiability and convergence rates of parameter estimation in finite mixtures.
\newblock {\em Electronic Journal of Statistics\/}~{\em 10\/}(1), 271--307.

\bibitem[\protect\citeauthoryear{Hoff}{Hoff}{2009}]{hoff2009simulation}
Hoff, P.~D. (2009).
\newblock {Simulation of the matrix Bingham--von Mises--Fisher distribution, with applications to multivariate and relational data}.
\newblock {\em Journal of Computational and Graphical Statistics\/}~{\em 18\/}(2), 438--456.

\bibitem[\protect\citeauthoryear{Hough, Krishnapur, Peres, et~al.}{Hough et~al.}{2009}]{hough2009zeros}
Hough, J.~B., M.~Krishnapur, Y.~Peres, et~al. (2009).
\newblock {\em {Zeros of Gaussian analytic functions and determinantal point processes}}, Volume~51.
\newblock American Mathematical Soc.

\bibitem[\protect\citeauthoryear{Hough, Krishnapur, Peres, and Vir{\'a}g}{Hough et~al.}{2006}]{hough2006determinantal}
Hough, J.~B., M.~Krishnapur, Y.~Peres, and B.~Vir{\'a}g (2006).
\newblock Determinantal processes and independence.
\newblock {\em Probability Surveys\/}~{\em 3}.

\bibitem[\protect\citeauthoryear{James, Lijoi, and Pr{\"u}nster}{James et~al.}{2009}]{james2009posterior}
James, L.~F., A.~Lijoi, and I.~Pr{\"u}nster (2009).
\newblock Posterior analysis for normalized random measures with independent increments.
\newblock {\em Scandinavian Journal of Statistics\/}~{\em 36\/}(1), 76--97.

\bibitem[\protect\citeauthoryear{Jauch, Hoff, and Dunson}{Jauch et~al.}{2021}]{jauch2021monte}
Jauch, M., P.~D. Hoff, and D.~B. Dunson (2021).
\newblock {Monte Carlo simulation on the Stiefel manifold via polar expansion}.
\newblock {\em Journal of Computational and Graphical Statistics\/}~{\em 30\/}(3), 622--631.

\bibitem[\protect\citeauthoryear{Johansson}{Johansson}{2001}]{Johansson2001DiscreteOPE}
Johansson, K. (2001).
\newblock {Discrete orthogonal polynomial ensembles and the Plancherel measure}.
\newblock {\em Annals of Mathematics\/}~{\em 153\/}(1), 259--296.

\bibitem[\protect\citeauthoryear{Kallenberg}{Kallenberg}{2021}]{kallenberg2021foundations}
Kallenberg, O. (2021).
\newblock {\em {Foundations of Modern Probability}\/} (3 ed.), Volume~99 of {\em Probability Theory and Stochastic Modelling}.
\newblock Springer.

\bibitem[\protect\citeauthoryear{Kappenman, Farrens, Zhang, Stewart, and Luck}{Kappenman et~al.}{2021}]{kappenman2021erp}
Kappenman, E.~S., J.~L. Farrens, W.~Zhang, A.~X. Stewart, and S.~J. Luck (2021).
\newblock {ERP CORE: An open resource for human event-related potential research}.
\newblock {\em NeuroImage\/}~{\em 225}, 117465.

\bibitem[\protect\citeauthoryear{Kulesza, Taskar, et~al.}{Kulesza et~al.}{2012}]{kulesza2012determinantal}
Kulesza, A., B.~Taskar, et~al. (2012).
\newblock Determinantal point processes for machine learning.
\newblock {\em Foundations and Trends{\textregistered} in Machine Learning\/}~{\em 5\/}(2--3), 123--286.

\bibitem[\protect\citeauthoryear{Lavancier, M{\o}ller, and Rubak}{Lavancier et~al.}{2015}]{lavancier2015determinantal}
Lavancier, F., J.~M{\o}ller, and E.~Rubak (2015).
\newblock Determinantal point process models and statistical inference.
\newblock {\em J. R. Stat. Soc. Ser. B\/}~{\em 77\/}(4), 853--877.

\bibitem[\protect\citeauthoryear{Lavancier and Rubak}{Lavancier and Rubak}{2023}]{lavancier2023simulation}
Lavancier, F. and E.~Rubak (2023).
\newblock On simulation of continuous determinantal point processes.
\newblock {\em Statistics and Computing\/}~{\em 33\/}(5), 120.

\bibitem[\protect\citeauthoryear{Luck}{Luck}{2014}]{Luck2014}
Luck, S.~J. (2014).
\newblock {\em {An Introduction to the Event-Related Potential Technique}\/} (2 ed.).
\newblock Cambridge, MA: MIT Press.

\bibitem[\protect\citeauthoryear{Lyons}{Lyons}{2014}]{lyons2014determinantal}
Lyons, R. (2014).
\newblock Determinantal probability: basic properties and conjectures.
\newblock {\em Proceedings of the International Congress of Mathematicians\/}, 137--161.

\bibitem[\protect\citeauthoryear{Macchi}{Macchi}{1975}]{macchi1975coincidence}
Macchi, O. (1975).
\newblock The coincidence approach to stochastic point processes.
\newblock {\em Advances in Applied Probability\/}~{\em 7\/}(1), 83--122.

\bibitem[\protect\citeauthoryear{Miller and Harrison}{Miller and Harrison}{2014}]{miller2014inconsistency}
Miller, J.~W. and M.~T. Harrison (2014).
\newblock {Inconsistency of Pitman-Yor process mixtures for the number of components}.
\newblock {\em The Journal of Machine Learning Research\/}~{\em 15\/}(1), 3333--3370.

\bibitem[\protect\citeauthoryear{Miller and Harrison}{Miller and Harrison}{2018}]{miller2018mixture}
Miller, J.~W. and M.~T. Harrison (2018).
\newblock Mixture models with a prior on the number of components.
\newblock {\em Journal of the American Statistical Association\/}~{\em 113\/}(521), 340--356.

\bibitem[\protect\citeauthoryear{M{\o}ller and O’Reilly}{M{\o}ller and O’Reilly}{2021}]{moller2021couplings}
M{\o}ller, J. and E.~O’Reilly (2021).
\newblock {Couplings for determinantal point processes and their reduced Palm distributions with a view to quantifying repulsiveness}.
\newblock {\em Journal of Applied Probability\/}~{\em 58\/}(2), 469--483.

\bibitem[\protect\citeauthoryear{M{\o}ller and Waagepetersen}{M{\o}ller and Waagepetersen}{2004}]{moller2004statistical}
M{\o}ller, J. and R.~P. Waagepetersen (2004).
\newblock {\em Statistical inference and simulation for spatial point processes}.
\newblock CRC press.

\bibitem[\protect\citeauthoryear{M{\"u}ller, Quintana, Jara, and Hanson}{M{\"u}ller et~al.}{2015}]{muller2015bayesian}
M{\"u}ller, P., F.~A. Quintana, A.~Jara, and T.~Hanson (2015).
\newblock {\em {{B}ayesian nonparametric data analysis}}, Volume~1.
\newblock Springer.

\bibitem[\protect\citeauthoryear{Neal}{Neal}{2000}]{neal2000markov}
Neal, R.~M. (2000).
\newblock {Markov chain sampling methods for Dirichlet process mixture models}.
\newblock {\em Journal of computational and graphical statistics\/}~{\em 9\/}(2), 249--265.

\bibitem[\protect\citeauthoryear{Nguyen}{Nguyen}{2013}]{nguyen2013convergence}
Nguyen, X. (2013).
\newblock Convergence of latent mixing measures in finite and infinite mixture models.
\newblock {\em The Annuals of Statistics\/}~{\em 41\/}(1), 370--400.

\bibitem[\protect\citeauthoryear{Papaspiliopoulos and Roberts}{Papaspiliopoulos and Roberts}{2008}]{papaspiliopoulos2008retrospective}
Papaspiliopoulos, O. and G.~O. Roberts (2008).
\newblock Retrospective markov chain monte carlo methods for dirichlet process hierarchical models.
\newblock {\em Biometrika\/}, 169--186.

\bibitem[\protect\citeauthoryear{Petralia, Rao, and Dunson}{Petralia et~al.}{2012}]{petralia2012repulsive}
Petralia, F., V.~Rao, and D.~Dunson (2012).
\newblock Repulsive mixtures.
\newblock {\em Advances in neural information processing systems\/}~{\em 25}.

\bibitem[\protect\citeauthoryear{Quinlan, Page, and Quintana}{Quinlan et~al.}{2018}]{quinlan2018density}
Quinlan, J.~J., G.~L. Page, and F.~A. Quintana (2018).
\newblock Density regression using repulsive distributions.
\newblock {\em Journal of Statistical Computation and Simulation\/}~{\em 88\/}(15), 2931--2947.

\bibitem[\protect\citeauthoryear{Quintana, M{\"u}ller, Jara, and MacEachern}{Quintana et~al.}{2022}]{quintana2022dependent}
Quintana, F.~A., P.~M{\"u}ller, A.~Jara, and S.~N. MacEachern (2022).
\newblock {The dependent Dirichlet process and related models}.
\newblock {\em Statistical Science\/}~{\em 37\/}(1), 24--41.

\bibitem[\protect\citeauthoryear{Regazzini, Lijoi, and Pr{\"u}nster}{Regazzini et~al.}{2003}]{regazzini2003distributional}
Regazzini, E., A.~Lijoi, and I.~Pr{\"u}nster (2003).
\newblock Distributional results for means of normalized random measures with independent increments.
\newblock {\em The Annals of Statistics\/}~{\em 31\/}(2), 560--585.

\bibitem[\protect\citeauthoryear{Rousseau and Mengersen}{Rousseau and Mengersen}{2011}]{rousseau2011asymptotic}
Rousseau, J. and K.~Mengersen (2011).
\newblock Asymptotic behaviour of the posterior distribution in overfitted mixture models.
\newblock {\em J. R. Stat. Soc. Ser. B\/}~{\em 73\/}(5), 689--710.

\bibitem[\protect\citeauthoryear{Xie and Xu}{Xie and Xu}{2020}]{xie2020bayesian}
Xie, F. and Y.~Xu (2020).
\newblock {Bayesian repulsive Gaussian mixture model}.
\newblock {\em Journal of the American Statistical Association\/}~{\em 115\/}(529), 187--203.

\bibitem[\protect\citeauthoryear{Xu, M{\"u}ller, and Telesca}{Xu et~al.}{2016}]{xu2016bayesian}
Xu, Y., P.~M{\"u}ller, and D.~Telesca (2016).
\newblock {Bayesian inference for latent biologic structure with determinantal point processes (DPP)}.
\newblock {\em Biometrics\/}~{\em 72\/}(3), 955--964.

\end{thebibliography}

\newpage
\appendix

\begin{center}
   \LARGE \textbf{ Supplementary materials to:\\
    ``Repulsive Mixture Model with Projection Determinantal Point Process''}
\end{center}



\renewcommand{\thesection}{\Alph{section}}
\renewcommand{\theequation}{\thesection\arabic{equation}}
\renewcommand{\themyalgorithm}{\thesection\arabic{myalgorithm}}
\renewcommand{\thedefinition}{\thesection\arabic{definition}} 
\renewcommand{\theremark}{\thesection\arabic{remark}} 
\renewcommand{\theproperty}{\thesection\arabic{property}} 
\renewcommand{\thefigure}{\thesection\arabic{figure}}
\renewcommand{\thetable}{\thesection\arabic{table}}
\renewcommand{\thelemma}{\thesection\arabic{lemma}}

\makeatletter
\@addtoreset{equation}{section}
\@addtoreset{myalgorithm}{section}
\@addtoreset{definition}{section}
\@addtoreset{remark}{section}
\@addtoreset{property}{section}
\@addtoreset{figure}{section}
\@addtoreset{table}{section}
\@addtoreset{lemma}{section}
\makeatother

This supplementary document provides additional technical details and results that support the main paper. Section~\ref{section: preliminaries} reviews background on determinantal point processes. Section~\ref{appendix: proofs} contains detailed proofs for the posterior and predictive distributions under projection DPP prior. Section~\ref{appendix: asymptotic_consistency_properties} provides proofs of the consistency results. Section~\ref{appendix: describe_repulsiveness_DPPs} discusses the repulsive properties of projection DPPs and global repulsiveness measurement. Section~\ref{appendix: more_algorithms} outlines the posterior sampling algorithms for our method. Section~\ref{appendix: mcmc_mixing_simulation} shows MCMC mixing performance for the proposed method in 1-dimensional simulation.  Finally, Sections~\ref{appendix: ERPs_clust_fPCA_DPM} and~\ref{appendix: one-step_method} present additional results from the ERP data analysis, including comparisons with the  Dirichlet process mixture model and an integrated one-step model combining probabilistic PCA with projection DPP prior.

\section{Details on determinantal point processes}\label{section: preliminaries}

Let $\Phi = \left\{\phi_1, \ldots, \phi_m\right\}$ be a simple point process defined on a probability space $(\Omega, \mathcal{A}, \mathsf{P})$, taking values in the set of locally finite counting measures on a Polish space $(\Theta, \mathcal{B}(\Theta))$, where $\mathcal{B}(\Theta)$ denotes the associated Borel $\sigma$-algebra. The atoms of $\Phi$ are given by a sequence of random variables $(\phi_i)_{i \geq 1}$ taking values in $\Theta$. Then $\Phi$ can be written as    
\[
\Phi(B)= \sum_{i = 1}^{m}\delta_{\phi_i} (B), \qquad B \in \mathcal{B}(\Theta),
\]
where the points $(\phi_i)_{i = 1}^m$ are almost surely pairwise distinct, meaning $\mathsf{P}(\phi_i = \phi_j) = 0$ for all $i \neq j$. We denote the probability distribution of $\Phi$ by $\bm{P}_{\Phi}$. For further measure theoretical details, see \citet{moller2004statistical} and \citet{baccelli2024random}.

For $k \in \mathds{N} = \{1,2,3,\cdots\}$, the $k$-th factorial measure $\Phi^{(k)}$ of the point process $\Phi$ on the product space $\Theta^k$ is defined by
\[
\Phi^{(k)} = \sum_{\left(j_1, \ldots, j_k\right) \in {\mathds{N}}^{(k)}} \delta_{\left(\phi_{j_1}, \ldots, \phi_{j_k}\right)},
\]  
where ${\mathds{N}}^{(k)}=\left\{\left(j_1, \ldots, j_k\right) \in {\mathds{N}}^k: j_m \neq j_n, \text{ for any } m \neq n\right\}$. An equivalent representation is
\[
\Phi^{(k)}\left(B_1 \times \cdots \times B_k\right)=\sum_{\phi_1, \ldots, \phi_k \in \Phi}^{\neq} \mathds{I}_{B_1}\left(\phi_1\right) \cdots \mathds{I}_{B_k}\left(\phi_k\right),
\]
for measurable sets $B_1, \cdots, B_k \in \mathcal{B}(\Theta)$. The summation is taken over all $k$-tuples of pairwise distinct points in $\Phi$. The $k$-th factorial moment measure $M_{\Phi^{(k)}}$ is defined by 
\[
M_{\Phi^{(k)}} \left(B_1 \times \cdots \times B_k\right) = \mathsf{E} \left[ \Phi^{(k)}\left(B_1 \times \cdots \times B_k\right) \right],
\]
where the expectation is taken with respect to $\Phi$. When $k = 1$, this coincides with the mean measure $M_{\Phi}$, defined as $M_{\Phi}(B) = \mathsf{E}\left[\Phi(B)\right]$ for all $B \in \mathcal{B}(\Theta)$. The notion of factorial moment measure is central to the development in the main paper. We use it to define the determinantal point process (DPP) in the following, which is fundamental to our proposed methodology.

\begin{definition}[Determinantal point process; see details in  \citet{hough2009zeros, lavancier2015determinantal, baccelli2024random}]\label{defn: DPP}
Let $\Theta$ be a compact subset of $\mathds{R}^d$ and $\omega$ be a locally finite measure on $\Theta$. Let $K: \Theta \times \Theta \to \mathds{C}$ be a continuous complex covariance function such that $\int_{\Theta} K(\phi, \phi)\,  \omega(\mathrm{d} \phi) < +\infty$. A point process $\Phi$ on $\Theta$ is said to be determinantal with background measure $\omega$ and kernel $K$ if, for all $k \in \mathds{N}$, its $k$-th factorial moment measure $M_{\Phi^{(k)}}$ admits a density with respect to the product measure $\omega^k$ satisfying
\begin{equation}
    M_{\Phi^{(k)}}(\mathrm d \phi_1 \cdots \mathrm d \phi_k) = \det \left\{ K(\phi_i, \phi_j) \right\}_{i, j=1}^k \omega(\mathrm d \phi_1) \cdots \omega(\mathrm d \phi_k),
\end{equation}
for $\omega^k$-almost all $(\phi_1, \cdots, \phi_k) \in \Theta^k$, where  $\left\{ K(\phi_i, \phi_j) \right\}_{i, j=1}^k$ denotes the $k \times k$ matrix with $(i,j)$-th entry $K(\phi_i, \phi_j)$, and $\det(\cdot)$ denotes the determinant. Then $\Phi$ is called a DPP with kernel $K$, and we write $\Phi \sim \text{DPP}_{\Theta} (K)$. 
\end{definition}

Intuitively, for any set of pairwise distinct points $\phi_1, \ldots, \phi_k \in \Theta$, the $M_{\Phi^{(k)}}(\mathrm d \phi_1 \cdots \mathrm d \phi_k)$ is the probability that, for each $i = 1, \ldots, k$, the point process $\Phi$ has a point in an infinitesimally small region around $\phi_i$ of volume $\mathrm d \phi_i$. We say that, the $\Phi$ has an $k$-th order joint intensity function $\rho(\phi_1, \ldots, \phi_k)$, also called the $k$-th order correlation function, given by the determinant
\begin{equation}\label{eq: joint_intensity_function}
\rho(\phi_1, \ldots, \phi_k) = \det \left\{ K(\phi_i, \phi_j) \right\}_{i, j=1}^k, \qquad \phi_1, \ldots, \phi_k \in \Theta .
\end{equation}

\begin{remark}\label{remark: general_dpp_defn}
If $\Phi$ is a DPP, then for $k = 1$, we have $M_{\Phi}(\Theta) = \int_{\Theta} K(\phi,\phi) \omega(\mathrm d \phi) < + \infty$ by definition. For any locally integrable function $f: \Theta^{k} \to \mathds{R}$, it holds that
\begin{align*}
& \mathsf{E} \left[ \sum_{(\phi_1, \ldots, \phi_k) \in \Phi}^{\neq} f(\phi_1, \ldots, \phi_k)\right] = \mathsf{E} \left[ \sum_{\left(j_1, \ldots, j_k\right) \in {\mathds{N}}^{(k)}} f(\phi_1, \ldots, \phi_k)\right] \\
& \qquad  = \mathsf{E} \left[\int_{\Theta^k} f(\phi_1, \ldots, \phi_k) \Phi^{(k)} \left(\mathrm d \phi_1, 
\cdots, \mathrm d \phi_k\right)\right] \\
& \qquad = \int_{\Theta^k} f(\phi_1, \ldots, \phi_k) M_{\Phi^{(k)}} \left(\mathrm d \phi_1, \cdots, \mathrm d \phi_k\right) \qquad \text{ (by Campbell averaging formula)} \\
& \qquad = \int_{\Theta^k} f(\phi_1, \ldots, \phi_k) \det \left\{ K(\phi_i, \phi_j) \right\}_{i, j=1}^k \omega(\mathrm d \phi_1) \cdots \omega(\mathrm d \phi_k) . 
\end{align*}
\end{remark}

\begin{remark}
By Mercer's Theorem, the kernel $K$ admits a spectral representation given by
\begin{equation}\label{eq: general_DPP_kernel_spectral}
K(x, y) = \sum_{j = 1}^{\infty} \lambda_j \varphi_j(x) \overline{\varphi_j(y)}, \qquad (x, y) \in \Theta \times \Theta,
\end{equation}
with absolute and uniform convergence of the series, where eigenfunctions $(\varphi_j)_{j \geq 1}$ form an orthonormal basis of the space of square integrable functions $\Theta \rightarrow \mathds{C}$, and the eigenvalues $(\lambda_j)_{j \geq 1}$ satisfy $\lambda_j \geq 0$ and $\sum_{j=1}^{\infty} \lambda_j < +\infty$. A necessary and sufficient condition for the existence of a DPP $\Phi$ on $\Theta$ with kernel $K$ is that $0 \leq \lambda_j \leq 1$ for all $j = 1,2,\ldots$ \citep{macchi1975coincidence, hough2006determinantal}.
\end{remark}

We list a few important properties of DPPs as follows since they are used in the main paper.

\begin{property}
The intensity function of a DPP $\Phi$ on $\Theta$ is $\rho(x) = K(x,x), x \in \Theta$. $\Phi$ is homogeneous if $\rho(x)$ is constant. Note that $\rho(x)$ is ensured to be real and non-negative, as by definition $K$ is a complex covariance function if and only if it is Hermitian and positive semi-definite. 
\end{property}

\begin{property}
If $\Phi \sim DPP_{\Theta}(K)$ where $K$ is of the form
\[
K(x,y) = K_0(x-y) \text{ for some function } K_0, \qquad x , y \in \Theta, 
\]
then $\Phi$ is called stationary, that is, its distribution is invariant under translations.
\end{property}

\begin{property}\label{property: pair_correlation_function}
The pair correlation function of a DPP $\Phi$ is 
\[
g(x,y) = 1 - \frac{|K(x, y)|^2}{K(x, x) K(y, y)}, \quad \text{if } K(x,x) >0 \text{ and } K(y,y) > 0,
\]
whereas it is 0 otherwise. In spatial statistics, the pair correlation function $g(x,y)$ is commonly defined as the ratio $\rho(x,y) / \left(\rho(x) \rho(y)\right)$, which reduces to the expression above for DPPs. For a homogeneous Poisson point process, $g(x,y) = 1$, indicating no interaction between points. In contrast, DPPs satisfy $g(x,y) \leq 1$, meaning that the presence of a point at $x$ lowers the likelihood of observing another nearby point at $y$, which reflects the repulsive nature of the process. 
\end{property}

\begin{property}\label{property: transformation_rule}
If we transform $\Phi \sim DPP_{\Theta}(K)$ by a one-to-one continuous differentiable mapping $T(x) = Ax+b, x \in \Theta$, such that its Jacobian matrix is invertible and $\operatorname{det}(A) \neq 0$, then $T(\Phi)$ is the DPP on $T(\Theta)$ with kernel $K\left(A^{-1}\left(x-b\right), A^{-1}\left(y-b\right)\right) /|\operatorname{det}(A)|$.
\end{property}

We consider the process $\Phi$ to be a particularly important class of DPPs, known as projection DPPs, defined as follows.

\begin{definition}[Projection DPPs; see details in \citet{lavancier2015determinantal, lavancier2023simulation}]
Suppose that, in the spectral representation~(\ref{eq: general_DPP_kernel_spectral}) of the kernel $K$, all non-zero eigenvalues $\lambda_j$ are equal to 1. Then $K$ is called a projection kernel, and the corresponding point process $\Phi$ is called a projection DPP. For simple notation, we write $\Phi \sim \text{pDPP}_{\Theta} (K)$.
\end{definition}

Since eigenvalues in projection DPPs take only the values 0 or 1, let $m = \sum_{j=1}^{\infty} \lambda_j < +\infty$ denote the finite number of non-zero eigenvalues. Without loss of generality, eigenfunctions associated with zero eigenvalues are discarded, and the projection kernel admits the spectral representation 
\begin{equation}\label{eq: projection_DPP_kernel_spectral}
K(x, y) = \sum_{j = 1}^{m} \varphi_j(x) \overline{\varphi_j(y)}, \qquad (x, y) \in \Theta \times \Theta .
\end{equation}
It is straightforward to verify that projection kernels satisfy
\begin{equation}\label{eq: projection_DPP_properties}
\int_{\Theta} K(x,z) K(z,y) \mathrm d z = K(x,y) \quad \text{and} \quad \int_{\Theta} K(x,x) \mathrm dx = m .
\end{equation}

\begin{remark}[Projection DPPs densities; see the proof in Appendix of \citet{lavancier2015determinantal}]
For a projection DPP $\Phi$ with kernel $K$ in the form of (\ref{eq: projection_DPP_kernel_spectral}), it consists of exactly $m$ points, with a joint density with respect to the product Lebesgue measure $\omega^m$ given by
\[
f_{\Phi}(\bm{\phi}) = \frac{1}{m !} \operatorname{det}\{K(\phi_i, \phi_j)\}_{\phi_i, \phi_j \in\left\{\phi_1, \ldots, \phi_m\right\}}, \qquad \bm{\phi} = (\phi_1, \ldots, \phi_m).
\]
\end{remark}

Projection DPPs are of particular interest for modeling our mixture components $\left\{\phi_1, \ldots, \phi_m\right\}$ for three reasons. First, they generate a fixed number of points $m$, naturally aligning with the structure of our mixture. Second, they represent the most repulsive class of DPPs, since all non-zero eigenvalues are 1, the maximum possible value. In Supplement Section~\ref{appendix: describe_repulsiveness_DPPs}, we further discuss a global repulsiveness measure for DPPs proposed by \citet{moller2021couplings}, which formalize the existence of a trade-off between repulsiveness and intensity. Under this measure, projection DPPs are the  most globally repulsive subclass within the DPP family. Third, their sufficient sampling algorithms are available even without requiring access to the spectral decomposition \citep{hough2006determinantal, lyons2014determinantal}.

To specify a projection kernel explicitly, we consider the special case where the orthonormal eigenfunctions $\varphi_j(x)$ in the spectral decomposition  (\ref{eq: projection_DPP_kernel_spectral}) correspond to the Fourier basis. The Fourier basis plays an important role in DPPs due to its simplicity, homogeneity, and repulsiveness it yields \citep{coeurjolly2021monte}. Additionally, it underpins the spectral approximation of general stationary DPP kernels discussed in \citet{lavancier2015determinantal}. Although exact spectral representations for general stationary DPP kernels are typically intractable and exists primarily in theory, practical approximations via Fourier expansions are feasible. Note that eigenvalues and eigenfunctions in representation~(\ref{eq: projection_DPP_kernel_spectral}) are indexed by positive integers; however, other countable index sets could also be used. In particular, the $d$-dimensional Fourier basis is indexed by $\mathds{Z}^d$, where $\mathds{Z} = \{\ldots, -2,-1,0,1,2,\ldots\}$.

Specifically, let $R = [-\frac{1}{2}, \frac{1}{2}]^d$ denote the unit hypercube. Consider the orthonormal Fourier basis on $R$, defined for $j \in \mathds{Z}^d$ and $x \in R$ by $\varphi_j(x) = \mathrm{e}^{2 i \pi j^T x}$, where $i^2 = -1$ and $j^T x$ denotes their inner product. To ensure the projection kernel $K$ such that $K(x,y) = K(y,x) \in \mathds{R}$ for all $x,y \in R$, which is crucial for statistical interpretation, let $L = \{-\ell, \ldots, 1, 0, 1, \ldots, \ell\}$ represent a symmetric set of successive frequencies centered around zero, and construct the projection kernel as:
\begin{equation}\label{eq: our_projection_kernel_unit}
K(x,y) = \sum_{j \in L^d} \varphi_j(x) \overline{\varphi_j(y)} = \sum_{j \in L^d} \mathrm{e}^{2 i \pi j^T (x-y)}, \qquad (x,y) \in R \times R . 
\end{equation}
The frequency index set $L^d \subset \mathds{Z}^d$ is a finite rectangular subset of $\mathds{Z}^d$ with cardinality $|L^d| = (1+2\ell)^d$, identifying $m$ eigenvalues equal to 1. Due to the invariance by translation of the Fourier basis, the kernel (\ref{eq: our_projection_kernel_unit}) is  stationary. This construction (\ref{eq: our_projection_kernel_unit})  implies that for any $x,y \in [-\frac{1}{2}, \frac{1}{2}]^d, K(x,y) = \prod_{t=1}^d K_t(x_t, y_t)$ where the $K_t$'s are the one-dimensional stationary kernels defined for any $x_t, y_t \in [-\frac{1}{2}, \frac{1}{2}]$ by $K_t (x_t, y_t) = \sum_{j = -\ell}^{\ell} \mathrm{e}^{2 i \pi j (x_t - y_t)}$.
The kernel in~(\ref{eq: our_projection_kernel_unit}) has previously been discussed and employed in \citet{coeurjolly2021monte,lavancier2023simulation}, to which we refer for further justification.

For mixture models defined over more general spaces $\Theta \subset \mathds{R}^d$, we need to apply an affine transformation $T(x) = Ax+b$ mapping from $R$ to $\Theta$ with $\operatorname{det}(A) \neq 0$, resulting in the volume of $\Theta$ to be $|\det(A)|$. Under this transformation, the resulting point process is $\Phi = \left\{\phi_1, \ldots, \phi_m\right\} \sim \operatorname{pDPP}_{\Theta}(K_{\operatorname{trans}})$, with the transformed kernel $K_{\text{trans}}$ expressed as 
\begin{equation}\label{eq: our_projection_kernel_transformed}
\begin{split}
K_{\operatorname{trans}}(x,y) &= K\left(A^{-1}\left(x-b\right), A^{-1}\left(y-b\right)\right) /|\operatorname{det}(A)| \\
&= \frac{1}{|\operatorname{det}(A)|} \sum_{j \in L^d} \mathrm{e}^{2 i \pi j^T A^{-1} (x-y)} \\
&= \frac{1}{|\operatorname{det}(A)|} \sum_{j \in L^d} \cos \left(2 \pi j^T \left(A^{-1} \left(x-y\right)\right)\right) , \qquad (x,y) \in \Theta \times \Theta,
\end{split}
\end{equation}
by applying Euler's Formula. For simplicity and clarity, we slightly abuse notation and henceforth denote the transformed kernel $K_{\operatorname{trans}}(\cdot, \cdot)$ simply as $K(\cdot, \cdot)$. This notation should not cause ambiguity, as the context and support space make clear whether a kernel has been transformed. The cardinality $(1+2\ell)^d$ of the frequency index set $L^d$ specifies the fixed number $m$ of points generated by $\Phi$ in $\Theta$, which correspond to the number of components in the mixture model.

\section{Proofs of the main results}
\label{appendix: proofs}

Before proceeding with the proofs, we need a bit more background on point processes and Palm calculus. 

Consider a simple point process $\Phi$, written as $\Phi (B) = \sum_{h \geq 1} \delta_{x_h} (B), B \in \Theta$, where $(x_h)_{h \geq 1}$ are the points of the process. Then, we associate with each point of $\Phi$ independent marks $s_h \in \mathds{R}_{+}$, such that each $s_h$ follows marginal density $\pi$, to define the marked point process $\Psi = \sum_{h \geq 1} \delta_{(x_h, s_h)}$ and set
\begin{equation}\label{eq: construct_random_mixing_measure}
    \widetilde{G}(B) = \int_{B \times \mathds{R}_+} s\,  \Psi (\mathrm{d}x \, \mathrm{d}s) = \sum_{h \geq 1} s_h \delta_{x_h} (B), \qquad B \in \Theta ,  
\end{equation}
which is a finite random mixing measure of our interest in the present paper.

To study the posterior of the point process $\Phi$, we require the characterization of its conditional distribution given that some atoms have already been observed. This is accomplished through Palm calculus (see \citet{kallenberg2021foundations, baccelli2024random}), which we present in an intuitive and accessible manner. Informally, the Palm version $\Phi_x$ of $\Phi$ at a point $x \in \Theta$ refers to the point process $\Phi$ conditioned on the event ``having an atom at $x$''. Its distribution is denoted by $\bm{P}_{\Phi}^{x}$, referred to as the Palm distribution of $\Phi$ at $x$ and regarded as the ``conditional distribution of $\Phi$'' given that it has an atom at $x$. When $x$ is an observed atom of $\Phi$, we can discard it and instead work with the reduced Palm version, defined as  $\Phi_x^{!} := \Phi_x - \delta_x$, whose law is denoted by  $\bm{P}_{\Phi^!}^{x}$. Informally, this represents the conditional distribution of the rest of the process, given that it includes $x$. The reduced Palm $\Phi_x^{!}$ is also a point process on $\Theta$ with $k$-th order intensity function $\rho_x^! (x_1, \ldots, x_k) = \rho(x, x_1, \ldots, x_k) / \rho(x)$. Combined with~(\ref{eq: joint_intensity_function}), it easily shows that $\Phi_x^{!}$ is a DPP with kernel 
\[
K_x^{!}(y,z) = K(y,z) - \frac{K(y,x) K(x,z)}{K(x,x)}, \qquad y, z \in \Theta,
\]
which is important to derive samplings from the posterior and predictive distributions. The argument outlined above can be extended to the case of multiple pairwise different points $\bm{x} = (x_1, \ldots, x_k)$, leading to the $k$-th Palm distribution $\{\bm{P}_{\Phi}^{\bm{x}}\}_{\bm{x} \in \Theta^{k}}$. Again, $\bm{P}_{\Phi}^{\bm{x}}$ can be understood as the law of $\Phi_{\bm{x}}$ conditional to $\Phi$ having atoms at $\{x_1, \ldots, x_k\}$ and removing the trivial atoms yields the reduced Palm distribution $\bm{P}_{\Phi^!}^{\bm{x}}$, that is, the law of $\Phi_{\bm{x}}^{!} := \Phi_{\bm{x}} - \sum_{h=1}^{k} \delta_{x_h}$. Likewise, $\Phi_{\bm{x}}^{!}$ is a DPP with kernel    
\[
K_{\bm{x}}^{!}(y,z) = K(y, z) - \bm{k}_{\bm{x}}^T (y) \bm{K}_{\bm{x}, \bm{x}}^{-1} \bm{k}_{\bm{x}}(z), \qquad y,z \in \Theta,
\]
with $\bm{k}_{\bm{x}}^T (y) = \left(K(y, x_1), \ldots, K(y, x_k)\right)$ and $\bm{K}_{\bm{x}, \bm{x}} = \left( K(x_t, x_r)\right)_{1 \leq t, r \leq k}$.

For a marked point process $\Psi$ with independent marks, the Palm version $\Psi_{\bm{x}, \bm{s}}$, with $\bm{x} = (x_1, \ldots, x_k)$ and $\bm{s} = (s_1, \ldots, s_k)$, does not depend on $\bm{s}$. Moreover, $\Psi_{\bm{x}, \bm{s}}^{!}$ has the same law of the point process obtained by considering $\Phi_{\bm{x}}^{!}$ and marking it with i.i.d. marks. We write $\Psi_{\bm{x}}^{!}$ in place of $\Psi_{\bm{x},\bm{s}}^{!}$ for simplicity. Since $\Psi_{\bm{x}}^{!}$ is a marked point process, we can define a random mixing measure on $\Theta$ as in the general construction~(\ref{eq: construct_random_mixing_measure}). Specifically, 
\[
\widetilde{G}_{\bm{x}}^{!}(B) := \int_{B \times \mathds{R}_+} s \Psi_{\bm{x}}^{!} (\mathrm{d}x \,  \mathrm{d}s) .  
\]
We write $\widetilde{G}_{\bm{x}}^{!} \sim \bm{P}_{\Psi^!}^{\bm{x}}$. Note that we can interpret $\bm{P}_{\Psi^!}^{\bm{x}}$ as the law of a random measure obtained as follows: (i) take the random measure $\widetilde{G}$ as in~(\ref{eq: construct_random_mixing_measure}), (ii) condition to $x_1, \ldots, x_k$ being atoms of $\widetilde{G}$, and then (iii) remove $x_1, \ldots, x_k$ from the support.

We need the following preparatory lemma for the proofs of our main results, which is the Lemma 11 borrowed from \citet{beraha2025bayesian}.

\begin{lemma}\label{lemma:laplace_marked}
Let $\widetilde{G}(B) = \int_{B \times \mathds{R}_+} s \Psi(\mathrm d x \,\mathrm d s)$ where $\Psi = \sum_{h \geq 1} \delta_{(x_h, s_h)}$ is a marked point process obtained by marking $\Phi$ with i.i.d. marks $s_h$ from a probability density $\pi(\cdot)$ on $\mathds{R}_+$. Then for any measurable function $f : \Theta \to \mathds{R}_+$, the Laplace functinoal is
\[
    \mathsf{E}\left[ \mathrm{e}^{- \int_{\Theta} f(x) \widetilde{G}(\mathrm{d} x)} \right] = \mathsf{E}\left[ \exp\left(\int_{\Theta} \log \psi(f(x)) \Phi(\mathrm{d} x) \right) \right] ,
\]
where $\psi(f(x)): = \int_{\mathds{R}_+} \mathrm{e}^{- s f(x)} \pi (s) \mathrm{d} s$ is the Laplace transform of $\pi(\cdot)$ evaluated at $f(x)$.
\end{lemma}

\subsection{Proof of Theorem~\ref{thrm: theorem_1}}\label{appendix: proof_of_theorem1}

Conditional on $u$ and $\bm{\theta} = \left\{\theta_1, \ldots, \theta_n\right\}$, as well as $\bm{\theta}^* = \left\{\theta_1^*, \ldots, \theta_k^*\right\}$, applying the Theorem 1 in \citet{beraha2025bayesian}, it follows that the distribution of the unnormalized mixing measure $G$ is equal to the distribution of 
\[
\sum_{h=1}^{k} s^{*}_h \delta_{\theta_h^*} + \widetilde{G} ,
\]
where $\left\{s_1^{*}, \ldots, s_{k}^{*}\right\}$ is a set of independent jumps, with posterior density $f_{s_h^{*}} (s) \propto s^{n_h} \mathrm{e}^{-u s} \pi(s) \mathrm{d}s$ for $h = 1, \ldots, k$; and $\widetilde{G}$ is a random measure with Laplace functional
\begin{equation}\label{eq: tildeG_laplace_functional}
    \mathsf{E}\left[\exp \int_{\Theta} - f(z) \widetilde{G}(\mathrm{d} z) \right] = \frac{\mathsf{E} \left[ \exp \left\{- \int_{\Theta} (f(z) + u) \widetilde{G}_{\bm{\theta}^{*}}^{!}(\mathrm{d} z) \right\} \right]}{\mathsf{E} \left[ \exp \left\{- \int_{\Theta} u \widetilde{G}_{\bm{\theta}^{*}}^{!} (\mathrm{d} z) \right\} \right]},
\end{equation}
and $\widetilde{G}_{\bm{\theta}^{*}}^{!}$ is defined as 
\[
\widetilde{G}_{\bm{\theta}^{*}}^{!}(B) := \int_{B \times \mathds{R}_+} s \Psi_{\bm{\theta}^{*}}^{!} (\mathrm{d}\phi \, \mathrm{d}s), \qquad B \in \Theta .  
\]
The marked point process $\Psi_{\bm{\theta}^{*}}^{!}$, in place of $\Psi_{\bm{\theta}^{*}, \bm{s}^{*}}^{!}$,  has the same law of the point process obtained by considering $\Phi_{\bm{\theta}^{*}}^{!} := \Phi_{\bm{\theta}^{*}} - \sum_{h=1}^{k} \delta_{\theta_{h}^{*}}$ and marking it with i.i.d. marks.

We say that, the random measure $\widetilde{G}$ consists of new, unobserved atoms $\bm{\phi}^\prime = \left\{\phi_1^{\prime}, \ldots, \phi_{m-k}^{\prime}\right\}$, sampled from the reduced Palm $\Phi_{\bm{\theta}^{*}}^{!}$. Using Palm calculus, it follows that the conditional distribution of the remaining atoms from $\Phi_{\bm{\theta}^{*}}^{!}$, given the existence of $\{\theta_1^{*}, \ldots, \theta_{k}^{*}\}$, has the $(m - k)$-th factorial moment measure
\[
M_{{\Phi_{\bm{\theta}^{*}}^{! (m - k)}}} \left(\mathrm{d} \phi_1^{\prime} \cdots \mathrm{d} \phi_{m - k}^{\prime}\right) = \frac{\operatorname{det}\{K(x, y)\}_{x, y \in\left\{\bm{\theta}^{*}, \bm{\phi}^{\prime}\right\}}}{\operatorname{det}\{K(x, y)\}_{x, y \in \bm{\theta}^{*}}}  
\prod_{h=1}^{m - k} \omega\left(\mathrm{d} \phi_h^{\prime}\right) ,
\]
as discussed in \citet{lavancier2023simulation}. Applying the matrix determinant identity
\[
\operatorname{det}\left(\begin{array}{ll}
A & B \\
C & D
\end{array}\right)=\operatorname{det}(A) \operatorname{det}\left(D-C A^{-1} B\right) ,
\]
where $\operatorname{det}(A) = \operatorname{det}\{K(x, y)\}_{x, y \in \bm{\theta}^{*}}$, we confirm that $\Phi_{\bm{\theta}^{*}}^{!}$ remains a projection DPP with updated kernel $K^!_{\bm{\theta}^{*}}$, defined as
\[
K^!_{\bm{\theta}^{*}}(\phi_i, \phi_j) := K(\phi_i, \phi_j) - \bm{k}_{\bm{\theta}^{*}}^T (\phi_i) \bm{K}_{\bm{\theta}^{*},\bm{\theta}^{*}}^{-1} \bm{k}_{\bm{\theta}^{*}}(\phi_j), \qquad \phi_i,\phi_j \in \Theta\backslash \bm{\theta}^{*} ,
\]
with $\bm{k}_{\bm{\theta}^{*}}^T (\phi_i) = \left(K(\phi_i, \theta_1^{*}), \ldots, K(\phi_i,\theta_k^*)\right)$ and $\bm{K}_{\bm{\theta}^{*},\bm{\theta}^{*}} = \left( K(\theta_t^{*}, \theta_r^{*})\right)_{1 \leq t, r \leq k}$. It can be verified that, given $\bm{\theta}^{*} = \left\{\theta_1^{*}, \ldots, \theta_k^*\right\}$, $K^{!}_{\bm{\theta}^*}(x,x) = 0$ for $x \in \bm{\theta}^{*}$ and $\int_{\Theta} K^!_{\bm{\theta}^{*}}(x,x) \mathrm{d}x = m - k$, implying that the projection DPP $\Phi_{\bm{\theta}^{*}}^{!}$ has cardinality $m - k$. In other words, we can write $\Phi_{\bm{\theta}^{*}}^{!} := \sum_{h=1}^{m-k} \delta_{\phi_h^{\prime}}$ and $\Phi_{\bm{\theta}^{*}}^{!} (\Theta) = m - k$.

For the numerator term in~(\ref{eq: tildeG_laplace_functional}), since $\widetilde{G}_{\bm{\theta}^{*}}^{!}(B) = \int_{B \times \mathds{R}_+} s \Psi_{\bm{\theta}^{*}}^{!} (\mathrm{d}\phi \, \mathrm{d}s)$ where $\Psi_{\bm{\theta}^{*}}^{!}$ is a marked point process obtained by marking the projection DPP $\Phi_{\bm{\theta}^{*}}^{!}$ with i.i.d. marks $s_h$ from a probability density $\pi(\cdot)$ on $\mathds{R}_+$, by applying Lemma~\ref{lemma:laplace_marked}, it follows that
\[
\mathsf{E} \left[ \exp \left\{- \int_{\Theta} (f(z) + u) \widetilde{G}_{\bm{\theta}^{*}}^{!}(\mathrm{d} z) \right\} \right] = \mathsf{E}\left[ \exp\left(\int_{\Theta} \log \left( \int_{\mathds{R}_+} \mathrm{e}^{- s (f(z) + u)} \pi(s) \mathrm{d} s \right) \Phi_{\bm{\theta}^{*}}^{!}(\mathrm{d} z) \right) \right] .
\]
We define 
\[
\pi^\prime(s) := \frac{\mathrm{e}^{- s u} \pi(s)}{\int_{\mathds{R}_+} \mathrm{e}^{- s u} \pi(s) \mathrm{d}s} = \frac{\mathrm{e}^{- s u} \pi(s)}{\psi(u)}
\]
to be the density of the exponential tilting of the jump variable. Then,
\[
\begin{split}
&\mathsf{E} \left[ \exp \left\{- \int_{\Theta} (f(z) + u) \widetilde{G}_{\bm{\theta}^{*}}^{!}(\mathrm{d} z) \right\} \right] = \mathsf{E}\left[ \exp\left(\int_{\Theta} \log \left( \psi(u) \int_{\mathds{R}_+} \mathrm{e}^{- s f(z)} \pi^\prime(s) \mathrm{d} s \right) \Phi_{\bm{\theta}^{*}}^{!}(\mathrm{d} z) \right) \right] \\
& \qquad = \mathsf{E}\left[ \exp\left\{\int_{\Theta} \log \left( \int_{\mathds{R}_+} \mathrm{e}^{- s f(z)} \pi^\prime(s)\mathrm{d} s \right) \Phi_{\bm{\theta}^{*}}^{!}(\mathrm{d} z) + \int_{\Theta} \log \left( \psi(u) \right) \Phi_{\bm{\theta}^{*}}^{!}(\mathrm{d} z) \right\} \right] \\
& \qquad = \mathsf{E}\left[ \exp\left\{\int_{\Theta} \log \left( \int_{\mathds{R}_+} \mathrm{e}^{- s f(z)} \pi^\prime(s) \mathrm{d} s \right) \Phi_{\bm{\theta}^{*}}^{!}(\mathrm{d} z) + \Phi_{\bm{\theta}^{*}}^{!}(\Theta) \log \left( \psi(u) \right) \right\} \right] \\
& \qquad = \mathsf{E}\left[ {\psi(u)}^{\Phi_{\bm{\theta}^{*}}^{!}(\Theta)} \exp\left\{\int_{\Theta} \log \left( \int_{\mathds{R}_+} \mathrm{e}^{- s f(z)} \pi^\prime(s) \mathrm{d} s \right) \Phi_{\bm{\theta}^{*}}^{!}(\mathrm{d} z) \right\} \right] \\
& \qquad = \mathsf{E}\left[ {\psi(u)}^{m - k} \exp\left\{\int_{\Theta} \log \left( \int_{\mathds{R}_+} \mathrm{e}^{- s f(z)} \pi^\prime(s)\mathrm{d} s \right) \Phi_{\bm{\theta}^{*}}^{!}(\mathrm{d} z) \right\} \right], \text{ since } \Phi_{\bm{\theta}^{*}}^{!} (\Theta) = m - k , \\
& \qquad = {\psi(u)}^{m - k} \mathsf{E}\left[ \exp\left\{\int_{\Theta} \log \left( \int_{\mathds{R}_+} \mathrm{e}^{- s f(z)} \pi^\prime(s) \mathrm{d} s \right) \Phi_{\bm{\theta}^{*}}^{!}(\mathrm{d} z) \right\} \right] .
\end{split}
\]
If $f(z) = 0$, then the denominator term in~(\ref{eq: tildeG_laplace_functional}) follows that 
\[
\mathsf{E} \left[ \exp \left\{- \int_{\Theta} u \widetilde{G}_{\bm{\theta}^{*}}^{!} (\mathrm{d} z) \right\} \right] = {\psi(u)}^{m - k} .
\]
Taking the ratio, the Laplace functional of the random measure $\widetilde{G}$ becomes
\[
\mathsf{E}\left[\exp \int_{\Theta} - f(z) \widetilde{G}(\mathrm{d} z) \right] = \mathsf{E}\left[ \exp\left\{\int_{\Theta} \log \left( \int_{\mathds{R}_+} \mathrm{e}^{- s f(z)} \pi^\prime(s)\mathrm{d} s \right) \Phi_{\bm{\theta}^{*}}^{!}(\mathrm{d} z) \right\} \right] .
\]
Again, by applying Lemma~\ref{lemma:laplace_marked}, it follows that the random measure $\widetilde{G} := \sum_{h=1}^{m-k} s^{\prime}_h \delta_{\phi_h^{\prime}}$ is such that 
\[
\widetilde{G}(B) = \sum_{h=1}^{m-k} s^{\prime}_h \delta_{\phi_h^{\prime}} (B) = \int_{B \times \mathds{R}_+} s \Psi_{\bm{\theta}^{*}}^{!} (\mathrm{d}\phi \, \mathrm{d}s), \qquad B \in \Theta,
\]
where $\Psi_{\bm{\theta}^{*}}^{!}$ is a marked point process obtained by marking $\Phi_{\bm{\theta}^{*}}^{!}$ with i.i.d. marks $s_h^{\prime}$ from a distribution density $\pi^\prime$ on $\mathds{R}_+$. 

Therefore, we conclude that, conditional on $\bm{\theta} = \left\{\theta_1, \ldots, \theta_n\right\}$ and $u$, the distribution of the unnormalized mixing measure $G$ is equal to the distribution of $\sum_{h=1}^{k} s^{*}_h \delta_{\theta_h^{*}} + \sum_{h=1}^{m-k} s^{\prime}_h \delta_{\phi_h^{\prime}}$, where the atoms $\bm{\phi}^{\prime} = \left\{\phi_1^{\prime}, \ldots, \phi_{m-k}^{\prime}\right\}$ are drawn from the reduced Palm projection DPP $\Phi_{\bm{\theta}^{*}}^{!}$ with kernel $K^!_{\bm{\theta}^{*}}$; the associated jumps $s_{h}^{\prime}$ are i.i.d. with exponentially tilted density $\pi^\prime(s) = \mathrm{e}^{-su} \pi(s) / \psi(u)$, where $\psi(u) = \int_{\mathds{R}_+}  \mathrm{e}^{-su} \pi(s) \mathrm{d} s$.

\subsection{Proof for the full conditional distribution of the auxiliary $u$}\label{appendix: proof_auxiliary_u}

Given the clustering configuration $\bm{c}$, the corresponding cluster sizes $(n_1, \ldots, n_{k})$, and $\bm{\theta}^{*}$ applying the Theorem 1 in \citet{beraha2025bayesian}, the full conditional distribution of $u$ has a density w.r.t to the Lebesgue measure given by
\[
f_{u \mid \bm{\theta}^{*}} (u) \propto u^{n-1} \mathsf{E} \left[ \exp \left\{- u \widetilde{G}_{\bm{\theta}^{*}}^{!} (\Theta) \right\} \right]  \prod_{h=1}^{k} \kappa(u, n_h), \qquad u > 0, 
\]
where $\kappa(u, n_h) := \int_{\mathds{R}_{+}} \mathrm{e}^{-us} s^{n_h} \pi(s) \mathrm{d} s$. We have shown in Section~\ref{appendix: proof_of_theorem1} that 
\[
\mathsf{E} \left[ \exp \left\{- u \widetilde{G}_{\bm{\theta}^{*}}^{!} (\Theta) \right\} \right] = \mathsf{E} \left[ \exp \left\{- \int_{\Theta} u \widetilde{G}_{\bm{\theta}^{*}}^{!} (\mathrm{d} z) \right\} \right] = {\psi(u)}^{m - k} ,
\]
where $\psi(u) = \int_{\mathds{R}_+}  \mathrm{e}^{-su} \pi(s) \mathrm{d} s$. It shows that the full conditional of auxiliary $u$ only depends on the clustering configuration $\bm{c}$ and the cluster sizes, independent with the observed active atoms $\bm{\theta}^{*}$. For $\pi(s) = \operatorname{Ga}(s \mid a_s, 1)$, it is easy to calculate that $\psi(u) = 1 / (1+u)^{a_s}$ and  
\[
\kappa(u, n_h) = \frac{\Gamma(n_h + a_s)}{\Gamma(a_s) (1+u)^{(n_h + a_s)}} .
\]
Therefore, the full conditional density of $u$ is given by 
\[
f_{u \mid \bm{c}} (u) \propto u^{(n-1)} (1 + u)^{-(a_s m + n)}, \qquad u > 0 .
\]

\subsection{Proof of Theorem~\ref{thrm: theorem_2}}

Given the allocation vector $\bm{c}$ and the latent samples $\bm{\theta} = \left\{\theta_1, \ldots, \theta_n\right\}$, let $\bm{\theta}^{*} = \left\{\theta_1^*, \ldots, \theta_{k}^{*}\right\}$ denote the vector of unique values in $\bm{\theta}$. We now characterize the predictive distribution of a new latent sample $\theta_{n+1}$, conditional on the observed samples $\bm{\theta}$.

Conditionally on $\bm{\theta}$ and the auxiliary $u$, with its conditional density $f_{u \mid \bm{c}}(u)$ as defined in Section~\ref{appendix: proof_auxiliary_u}, by applying the Theorem 3 in \citet{beraha2025bayesian}, the predictive distribution of a new latent sample $\theta_{n+1}$ is, when $k \leq m-1$, 
\[
\begin{split}
\operatorname{Pr}(\theta_{n+1} \in B \mid \bm{\theta}, u) & \propto \sum_{h=1}^{k} \frac{\kappa(u, n_h + 1)}{\kappa(u, n_h)} \delta_{\theta_{h}^{*}} (B)   \\
& \qquad + \int_{B} \kappa(u,1) \frac{\mathsf{E} \left[ \exp \left\{- u \widetilde{G}_{(\bm{\theta}^*, \theta)}^{!} (\Theta) \right\} \right]}{\mathsf{E} \left[ \exp \left\{- u \widetilde{G}_{\bm{\theta}^*}^{!} (\Theta) \right\} \right]} \frac{m_{\Phi^{(k+1)}} (\bm{\theta}^*, \theta)}{m_{\Phi^{(k)}} (\bm{\theta}^*)} P_0(\mathrm{d} \theta),
\end{split}
\]
where $B \in \mathcal{X}$.

Assume that the factorial moment measure $M_{\Phi^{(k)}}$ is absolutely continuous w.r.t. the product measure $P_0^k$, where $P_0$ is a non-atomic probability on $(\Theta, \mathcal{X})$ and let $m_{\Phi^{(k)}}$ be the associated Radon-Nikodym derivative. As introduced in Section~\ref{section: preliminaries}, we have the factorial moment measure    
\[
M_{\Phi^{(k)}}(\mathrm{d} \bm{\theta}^{*}) = \det \left\{ K(\theta_i^*, \theta_j^*) \right\}_{i, j=1}^{k} \omega (\mathrm{d} \theta_1^*) \cdots \omega (\mathrm{d} \theta_{k}^*) ,
\]
where $\omega$ is the Lebesgue measure. Denote the non-atomic probability measure $P_0$ on $(\Theta, \mathcal{X})$ as $P_0 (\mathrm{d} \theta) = \omega (\mathrm{d} \theta) / \omega (\Theta)$, then we have
\[
M_{\Phi^{(k)}}(\mathrm{d} \bm{\theta}^{*}) = \det \left\{ K(\theta_i^*, \theta_j^*) \right\}_{i, j=1}^{k} \omega^{k} (\Theta) P_0^{k} (\mathrm{d} \bm{\theta}^*) .
\]
Thus, the associated Radon-Nikodym derivative $m_{\Phi^{(k)}}$ is given by
\[
m_{\Phi^{(k)}} (\bm{\theta}^*) = \det \left\{ K(\theta_i^*, \theta_j^*) \right\}_{i, j=1}^{k} \omega^{k} (\Theta) .
\]
Likewise, if given one more new latent sample $\theta_{n+1} = \theta$, the corresponding factorial moment measure and Radon-Nikodym derivative are as follows,
\[
\begin{split}
& M_{\Phi^{(k+1)}}(\mathrm{d} \bm{\theta}^{*} \mathrm{d} \theta) = \det \left\{ K(\theta_i^*, \theta_j^*) \right\}_{i, j=1}^{k+1} \omega^{k+1} (\Theta) P_0^{k+1} (\mathrm{d} \bm{\theta}^* \mathrm{d} \theta) , \\
& m_{\Phi^{(k+1)}} (\bm{\theta}^*, \theta) = \det \left\{ K(\theta_i^*, \theta_j^*) \right\}_{i, j=1}^{k+1} \omega^{k+1} (\Theta) .
\end{split}
\]
The ratio of Radon-Nikodym derivatives becomes
\[
\begin{split}
\frac{m_{\Phi^{(k+1)}} (\bm{\theta}^*, \theta)}{m_{\Phi^{(k)}} (\bm{\theta}^*)} P_0(\mathrm{d} \theta) &= \frac{\det \left\{ K(\theta_i^*, \theta_j^*) \right\}_{i, j=1}^{k+1} \omega^{k+1} (\Theta)}{\det \left\{ K(\theta_i^*, \theta_j^*) \right\}_{i, j=1}^{k} \omega^{k} (\Theta) } P_0 (\mathrm{d} \theta) \\
&= \det \left(K(\theta,\theta) -  \bm{k}_{\bm{\theta}^{*}}^T (\theta) \bm{K}_{\bm{\theta}^{*},\bm{\theta}^{*}}^{-1} \bm{k}_{\bm{\theta}^{*}}(\theta) \right) \omega(\mathrm{d} \theta), \\
&= \left(K(\theta,\theta) -  \bm{k}_{\bm{\theta}^{*}}^T (\theta) \bm{K}_{\bm{\theta}^{*},\bm{\theta}^{*}}^{-1} \bm{k}_{\bm{\theta}^{*}}(\theta) \right) \omega(\mathrm{d} \theta) ,
\end{split}
\]
by the matrix determinant identity and the fact that determinant of a constant is the constant itself, where $\bm{k}_{\bm{\theta}^{*}}(\theta) = \left(K(\theta, \theta_1^*), \ldots, K(\theta, \theta_{k}^{*})\right)$ and $\bm{K}_{\bm{\theta}^{*},\bm{\theta}^{*}} = \left( K(\theta_t^{*}, \theta_r^{*})\right)_{1 \leq t, r \leq k}$. As discussed in Section~\ref{appendix: proof_of_theorem1}, it follows that 
\[
\frac{\mathsf{E} \left[ \exp \left\{- u \widetilde{G}_{(\bm{\theta}^*, \theta)}^{!} (\Theta) \right\} \right]}{\mathsf{E} \left[ \exp \left\{- u \widetilde{G}_{\bm{\theta}^*}^{!} (\Theta) \right\} \right]}  = \frac{{\psi(u)}^{m - k-1}}{{\psi(u)}^{m - k}} = \frac{1}{\psi(u)} .
\]
Therefore, when $k \leq m-1$, the predictive distribution of a new latent sample $\theta_{n+1}$ is
\[
\begin{split}
\operatorname{Pr}(\theta_{n+1} \in B \mid \bm{\theta}, u) & \propto \sum_{h=1}^{k} \frac{\kappa(u, n_h + 1)}{\kappa(u, n_h)} \delta_{\theta_{h}^{*}} (\theta)   \\
& \qquad + \frac{\kappa(u,1)}{\psi(u)} \left(K(\theta,\theta) -  \bm{k}_{\bm{\theta}^{*}}^T (\theta) \bm{K}_{\bm{\theta}^{*},\bm{\theta}^{*}}^{-1} \bm{k}_{\bm{\theta}^{*}}(\theta) \right), \qquad B \in \mathcal{X} .
\end{split}
\]When $k = m$, meaning all $m$ components have already been used, the predictive distribution simplifies to, 
\[
\operatorname{Pr}(\theta_{n+1} \in B \mid \bm{\theta}, u) \propto \sum_{h=1}^{m} \frac{\kappa(u, n_h + 1)}{\kappa(u, n_h)} \delta_{\theta_{h}^{*}} (B), \quad B \in \mathcal{X}.
\]

\section{Proofs for asymptotic consistency properties}\label{appendix: asymptotic_consistency_properties}

\subsection{Notation and preliminaries}

For a set of densities $\mathcal P$ and $\|\cdot\|_1$ the $L_1$ metric, we denote by $D(\varepsilon, \mathcal P, \|\cdot\|_1)$ the corresponding $\varepsilon$-packing number and let $B(\varepsilon, f_0) = \{f \colon KL(f, f_0) \leq \varepsilon^2, V(f, f_0) \leq \varepsilon^2\}$ for $KL$ the Kullback-Leibler divergence and $V(f, f_0) = \int (\log f(x)/f_0(x))^2 f_0(dx)$.

\subsection{Proof of Theorem \ref{thm:l1_consistency}}

The proof of Theorem \ref{thm:l1_consistency} follows by an application of Theorem 2.1 in \cite{ghosal2001entropies}.
In particular, to apply Theorem 2.1 in \cite{ghosal2001entropies} we need to check the following three assumptions, for $\varepsilon_n = \sqrt{\log n/n}$ and $\mathcal{P}_n \subseteq \mathcal P$:
\begin{align}
    \log D(\varepsilon_n, \mathcal P_n, \|\cdot\|_1) &\leq c_1 n \varepsilon_n^2 \label{eq:a1}\\
    \Pi(\mathcal P \setminus \mathcal P_n) &\leq c_3 e^{-(c_2 + 4) n \varepsilon_n^2} \label{eq:a2}\\
    \Pi(\mathcal B(\varepsilon_n, f_0)) &\geq c_4 e^{-c_2 n \varepsilon_n^2} \label{eq:a3}
\end{align}

Assumption \eqref{eq:a1} is easily verified since we consider a finite-dimensional parametric model, with bounded support $\Theta$ for the means and strictly positive definite covariance matrices with bounded eigenvalues. Hence, letting $\mathcal{P}_n = \mathcal P$ for any $n$, $\log D(\varepsilon_n, \mathcal P_n, \|\cdot\|_1) = O(\log(1/\varepsilon_n)) = O(n \varepsilon_n^2)$.
Similarly, since the model is finite dimensional, \eqref{eq:a2} is trivially satisfied for $\mathcal{P}_n = \mathcal P$.
Hence, we need to check \eqref{eq:a3}. 

By a straightforward generalization of Lemma 4.1 in \cite{ghosal2001entropies} we have that for two mixture densities $f$ and $g$ with $d_H(f, g) = \varepsilon$ sufficiently small
\begin{equation}\label{eq:kl_bound}
     KL(f, g) \leq C_1 \varepsilon^2 \log \frac{1}{\varepsilon}, \quad \text{and} \quad V(f, g) \leq C_2 \varepsilon^2 \left(\log \frac{1}{\varepsilon}\right)^2
\end{equation}
for some constants $C_1$ and $C_2$. Moreover, by triangle inequality, 
\[
    \|f_P - f_0\|_1 \leq C_3 \left(\sum_{h=1}^{k_0} |w_h - w_{0, h}| + \|\phi_h - \phi_{0,h}\| + \|\Sigma_h - \Sigma_{0, h}\|_F \right) + \sum_{h=k_0+1}^m w_h.
\]
Define a neighborhood $\mathcal U(\rho)$ of $f_0$ consisting of all $m$-components Gaussian mixture densities $\tilde f = \sum_{h=1}^m w_h \mathcal N(\phi_h, \Sigma_h)f$ such that $|\tilde{w}_{0, h} -  w_{0, h}| \leq \rho$ for $h=1, \ldots, k_0$ and $\tilde{w}_{0, h} \leq \rho$ for $h > k_0$, $ \|\Sigma_h - \Sigma_{0, h}\|_F \leq \rho$ for $h=1, \ldots, k_0$, and $ \|\phi_h - \phi_{0, h}\| \leq \rho$ for $h=1, \ldots, k_0$ and $\phi_{h+1}, \ldots \phi_{m}$ such that $\|\phi_{j} - \phi_h\| \geq r$. Then for any such $\tilde f \in \mathcal U(\rho)$ we have $\|\tilde f - f_0\| \leq C_4 \rho$.
Choosing $\rho = \rho_n = c_0/(n \log n)$ for a sufficiently small $c_0$ so that $d_H(\tilde f, f_0)$ is small, by \eqref{eq:kl_bound} and the fact that $d_h(\tilde f, f_0)^2 \leq \|\tilde f - f_0\|_1$, we have
\[
    KL(\tilde f, f_0) \leq C_5 \varepsilon_n^2, \qquad V(\tilde f, f_0) \leq C_6\varepsilon_n^2.
\]
Hence, to check \eqref{eq:a3} we need to control the prior mass on the neighborhood $\mathcal U(\rho_n)$. By the independence properties of the prior
\begin{align*}
     \Pi(\mathcal U(\rho_n)) &= \Pi(|\tilde{w}_{0, h} -  w_{0, h}| \leq \rho \text{ for } h=1, \ldots, k_0 \text{  and } \tilde{w}_{0, h} \leq \rho \text{ for }h > k_0) \\
     & \times \Pi(\|\Sigma_h - \Sigma_{0, h}\|_F \leq \rho \text{ for }h=1, \ldots, k_0) \\
     & \times \Pi((\phi_{1}, \ldots, \phi_m) \in U^*)
\end{align*}
where $U^*$ is the set of configurations with $\|\phi_h - \phi_{0, h}\| \leq \rho$ for $h=1, \ldots, k_0$ and $\phi_{h+1}, \ldots \phi_{m}$ such that $\|\phi_{j} - \phi_h\| \geq r$.
Observe that the Dirichlet prior has a continuous, strictly positive density in a neighbourhood of $w_0$. The set $\{|w_h-w_{0,h}|\le \rho_n, h\le k_0;  0\le w_j\le \rho_n, j>k_0\}\subset\Delta_{m-1}$ has $(m-1)$-dimensional volume $\asymp \rho_n^{m-1}$, and the factors $w_j^{\alpha-1}$ for $j>k_0$ contribute an extra $\rho_n^{(m - k_0)\alpha}$. Hence
\[
  \Pi(|\tilde{w}_{0, h} -  w_{0, h}| \leq \rho \text{ for } h=1, \ldots, k_0 \text{  and } \tilde{w}_{0, h} \leq \rho \text{ for }h > k_0) \ge C_w \rho_n^{ d_w},
\]
for $d_w:=(m-1)+\alpha(m-k_0)$.
The inverse–Wishart prior has a positive continuous density on the spectral shell $[\lambda_{\ast},\lambda^\ast]$. Therefore each ball $\{\|\Sigma_h-\Sigma_{0,h}\|_F\le \rho_n\}$ has prior mass $\ge C_v \rho_n^{P}$ with $P=\frac12 d(d+1)$. For $h\le k_0$,
\[
  \Pi(\|\Sigma_h - \Sigma_{0, h}\|_F \leq \rho \text{ for }h=1, \ldots, k_0) \ge C'_v \rho_n^{ P k_0}.
\]
Finally, the joint projection DPP density is strictly positive on the set $U^*$ since all means $\phi_j$ are at least $r$-apart, and $\Theta$ is compact, so that
\[
   \Pi((\phi_{1}, \ldots, \phi_m) \in U^*) \geq C_x \rho_n^{dm}. 
\]
Putting things together and simple algebra leads to \eqref{eq:a3}, which allows us to invoke Theorem 2.1 in \cite{ghosal2001entropies} and conclude the proof of Theorem \ref{thm:l1_consistency}.

\subsection{Proof of Theorem \ref{thm:W1_contraction}}

Fix $\varepsilon>0$ and set
\[
\varepsilon_n:=\sqrt{\frac{\log n}{n}},\qquad 
\delta_n:=n^{-1/2+\varepsilon},\qquad 
D:=\mathrm{diam}(\Xi).
\]
We work on events whose posterior probability tends to $1$ (under $\mathds{P}_0^{(n)}$) on which,  by Theorem \ref{thm:l1_consistency}, for some $M<\infty$, $\|f_P-f_{P_0}\|_{1}\le M\epsilon_n$ and, by Theorem \ref{thm:extra_components}, there exists a set $I\subset{1,\dots,m}$ with $|I|=m-k_0$ and total weight $\sum_{i\in I}p_i\le \delta_n$.
Let $S:=I^c$, so $|S|=k_0$, and decompose $P=(1-\delta)\bar P_S+\delta,\bar P_I$ such that
$$
(1 - \delta)=\sum_{i\in S}p_i=1-\delta,\qquad 
\bar P_S:=\frac{1}{1-\delta}\sum_{i\in S}p_i\delta_{\theta_i}.
$$
and define $\bar P_I$ analogously. Writing $f_P=(1 -\delta) f_{\bar P_S}+\delta f_{\bar P_I}$,by triangle inequality
\[
\|f_{\bar P_S}-f_P\|_1 =  \|(1 - \delta -1)f_{\bar P_S}+\delta f_{\bar P_I}\|_1
\le 2\delta,
\]
which implies 
\begin{equation}\label{eq:l1_triangle}
    \|f_{\bar P_S}-f_{P_0}\|_1\le \|f_P-f_{P_0}\|_{1}+2\delta.
\end{equation}
We further claim
\begin{equation}\label{eq:peel}
    W_1(P,P_0)\le W_1(\bar P_S,P_0)+ D\delta,
\end{equation}
where $D=\text{diam}(\Xi)$. Indeed, the right-hand-side of \eqref{eq:peel} corresponds to $\int_{\theta^2} \rho(\theta, \theta^\prime) \tilde \pi(\ddr \theta \, \ddr \theta^\prime)$ where $\tilde \pi$ is a coupling  transporting the $(1-\delta)$-mass of $P$ using an optimal coupling between $\bar P_S$ and $P_0$, and transporting the remaining $\delta$ mass arbitrarily.

Next, observe that by Theorem 1 in \citet{nguyen2013convergence} and the remark that follows it, since $\|f_{\bar P_S}-f_{P_0}\|_{1}$ is small enough, the $k_0$ components in $S$ can be matched one-to-one with the $k_0$ true components and all parameters lie in a fixed neighborhood of the truth. In this regime, Corollary 1 of \citet{ho2016strong} yields a constant $c_0>0$  such that
\begin{equation}\label{eq:local-inv}
    W_1(\bar P_S,P_0)\le c_0^{-1} \|f_{\bar P_S}-f_{P_0}\|_{1}.
\end{equation}
Combining this with \eqref{eq:l1_triangle}, we have
\[
  W_1(\bar P_S,P_0)\le c_0^{-1}\|f_P-f_{P_0}\|_{L^1}+2c_0^{-1}\delta.
\]
Plugging it into \eqref{eq:peel} yields
\begin{align*}
     W_1(P,P_0) &\le c_0^{-1}\|f_P-f_{P_0}\|_{L^1}+\bigl(2c_0^{-1}+D\bigr)\delta \\
     & \le c_0^{-1}M \varepsilon_n+\bigl(2c_0^{-1}+D\bigr)\delta_n.
\end{align*}
Taking $\delta_n = n^{-1/2 + \epsilon}$ (thanks to Theorem \ref{thm:extra_components}) and $\varepsilon_n = \sqrt{\log n / n}$ we obtain the proof.

\section{Describing repulsiveness in DPPs}\label{appendix: describe_repulsiveness_DPPs}

Whilst DPPs are employed in our mixture method for repulsiveness, there exists a trade-off between repulsiveness and intensity from using them. \citet{moller2021couplings} introduces a global measure to quantify the repulsiveness in DPPs. Consider a DPP $\Phi$ on $\Theta$ with kernel $K$ defined in Definition~\ref{defn: DPP}, for any $x \in \Theta$ with the intensity $\rho(x) = K(x,x) > 0$, there exists a coupling of $\Phi$ and the reduced Palm $\Phi_{x}^{!}$ such that, almost surely, $\Phi_{x}^{!} \subseteq \Phi$ and $\zeta_{x} := \Phi \backslash \Phi_{x}^{!}$ consists of at most one point. As $\Phi$ and $\Phi_{x}^{!}$ have intensities $\rho(\cdot)$ and $\rho(x, \cdot)/\rho(x)$, respectively, the difference $\zeta_x$ has intensity function 
\[
\rho_x(y) := |K(x,y)|^2 / K(x,x), \qquad y \in \Theta,
\]
which is the intensity for the points in $\Phi$ ``pushed out'' by $x$ under the Palm distribution. The probability that $\zeta_x \neq \emptyset$ is thus 
\begin{equation}\label{eq: repulsiveness_measurement}
p_x := \operatorname{Pr}\left(\zeta_x \neq \emptyset \right) = \int_\Theta \rho_x (y) \omega(\mathrm{d} y) = \frac{1}{K(x,x)} \int_{\Theta} |K(x,y)|^2 \omega(\mathrm{d} y) .
\end{equation}
\citet{moller2021couplings} suggest to employ this probability $p_x$ as a global measure of repulsiveness in $\Phi$ when having a point of $\Phi$ at $x$. It is trivial to verify that 
\begin{equation}\label{eq: coupling_intensity}
\rho_x(y) = \rho(y) \left(1 - g(x,y)\right), \qquad y \in \Theta,
\end{equation}
with $g(x,y)$ the pair correlation function introduced in Property~\ref{property: pair_correlation_function}, from which we can see the existence of a trade-off between intensity and repulsiveness. If $p_x$ is fixed, increasing the intensity $\rho(\cdot)$ will lead to an increasing in the pair correlation $g(x,\cdot)$. Therefore, when using $p_x$ as a measure to compare repulsiveness in two DPPs, they should share the same intensity function $\rho$. Then, small (resp. large) values of $p_x$ correspond to a small (resp. large) degree of repulsiveness. For two stationary DPPs with kernels $K_1$ and $K_2$, 
\[
p_x = \rho \int_{\Theta} \left(1 - g(x,y)\right) \omega(\mathrm{d}y) ,
\]
we say that $DPP(K_1)$ is more repulsive than $DPP(K_2)$ if $\rho_1 = \rho_2$ and ${p_x}_1 \geq {p_x}_2$.

For kernels with the spectral representation in~(\ref{eq: general_DPP_kernel_spectral}), it follows that  
\[
\begin{split}
\int_{\Theta} |K(x,y)|^2 \omega (\mathrm{d} y) &= \sum_{j} \sum_{k} \lambda_j \lambda_k \varphi_j(x) \overline{\varphi_k(x)} \int_{\Theta} \overline{\varphi_j(y)} \varphi_k(y) \omega (\mathrm{d} y) \\
&= \sum_{j} \lambda_j^2 \varphi_j(x) \overline{\varphi_j(x)} \leq \sum_{j} \lambda_j \varphi_j(x) \overline{\varphi_j(x)} = K(x,x).
\end{split}
\]
Thus, the projection DPPs, with kernels defined in~(\ref{eq: projection_DPP_kernel_spectral}) with all non-zero eigenvalues $\lambda_j$ equal to 1, have $p_x = 1$, and are the globally most repulsive class within the general DPP family.

\citet{beraha2025bayesian} consider the general class of Gaussian DPPs, employing a Fourier-based spectral approximation over $R = [-\frac{1}{2}, \frac{1}{2}]^d$. The kernel takes the form
\[
K(x,y) = \sum_{j \in \mathds{Z}^d} \lambda_j  \mathrm{e}^{2 i \pi j^T (x-y)}, \qquad (x,y) \in R \times R , 
\]
where eigenvalues $\lambda_j$ are defined by 
\[
\lambda_j  = \rho \frac{\alpha^d}{\pi^{d/2}} \exp{\left(- ||\alpha j||^2\right)}, \qquad \rho > 0.   
\]
Such a DPP exists if $0 < \alpha \leq \alpha_{\max}$, where $\alpha_{\max}^d = \rho^{-1} \pi^{d/2}$. 
In the univariate case where $d = 1$, letting $\alpha = s \alpha_{\max} = s \rho^{-1} \pi^{1/2}$, it can be verified that for this class of Gaussian DPPs, the measurement $p_x$ defined in~(\ref{eq: repulsiveness_measurement}) satisfies
\begin{equation}\label{eq: Gaussian_DPP_measurement}
p_x = \frac{\sum_{j \in \mathds{Z}} \lambda_j^2}{\sum_{j \in \mathds{Z}} \lambda_j} = \frac{s \sum_{j \in \mathds{Z}} \mathrm{e}^{-2 \pi s^2 j^2 / \rho^2}}{\sum_{j \in \mathds{Z}} \mathrm{e}^{- \pi s^2 j^2 / \rho^2}} < 1,
\end{equation}
indicating that Gaussian DPPs are globally less repulsive than projection DPPs.

\subsection{The Gaussian DPPs prior and hyperparameter selection}\label{appendix: Gaussian_DPPs_prior}

\citet{beraha2025bayesian} consider the general class of Gaussian DPPs, employing a Fourier-based spectral approximation over $R = [-\frac{1}{2}, \frac{1}{2}]^d$. The kernel takes the form
\[
K(x,y) = \sum_{j \in \mathds{Z}^d} \lambda_j  \mathrm{e}^{2 i \pi j^T (x-y)}, \qquad (x,y) \in R \times R , 
\]
where eigenvalues $\lambda_j$ are defined by 
\[
\lambda_j  = \rho \frac{\alpha^d}{\pi^{d/2}} \exp{\left(- ||\alpha j||^2\right)}, \qquad \rho > 0.   
\]
Such a DPP exists and has a density with respect to the unit-rate Poisson process on $R$ if $0 < \alpha \leq \alpha_{\max}$, where $\alpha_{\max}^d = \rho^{-1} \pi^{d/2}$. It is easily checked that $\rho$ corresponds to the intensity of the model. Having fixed $\alpha$, increasing the intensity $\rho$ reduces the repulsiveness of the process. For fixed $\rho$, higher values of $\alpha$ increase the repulsiveness of the process. See more similar discussions in \citet{lavancier2015determinantal,ghilotti2025bayesian,beraha2025bayesian}.

Focusing on the case $d = 1$, we let $\alpha = s \alpha_{\max} = s \rho^{-1} \pi^{1/2}$, then the $s \in (0,1)$ takes the interpretation of the repulsiveness of the DPP. Denoting $K(x ,y) = K(x-y
)$ for the stationary kernel $K$, then we have
\[
\begin{split}
\int_{-1/2}^{1/2} |K(x,y)|^2 \omega(\mathrm{d} y) &= \int_{-1/2}^{1/2} |K(x-y)|^2 \omega(\mathrm{d} y) = \int_{x-1/2}^{x+1/2} |K(t)|^2 \mathrm{d} t  \\
&= \int_{-1/2}^{1/2} |K(t)|^2 \mathrm{d} t, \quad K(t) \text{ is 1-periodic}, \\
&= \int_{-1/2}^{1/2} K(t) \overline{K(t)} \mathrm{d} t = \int_{-1/2}^{1/2} \left(\sum_{j \in \mathds{Z}} \lambda_j \mathrm{e}^{2 \pi i j t}\right) \left(\sum_{k \in \mathds{Z}} \lambda_k \mathrm{e}^{-2 \pi i k t}\right) \mathrm{d} t \\
&= \sum_{j \in \mathds{Z}} \sum_{k \in \mathds{Z}} \lambda_j \lambda_k \int_{-1/2}^{1/2} \mathrm{e}^{2 \pi i (j-k) t} \mathrm{d} t 
= \sum_{j \in \mathds{Z}} \lambda_j^2 ,
\end{split}
\]
and the intensity function
\[
\rho = K(x,x) = \sum_{j \in \mathds{Z}} \lambda_j .
\]
Therefore, for this class of Gaussian DPPs, the global measurement $p_x$ defined in~(\ref{eq: repulsiveness_measurement}) satisfies
\[
p_x = \frac{\sum_{j \in \mathds{Z}} \lambda_j^2}{\sum_{j \in \mathds{Z}} \lambda_j} = \frac{s \sum_{j \in \mathds{Z}} \mathrm{e}^{-2 \pi s^2 j^2 / \rho^2}}{\sum_{j \in \mathds{Z}} \mathrm{e}^{- \pi s^2 j^2 / \rho^2}} < 1 ,
\]
where $\lambda_j = s \mathrm{e}^{- \pi s^2 j^2 / \rho^2}$.

\begin{figure}[H]
\centering
\includegraphics[width=1\textwidth]{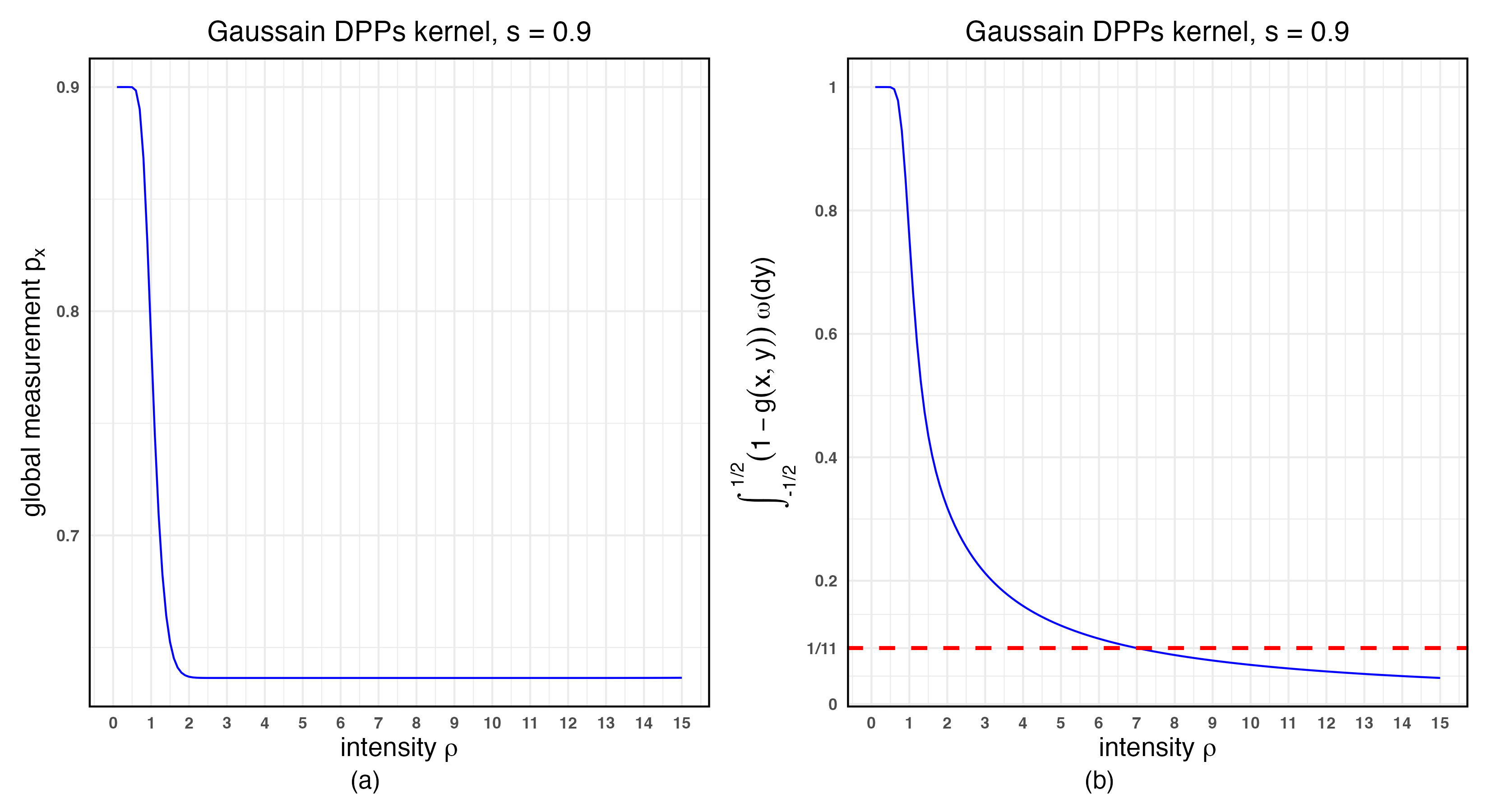}
\caption{Comparison of repulsiveness metrics for Gaussian DPPs with $s = 0.9$. Panel (a): The global repulsiveness measure $p_x$ rapidly decreases and stabilizes around 0.63 as the intensity $\rho$ increases. Panel (b): The integral $\int_{-1/2}^{1/2} \left(1 - g(x,y)\right) \omega(\mathrm{d}y)$, representing net spatial inhibition around location $x$, decreases with $\rho$. The red dashed line indicates the value $1/11$, corresponding to our proposed model with kernel intensity $K(x,x) = 11$. Matching this level of inhibition with Gaussian DPPs requires setting approximately $\rho = 7$.}
\label{figure: Gaussian_DPPs_kernel_para}
\end{figure}

Letting $s = 0.9$ to introduce stronger repulsiveness, Figure~\ref{figure: Gaussian_DPPs_kernel_para} panel (a) illustrates that the global repulsive measurement $p_x$ rapidly converges to approximately 0.63 as the intensity $\rho$ increases. As noted in Section~\ref{sec: Simulation_1}, our proposed method uses $\ell = 5$, resulting in the corresponding kernel intensity $K(x,x) = 1+2\ell = 11$, and hence 
\[
\int_{-1/2}^{1/2} \left(1 - g(x,y)\right) \omega(\mathrm{d} y) = \frac{p_x}{K(x,x)} = \frac{1}{11} ,
\]
which quantifies the net spatial inhibition around location $x$ across the region $R$. To align the Gaussian DPPs with this level of spatial inhibition, Figure~\ref{figure: Gaussian_DPPs_kernel_para} panel (b) suggests selecting its intensity $\rho = 7$, such that 
\[
\int_{-1/2}^{1/2} \left(1 - g(x,y)\right) \omega(\mathrm{d} y) = \frac{\sum_{j \in \mathds{Z}} \mathrm{e}^{-2 \pi s^2 j^2 / \rho^2}}{\left(\sum_{j \in \mathds{Z}} \mathrm{e}^{- \pi s^2 j^2 / \rho^2}\right)^2} \approx \frac{1}{11},
\]
yielding the same integrated repulsion as achieved by our proposed method.

\section{More sampling algorithms in details}\label{appendix: more_algorithms}

\subsection{Projection DPP sampling}\label{appendix: projection_DPP_sampling}

Consider a projection DPP $\Phi \sim \mathrm{pDPP}_{\Theta}(K)$ characterized by a kernel $K$, defined as per~(\ref{eq: our_projection_kernel_transformed}), with cardinality $m = |L^d|$. 
Let $\Theta \subset \mathds{R}^d$ be compact with volume $|\det(A)|$. Algorithm~\ref{algo: projection_DPP_sampling} below describes the step-by-step procedure to simulate points $\{\phi_1, \ldots, \phi_m\}$ from $\Phi$. 

\begin{myalgorithm}[Projection DPP sampling]\label{algo: projection_DPP_sampling} 
\leavevmode
\begin{enumerate}
\item[(1)] Sample $\phi_m$ from the distribution with density $p_m(\phi) = K(\phi,\phi)/m, \phi \in \Theta$.
\item[(2)] Sequentially, for $j = m-1, \ldots, 1:$
\begin{enumerate}
    \item let $\bm{k}_j^T(\phi) = \left(K(\phi, \phi_{j+1}), \ldots, K(\phi, \phi_m)\right)$ and $\bm{K}_{j,j} = \left(K(\phi_t, \phi_r)\right)_{j+1 \leq t,r \leq m}$,
    \item sample $\phi_j$ from the distribution with density
    \begin{equation}\label{eq: projection_DPP_sampling}
        p_j(\phi) = \frac{1}{j} \left(K(\phi,\phi) - \bm{k}_j^T(\phi) \bm{K}_{j,j}^{-1} \bm{k}_j (\phi)\right), \qquad \phi \in \Theta .
    \end{equation}
\end{enumerate}
\item[(3)] Return the point set $\{\phi_1, \ldots, \phi_m\}$. 
\end{enumerate}
\end{myalgorithm}

To practically sample $\phi_j$ from the target density $p_j(\phi)$ in~(\ref{eq: projection_DPP_sampling}),  standard rejection sampling is employed using a uniform proposal distribution $g(\phi) = 1/|\det(A)|$ defined on $\Theta$. Determine the finite constant $M = K(\phi,\phi) / \left(j g(\phi)\right) = m/j$ such that $\sup_{\phi} p_j(\phi) / g(\phi) \leq M$. Generate a proposal $Z \sim g(\cdot)$ and an independent $U \sim \operatorname{Unif}([0,1])$, accepting $Z$ if 
\[
U < \frac{p_j(Z)}{M g(Z)} = 1 - \frac{|\det(A)| \bm{k}_j^T(Z) \bm{K}_{j,j}^{-1} \bm{k}_j (Z)}{m} .
\]
Otherwise, resample $(Z,U)$ until acceptance. 
This Algorithm~\ref{algo: projection_DPP_sampling} can be adapted for use with various projection kernels.

\subsection{Sampling from inhomogeneous projection DPP}\label{appendix: inhomogeneous_projection_DPP_sampling}

To sample non-active atoms $\bm{\phi}^{(na)} = \{\phi_1^{(na)}, \ldots, \phi_{m-|\bm{c}|}^{(na)}\}$ from the inhomogeneous projection DPP $\Phi_{\bm{\phi}^{(a)}}^{!}$ with kernel $K^!_{\bm{\phi}^{(a)}}$ previously defined in~(\ref{eq: conditional_DPP_kernel}) on $\Theta \backslash \bm{\phi}^{(a)}$, we adapt Algorithm~\ref{algo: projection_DPP_sampling} and have the following rejection sampling steps:

\begin{myalgorithm}[Inhomogeneous projection DPP $\Phi_{\bm{\phi}^{(a)}}^{!}$ sampling]\label{algo: conditional_nonactive_atoms_samplings} The proposal is the uniform distribution on $\Theta \backslash \bm{\phi}^{(a)}$,
\begin{enumerate}
\item[(1)] Initialize by sampling $Z$ uniformly on $\Theta \backslash \bm{\phi}^{(a)}$ and $U \sim \operatorname{Unif}([0,1])$. Accept $Z$ and let $\phi_{m-|\bm{c}|}^{(na)} = Z$, if 
\[
U < |\det(A)| K^!_{\bm{\phi}^{(a)}}(Z,Z)/m 
= 1 - |\det(A)| \bm{k}_{\bm{\phi}^{(a)}}^T (Z) \bm{K}_{\bm{\phi}^{(a)},\bm{\phi}^{(a)}}^{-1} \bm{k}_{\bm{\phi}^{(a)}}(Z) / m .
\]
Otherwise, draw another new pair of $(Z, U)$ until acceptance.
\item[(2)] Sequentially sample remaining atoms, for $j = m-|\bm{c}|-1, \ldots, 1:$
\begin{enumerate}
    \item Let ${\bm{k}_j^{\prime}}^{T}(\phi) = \left(K^!_{\bm{\phi}^{(a)}}(\phi, \phi_{j+1}^{(na)}), \ldots, K^!_{\bm{\phi}^{(a)}}(\phi, \phi_{m-|\bm{c}|}^{(na)})\right)$ and 
    
    $\bm{K}^\prime_{j,j} = \left(K^!_{\bm{\phi}^{(a)}}(\phi_t^{(na)}, \phi_r^{(na)})\right)_{j+1 \leq t,r \leq m-|\bm{c}|}$ .
    \item Generate $Z$ uniformly on $\Theta \backslash \bm{\phi}^{(a)}$ and $U \sim \operatorname{Unif}([0,1])$ independent of $Z$. Accept this $Z$ and let $\phi_{j}^{(na)} = Z$, if
    \[
    U < |\det(A)| \left(K^!_{\bm{\phi}^{(a)}}(Z,Z) - {\bm{k}_j^{\prime}}^{T}(Z) {\bm{K}^\prime_{j,j}}^{-1} \bm{k}_j^{\prime}(Z) \right) / m .
    \]
    Otherwise, draw another new pair of $(Z, U)$ until acceptance.
\end{enumerate}
\item[(3)] Return the set $\bm{\phi}^{(na)} = \{\phi_1^{(na)}, \ldots, \phi_{m-|\bm{c}|}^{(na)}\}$. 
\end{enumerate}
\end{myalgorithm}

\subsection{Updating active location atoms in conditional algorithm}\label{appendix: update_active_location_atoms}

In the step for updating the active location atoms $\bm{\phi}^{(a)}$, we employ a MH strategy to update each $\phi_h^{(a)}$, conditional on the remaining active atoms $\bm{\phi}^{(a)} \backslash \{\phi_h^{(a)}\}$ and all non-active atoms $\bm{\phi}^{(na)}$, for $h = 1, \ldots, |\bm{c}|$. As a symmetric proposal distribution,  we use a mixture comprising two components: a normal distribution centered at the current value of $\phi_h^{(a)}$ with a small variance, and a reduced Palm version $\Phi^!_{\bm{\phi}\backslash \{\phi_h^{(a)}\}}$, which is a projection DPP conditional on all remaining atoms. Specifically, with probability 0.9, a candidate $\phi_h^{\dagger}$ is drawn from $\mathcal{N}_d (\phi_h^{\dagger} \mid \phi_h^{(a)}, 0.01 I_d)$; with probability 0.1, the candidate is sampled from the reduced Palm $\Phi^!_{\bm{\phi}\backslash \{\phi_h^{(a)}\}}$.

\begin{myalgorithm}[Active atoms MH updating]\label{algo: active_atoms_MH_updating} To update active atoms $\bm{\phi}^{(a)} = \{\phi_1^{(a)}, \ldots, \phi_{|\bm{c}|}^{(a)}\}$, perform the following steps for $h = 1, \ldots, |\bm{c}|$: 
\begin{enumerate}
\item[(1)] Sample a candidate location $\phi_h^\dagger$ from the symmetric proposal
\[
\phi_h^\dagger \sim 0.9 \mathcal{N}_d(\phi_h^\dagger \mid \phi_h^{(a)}, 0.01 I_d) + 0.1 \Phi^!_{\bm{\phi}\backslash \{\phi_h^{(a)}\}} , \qquad \phi_h^\dagger \in \Theta,  
\]
where the reduced Palm $\Phi^!_{\bm{\phi}\backslash \{\phi_h^{(a)}\}}$ is a projection DPP of cardinality one, and can be sampled using the similar procedure outlined in Algorithm~\ref{algo: conditional_nonactive_atoms_samplings}.
\item[(2)] Let ${\bm{\phi}^{(a)}}^\dagger = \{\phi_1^{(a)}, \ldots, \phi_h^\dagger, \ldots, \phi_{|\bm{c}|}^{(a)}\}$, and compute  the acceptance ratio:
\[
Ratio = \frac{\prod_{i: c_i = h} \mathcal{N}_d \left(y_i \mid \phi_h^{\dagger}, \Sigma_h^{(a)} \right) \times \det\left\{K\left(\phi_t, \phi_r \right)\right\}_{\phi_t, \phi_r \in {\bm{\phi}^{(a)}}^\dagger \cup \bm{\phi}^{(na)}}}{\prod_{i: c_i = h} \mathcal{N}_d \left(y_i \mid \phi_h^{(a)}, \Sigma_h^{(a)} \right) \times \det\left\{K\left(\phi_t, \phi_r \right)\right\}_{\phi_t, \phi_r \in {\bm{\phi}^{(a)}} \cup \bm{\phi}^{(na)}}} .
\]
\item[(3)] Draw $U \sim \operatorname{Unif}([0,1])$. If $U < Ratio$, accept the proposal and set $\phi_h^{(a)} = \phi_h^\dagger$; otherwise, retain the current value $\phi_h^{(a)}$ . 
\end{enumerate}
\end{myalgorithm}

To further illustrate the intuition behind the proposal mechanism, consider the scenario where two active location atoms $\phi_1^{(a)}$ and $\phi_2^{(a)}$ are close to each other and distant from the remaining atoms in $\bm{\phi}$, and suppose we are updating $\phi_1^{(a)}$. If the number of observations allocated to $\phi_1^{(a)}$ is small, we prefer a proposal distribution that places higher probability on locations far from $\phi_2^{(a)}$. Due to the repulsiveness of the projection DPP, such proposals are more likely to be accepted. This behavior is facilitated by drawing from the reduced Palm DPP $\Phi^!_{\bm{\phi}\backslash \{\phi_h^{(a)}\}}$, which inherently avoids placing new atoms near existed ones. On the other hand, if a large number of observations are allocated to $\phi_1^{(a)}$, we desire proposals concentrated near its current location to maintain a good fit to the data. This is achieved through the proposal $\mathcal{N}_d (\phi^{\dagger} \mid \phi_h^{(a)}, 0.01 I_d)$, which encourages small, local moves that preserve the data likelihood.

\subsection{Marginal algorithm in Neal 8 version}\label{appendix: marginal_algorithm_Neal8}

The integral  
\[
\int_{\Theta} \frac{K^!_{\bm{\theta}^*_{-i}}(\theta,\theta)}{|\Omega + (y_i - \theta)(y_i - \theta)^T|^{(1+\tau)/2}} \mathrm{d} \theta
\]
in Algorithm~\ref{algo: marginal_algorithm_Neal2} can become computationally expensive, particularly for multivariate data. To circumvent this difficulty, we adopt the auxiliary variable strategy devised by Neal Algorithm 8 \citep{neal2000markov}. Specifically, we introduce $T$ auxiliary variables to avoid direct evaluation of the integral. The resulting modified algorithm is presented below:

\begin{myalgorithm}[Marginal algorithm: Neal 8 version]\label{algo: marginal_algorithm_Neal8} 
\leavevmode
\begin{enumerate}
\item[(1)] For each observation $i = 1, \ldots, n$, update $(\theta_i, \Delta_i)$ sequentially while keeping $(\bm{\theta}, \bm{\Delta})_{-i}$ fixed.
\begin{enumerate}
    \item[(a)] Sample $ T$  i.i.d. auxiliary variables, where $q$ denotes the number of unique values in $(\bm{\theta}, \bm{\Delta})_{-i}$. For $t = 1, \ldots, T$, sample $\theta_{q+t}^*$ i.i.d. from 
    \[
    \operatorname{Pr}(\theta_{q+t}^* \in \mathrm{d} \theta \mid \bm{\theta}_{-i}^*) \propto K^!_{\bm{\theta}^*_{-i}}(\theta,\theta) \mathrm{d} \theta ,
    \]
    using rejection sampling, and sample corresponding covariances as 
    \[
    \Delta_{q+t}^* \simiid \operatorname{InvWi}(\Delta \mid \tau, \Omega) .
    \]
    \item[(b)] Set $(\theta_i, \Delta_i)$ equal to $(\theta_j^*, \Delta_j^*)$ with probability proportional to :

    \begin{equation*}
    \begin{cases}
    \left(n_j^{(-i)} + a_s\right) \cdot \mathcal{N}_d (y_i \mid \theta_j^*, \Delta_j^*), & \text{ for } j = 1, \ldots, q, \\
    \frac{1}{T} a_s \cdot \mathcal{N}_d (y_i \mid \theta_j^*, \Delta_j^*), & \text{ for } j = q+1, \ldots, q+T.
    \end{cases}
    \end{equation*}
\end{enumerate}
\item[(2)] Repeat step (2) from Algorithm~\ref{algo: marginal_algorithm_Neal2} as previously defined.
\end{enumerate}
\end{myalgorithm}

\section{MCMC mixing in 1-dimensional simulation}\label{appendix: mcmc_mixing_simulation}

The trace plots in Figure~\ref{figure: simulation_mcmc_mixing_efficiency} illustrate the evolution of the estimated number of clusters and the entropy of the partitions across iterations. Table~\ref{table: Simulation_1_ESS} reports the effective sample size (ESS) per iteration
and per second for each algorithm or method, focusing on the estimated number of clusters
and the partition entropy. 

\begin{figure}[H]
\centering
\includegraphics[width=\textwidth]
{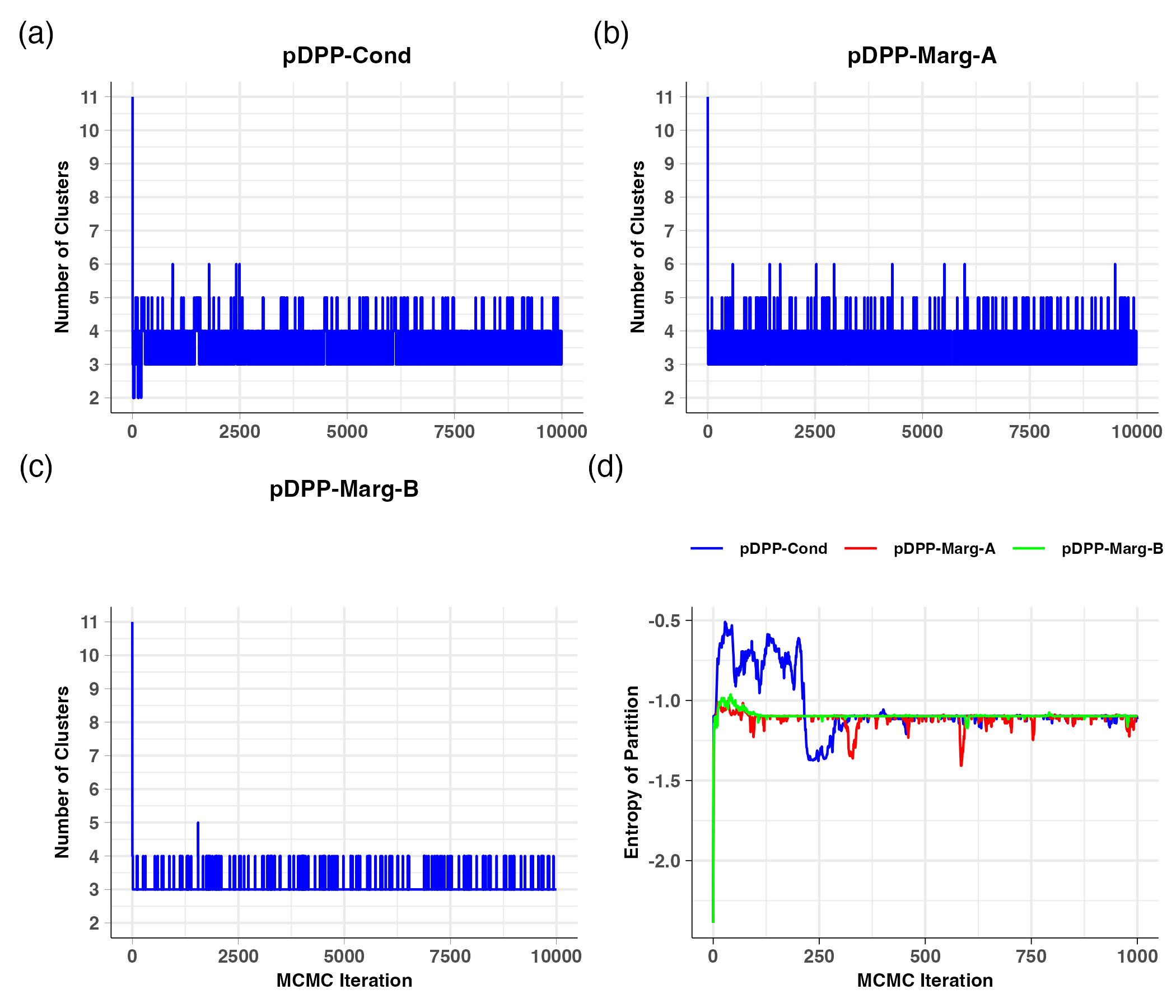}
\caption{Simulation study 1: Trace plots for the estimated number of clusters (a, b, c) and partition entropy (d) for the proposed MCMC Algorithms~\ref{algo: algorithm_1},~\ref{algo: marginal_algorithm_Neal2},~\ref{algo: marginal_algorithm_Neal8}. See Section~\ref{sec: Simulation_1}.}
\label{figure: simulation_mcmc_mixing_efficiency}
\end{figure}

\begin{table}[H]
\centering
\small
\resizebox{\linewidth}{!}{%
\begin{tabular}{lcccc}
\toprule
\textbf{Algorithm} & \multicolumn{2}{c}{\textbf{Number of Clusters}} & \multicolumn{2}{c}{\textbf{Partition Entropy}} \\
\cmidrule(lr){2-3} \cmidrule(lr){4-5}
& ESS/Iter & ESS/Sec & ESS/Iter & ESS/Sec \\
\midrule
\emph{pDPP-Cond}   & 0.105 (0.087) & 51.992 (43.817) & 0.022 (0.019) & 10.948 (9.121) \\
\emph{pDPP-Marg-A}  & 0.131 (0.103) & 3.065 (2.359) & 0.024 (0.022) & 0.572 (0.514) \\
\emph{pDPP-Marg-B}  & 0.137 (0.142) & 8.216 (8.358) & 0.036 (0.044) & 2.186 (2.660)
\\
\midrule
\emph{L-ensembles}  & 0.017 (0.034) & 1.998 (4.662) & 0.009 (0.007) & 0.936 (0.968)
\\
\emph{Gaussian DPP}  & 0.003 (0.002) & 4.866 (2.631) & 0.004 (0.002) & 5.819 (3.579)
\\
\bottomrule
\end{tabular}
}
\caption{Effective Sample Size (ESS) computed per iteration and per second for each algorithm, comparing statistical efficiency and computational speed in estimating the number of clusters and partition entropy, based on 5,000 MCMC samples following the first 5,000 samples as burn-in. Each entry represents the mean (standard error) across 100 replications. Results above the horizontal line correspond to Simulation Study 1; results below correspond to Simulation Study 2. See Section~\ref{sec: Simulation_1}.}
\label{table: Simulation_1_ESS}
\end{table}

\section{ERP clustering using DPM model on functional PCA scores}\label{appendix: ERPs_clust_fPCA_DPM}

In this section, we present the clustering results obtained using the DPM model applied to the ERP component scores extracted via functional PCA, as discussed in Section~\ref{sec: ERPs_data_analysis} of the main paper. The three figures below correspond to different values of the total mass parameter in the DPM prior, which affects the number of inferred clusters. As shown, varying this parameter leads to differences in the number and composition of clusters; however, in all cases, the resulting configurations remain meaningless and lack interpretability. These results provide a useful baseline for comparison with the projection DPP prior mixture models presented in the main paper, which offers more robust and interpretable clustering results.

\begin{figure}[H]
\centering
\includegraphics[width=\textwidth]{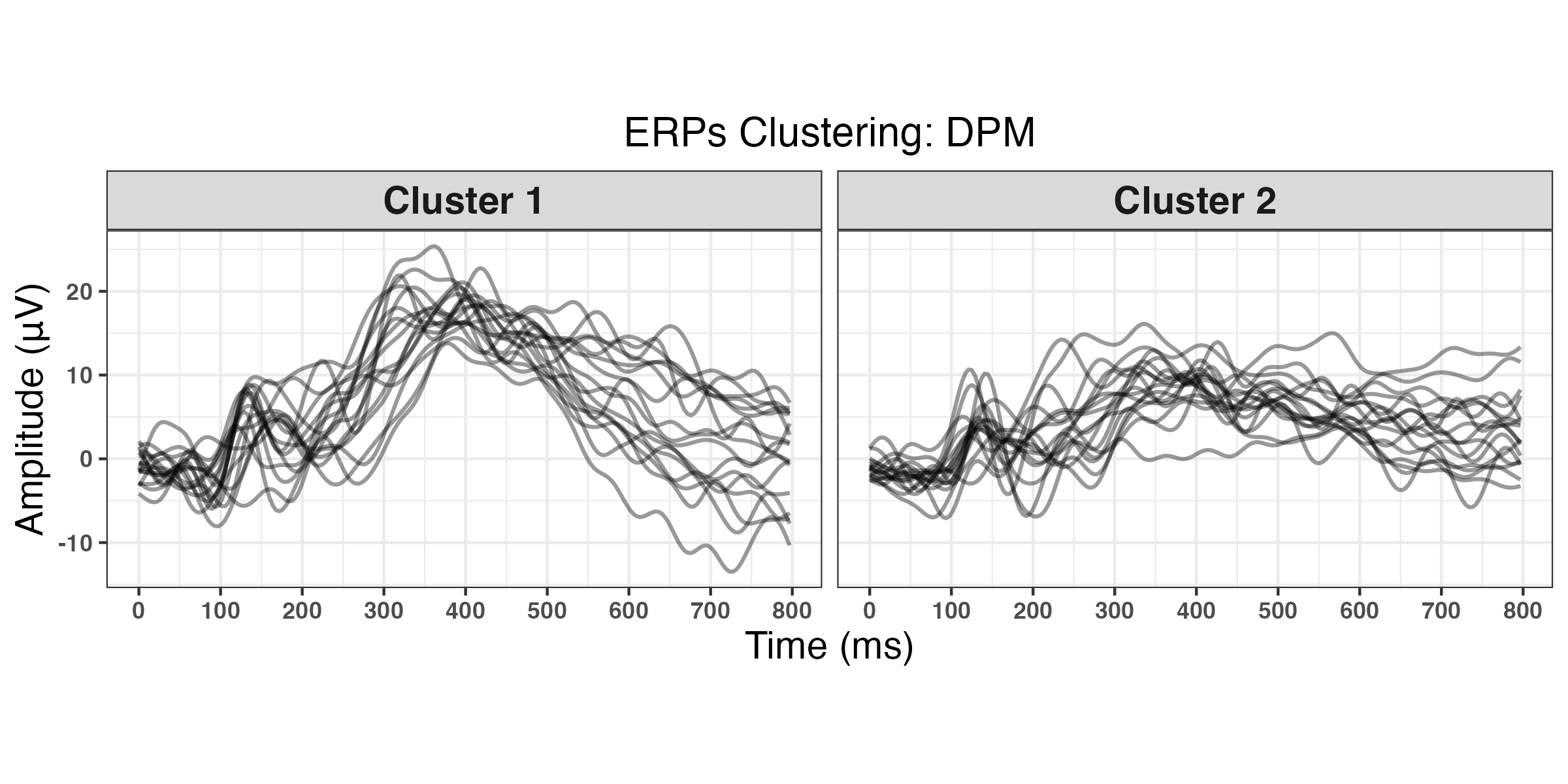}
\caption{\emph{Real data analysis}, clustering of 34 ERPs by applying the DPM model to the functional PCA scores with total mass parameter 0.01. The model partitions the ERPs into two clusters with sizes 17 and 17. See details in Section~\ref{sec: ERPs_data_analysis} of the main paper.}
\label{figure: fPCA_DPM_mass0_0_1}
\end{figure}

\begin{figure}[H]
\centering
\includegraphics[width=\textwidth]{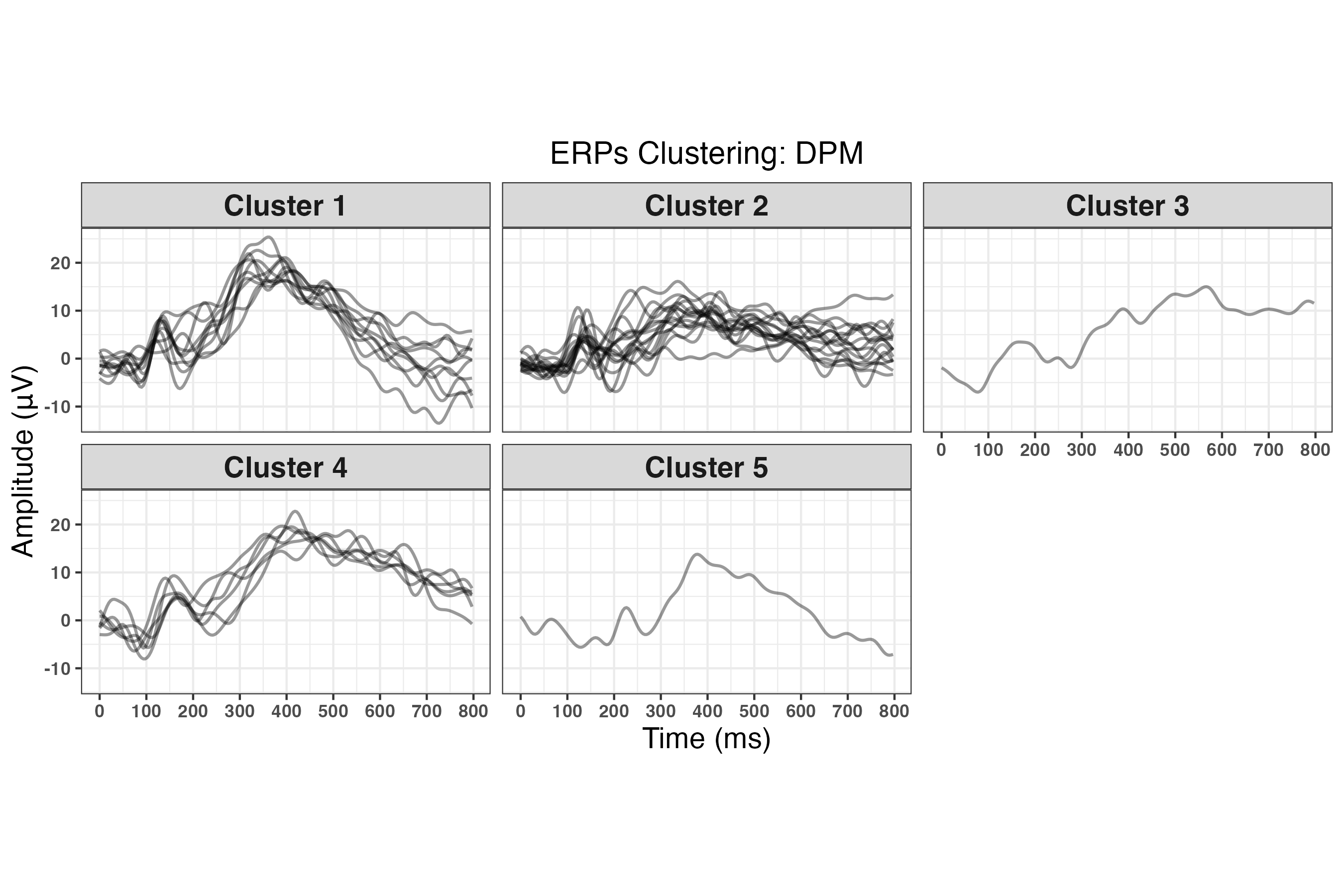}
\caption{\emph{Real data analysis}, clustering of 34 ERPs by applying the DPM model to the functional PCA scores with total mass parameter 0.05. The model partitions the ERPs into five clusters with sizes 10, 16, 1, 6, and 1. See details in Section~\ref{sec: ERPs_data_analysis} of the main paper.}
\label{figure: fPCA_DPM_mass0_0_5}
\end{figure}

\begin{figure}[H]
\centering
\includegraphics[width=\textwidth]{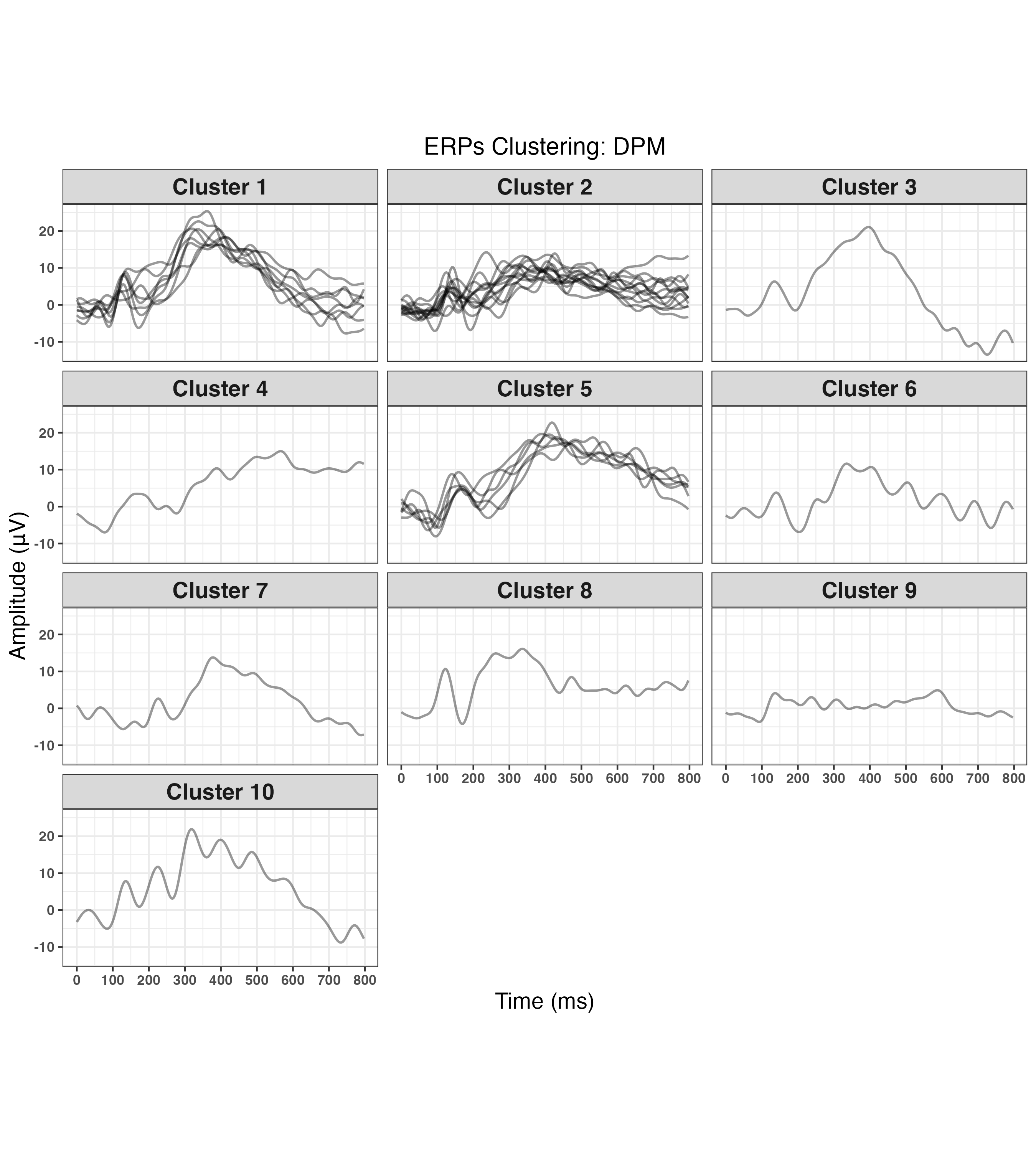}
\caption{\emph{Real data analysis}, clustering of 34 ERPs by applying the DPM model to the functional PCA scores with total mass parameter 0.2. The model partitions the ERPs into ten clusters with sizes 8, 13, 1, 1, 6, 1, 1, 1, 1, and 1. See details in Section~\ref{sec: ERPs_data_analysis} of the main paper.}
\label{figure: fPCA_DPM_mass0_2}
\end{figure}

\section{Probabilistic PCA with projection DPP prior}\label{appendix: one-step_method}

In Section~\ref{sec: ERPs_data_analysis} of the main paper, we adopt a two-step procedure for the ERP data analysis. We first apply functional principal component analysis (fPCA) to the ERP waveforms to extract low-dimensional component scores, and then use these scores as inputs to our proposed mixture model with projection DPP priors. In this supplementary section, we introduce a more coherent, fully Bayesian alternative, a one-step approach that integrates probabilistic functional PCA directly into the projection DPP mixture framework. Rather than treating the component scores as fixed, we model them as latent variables and jointly infer them along with the clustering structure. We describe the model formulation, outline the corresponding MCMC algorithm, and present the results of applying this integrated method to the ERP data. For comparison, we also implement an analogous one-step model that combines probabilistic PCA with the Dirichlet process mixture (DPM) model, and report the results alongside ours.

We let an $p \times n$ data matrix $Y$ denote the functional data collected from $n = 34$ subjects recorded at $p = 205$ time points. The data $Y$ are centered by removing both row and column means. As in model-based probabilistic principal component analysis (PCA), we model the functional data as, for $i = 1, \ldots, n$, 
\[
y_i = Q x_i + \epsilon_i, \qquad \epsilon_i \sim \mathcal{N}_p (0, \sigma_{\epsilon}^{2} I_p) ,
\]
where $Q$ is an orthonormal matrix on the Stiefel manifold $\mathcal{V}(d, p)=\left\{\bm{Q} \in \mathds{R}^{p \times d} \mid \bm{Q}^{\top} \bm{Q}=\bm{I}_d\right\}$ with $d \leq p$; see examples in \citet{hoff2009simulation} and \citet{jauch2021monte}. We assume that the orthonormal matrix $Q$ is uniformly distributed on the Stiefel manifold, whose probability measure is simply proportional to surface area on the $pd-d(d+1)/2$ dimensional surface of $\mathcal{V}(d,p)$ in $\mathds{R}^{pd}$. The columns in $Q$ are orthonormal bases spanning a latent subspace of the original data space, and $x_i$ are coordinates of subject $i$ in that latent subspace. Besides $\sigma^2_{\epsilon} \sim \text{InvGa} (a_\epsilon, b_\epsilon)$, we further assume the latent principal components $\{x_1, \ldots, x_n\}$ follow our proposed mixture of projection DPPs discussed in Section~\ref{sec: thms_asymptotic_consistency} and~\ref{sec: algorithms}. That is, $x_i \simind \mathcal{N}_d (\cdot \mid \theta_i, \Delta_i)$ and the component locations follow a projection DPP prior.

The MCMC iterates as following steps:

\begin{myalgorithm}[Combining probabilistic PCA and the mixture of projection DPPs]\label{algo: probPCA_projDPP_marginal} 
\leavevmode
\begin{enumerate}
\item[(1)] Sample $Q$ from a matrix von Mises-Fisher distribution with parameter $\frac{1}{\sigma_{\epsilon}^{2}} Y X^T$,
\[
p(Q | X, Y, \sigma_\epsilon^2) \propto \exp\{\frac{1}{\sigma_{\epsilon}^{2}}\mathrm{tr}(X Y^T Q)\} ,
\]
where $X$ is a $d \times n$ matrix with columns $\{x_1, \ldots, x_n\}$.
\item[(2)] Let $\bm{z}$ denote the vector of membership, where $x_i$ is assigned with membership $z_i$. For $i = 1, \ldots, n$, we sample $x_i$ from 
\[
x_i | \text{rest} \simind \mathcal{N}_{d} \left(\left(\frac{1}{\sigma_{\epsilon}^2} I_d + \Delta_{z_i}^{-1}\right)^{-1} \left(\frac{1}{\sigma_{\epsilon}^2} Q^T y_i + \Delta_{z_i}^{-1} \theta_{z_i} \right), \left(\frac{1}{\sigma_{\epsilon}^2} I_d + \Delta_{z_i}^{-1}\right)^{-1}\right) .
\]
\item[(3)] Update $(\theta_i, \Delta_i)$ for each observation $i = 1, \ldots, n$, following Algorithm~\ref{algo: marginal_algorithm_Neal8}. 
\item[(4)] Sample the global error variance $\sigma_\epsilon^2$ from its posterior
\[
\sigma_\epsilon^2 \sim \text{InvGa} \left(a_\epsilon + \frac{np}{2}, b_\epsilon + \frac{1}{2} \left \lVert Y - Q X \right \lVert_F^2 \right) ,
\]
given its prior as $\sigma_\epsilon^2 \sim \text{InvGa}(a_\epsilon, b_\epsilon)$ , where $\left \lVert \cdot \right \lVert_F$ is the Frobenius norm.  
\end{enumerate}
\end{myalgorithm}

We reduce the number of dimensions by letting $d = 4$ in the probabilistic PCA. To set up the hyperparameters, we first apply standard functional PCA to the data matrix $Y$ and retain the first $d$ principal component score vectors, denoted as $x^\dagger_i \in \mathbb{R}^d, i = 1, \ldots, n$. Then, we transform the unit hypercube $R = [-\frac{1}{2}, \frac{1}{2}]^d$ to the space $\otimes_{e=1}^{d} [\bar{x}^\dagger_e - \zeta^\dagger_e,  \bar{x}^\dagger_e + \zeta^\dagger_e]$ that covers all the principal component scores,  where $\bar{x}^\dagger_e = \frac{1}{n} \sum_{i=1}^{n} x^\dagger_{ie}$ and $\zeta^\dagger_e = \max_{i=1,\ldots,n} |x^\dagger_{ie} - \bar{x}^\dagger_{e}|$ for $e = 1, \ldots, d$. For the rest hyperparameters, we set $\sigma_\epsilon^2 \sim \mathrm{InvGa}(a_\epsilon = 2, b_\epsilon = 1), a_s = 5,$ and $\Sigma_h \simiid \mathrm{InvWi}\left(\tau, \Omega\right)$, where $\tau = d + 2$ and $\Omega$ is the sample covariance matrix of the functional PCA score $x_i^\dagger$'s. We let $\ell = 1$ that results in $m = (2 \ell + 1)^d$ atoms in total. We run 10,000 MCMC iterations and discard the first 5,000 as burn-in. Based on the MCMC posterior samples, we compute the point estimate of the clustering by using the SALSO greedy search method to minimize the posterior expectation of the Variation of Information loss \citep{dahl2022search}.

\begin{figure}[H]
\centering
\includegraphics[width=\textwidth]{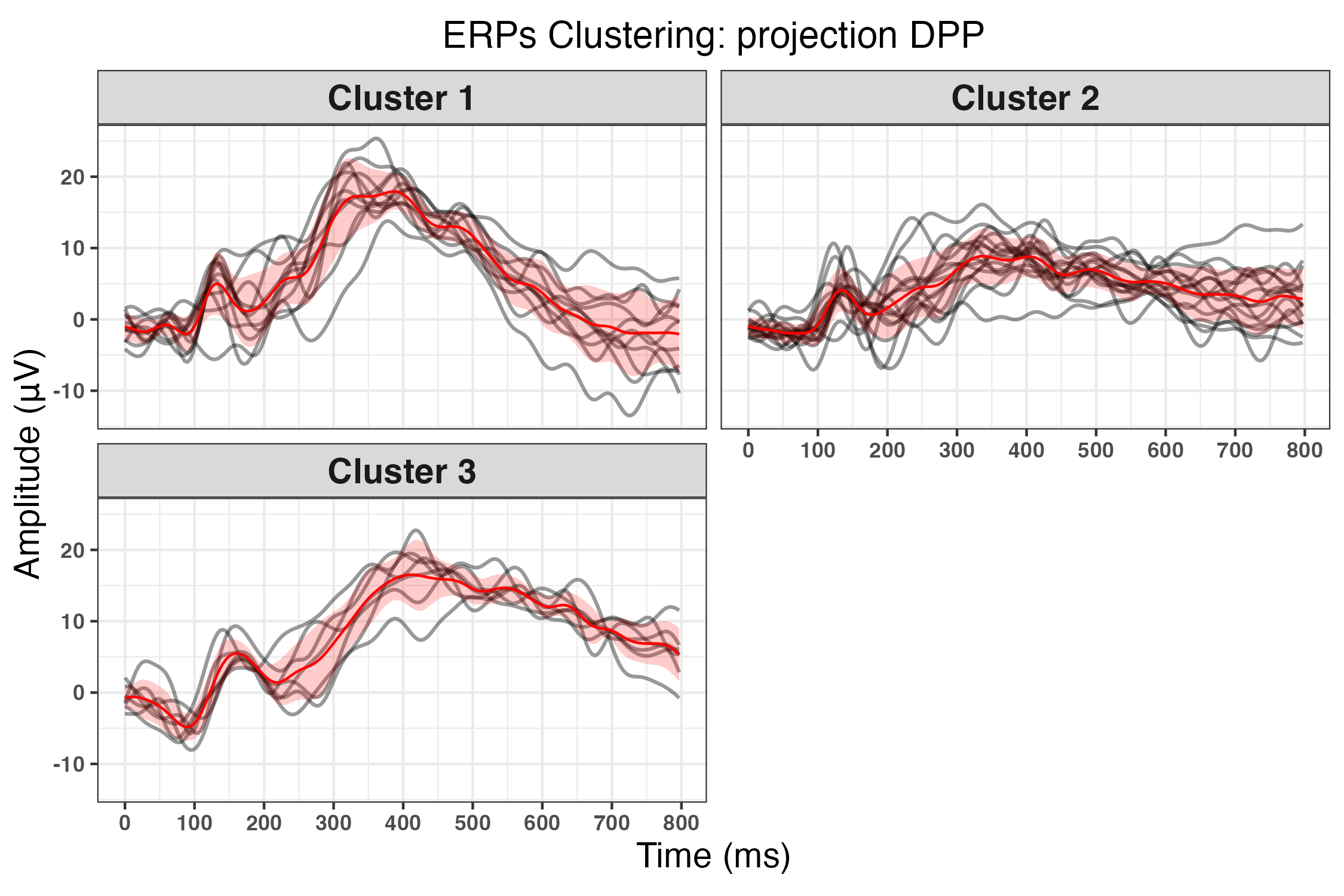}
\caption{Clustering of 34 ERPs by combining probabilistic functional PCA and projection DPP. The model partitions the ERPs into three clusters with sizes 11, 16, and 7. The red curve represents the averaged ERP waveform within each cluster, and the red shaded area indicates the $\pm$ 1 standard deviation bandwidth.}
\label{figure: clusts_probPCA_pDPP_34subjects}
\end{figure}

Figure~\ref{figure: clusts_probPCA_pDPP_34subjects} displays the clustering result. The proposed method identifies three well-separated clusters, with sizes 11, 16, and 7, respectly. Importantly, no singleton or overly small clusters were generated, illustrating the repulsive effect of the projection DPP prior in discouraging redundant clusters. For each cluster, Figure~\ref{figure: clusts_probPCA_pDPP_34subjects} also displays the average ERP waveform (red curve) and a $\pm$ 1 standard deviation bandwidth (red shade). The clustering reveals meaningful neuro-cognitive heterogeneity. As shown, Cluster 1 exhibits a strong positive peaking around 300 ms, resembling the well-known P300 component, typically associated with stimulus evaluation and attention allocation, suggesting that subjects in this group demonstrate high cognitive engagement and efficient processing. Cluster 2 displays a much flatter trajectory with subtle positive shifts, indicative of diffuse or attenuated responses, reflecting reduced arousal or slower cognitive integration. Cluster 3 shows a clear positive peak emerging slightly later, around 400 ms, with a pronounced rise but a more gradual and shallow decline compared to Cluster 1, possibly reflecting a prolonged engagement or delayed disengagement with the stimulus. These findings highlight the strength of the proposed method
in uncovering interpretable clusters from complex brain data, offering insights for subsequent neuro-cognitive analysis.

For comparison, we also combine the probabilistic PCA and DPM model and apply it to the same set of 34 ERP waveforms. As shown in Figure~\ref{figure: clusts_probPCA_DPM_34subjects}, it produces a cluster of 31 waveforms along with three singletons. This clustering configuration does not
well-separate the waveforms and fails to provide us an interpretable result.

\begin{figure}[H]
\centering
\includegraphics[width=\textwidth]{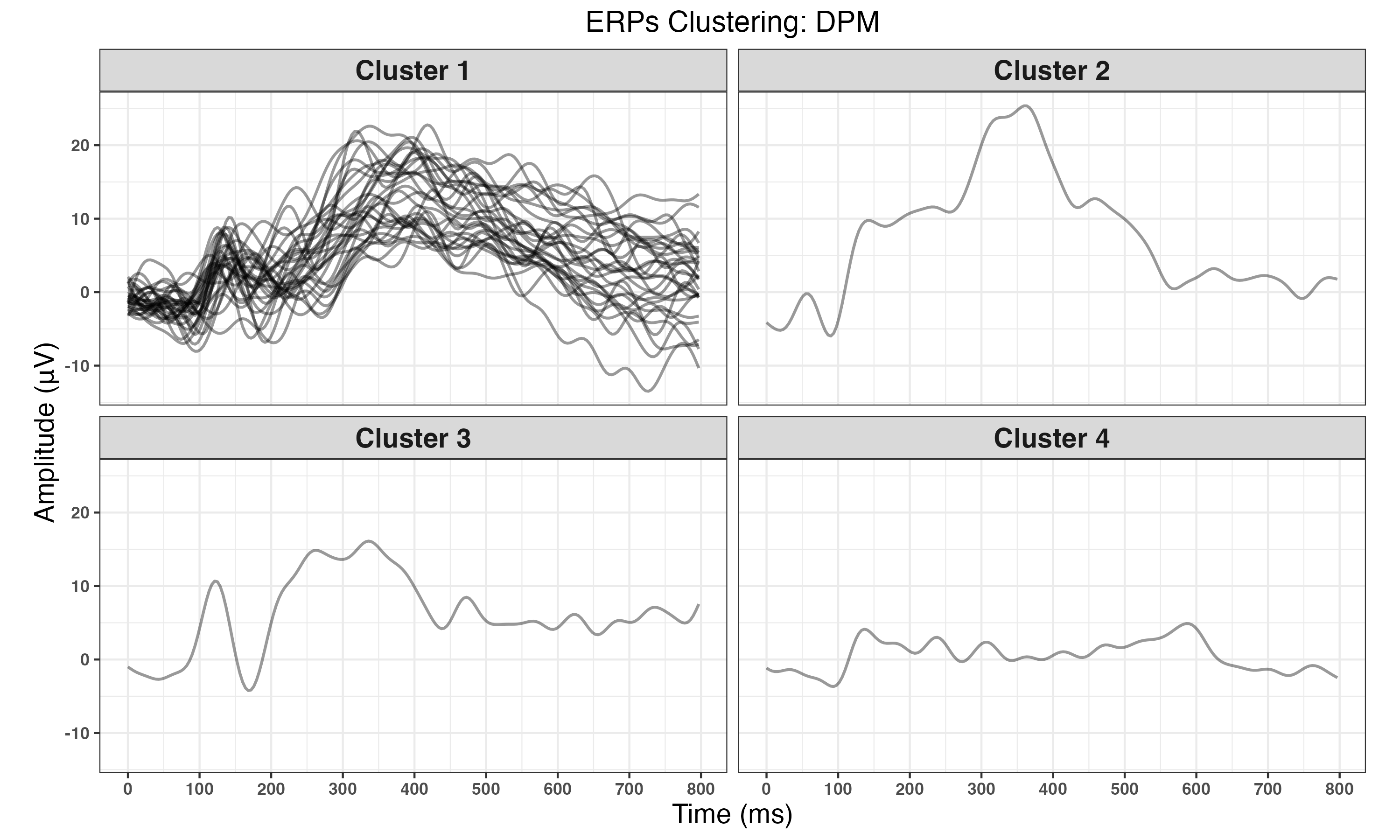}
\caption{Clustering of 34 ERPs by combining probabilistic functional PCA and DPM model, with total mass parameter as 40. The model partitions the ERPs into four clusters with sizes 31, 1, 1, and 1.}
\label{figure: clusts_probPCA_DPM_34subjects}
\end{figure}

\end{document}